\documentclass[12pt]{article}
\usepackage{epsf}
\newcommand{\tr}{\hbox{tr}}

\newcommand{\bm}[1]{\mbox{\boldmath #1}}

\newcommand{\plaq}{\displaystyle \prod _{\tilde{l} \in \partial
\tilde{P}} \hspace*{-6.4mm} \bigcirc \ U(\tilde{l})}
\newcommand{\simpplaq}{\displaystyle \prod \hspace*{-5.1mm} \bigcirc \ U(\tilde{l})}
\newcommand{\sixj}[6]{
\renewcommand{\arraystretch}{1}
  \left\{
   \begin{array}{ccc}
    #1 & #2 & #3 \\
    #4 & #5 & #6
   \end{array}
\right\}}
\newcommand{\threej}[6]{
\renewcommand{\arraystretch}{1}
 \left(
  \begin{array}{ccc}
    #1 & #2 & #3 \\
    #4 & #5 & #6
  \end{array}
 \right)}
\renewcommand{\theequation}{\arabic{section}.\arabic{equation}}

\begin{document}
\renewcommand{\theequation}{\arabic {section}.\arabic{equation}}
\setlength{\baselineskip}{7mm}
\begin{titlepage}
\begin{flushright}
EPHOU-99-0015 \\
November, 1999
\end{flushright}
 
\vspace{15mm}
\begin{center} 
{\Large 4-dimensional $BF$ Gravity on the Lattice}

\vspace{1cm}

 {\sc Noboru Kawamoto}\footnote{kawamoto@particle.sci.hokudai.ac.jp},
 {\sc Noriaki Sato}\footnote{noriaki@particle.sci.hokudai.ac.jp}
 and {\sc Yukiya Uchida}\footnote{uyukiya@particle.sci.hokudai.ac.jp}
 \\
{\it{ Department of Physics, Hokkaido University }}\\
{\it{ Sapporo, 060-0810, Japan}}\\
\end{center}
\vspace{2cm}

\begin{abstract}

We propose the lattice version of $BF$ gravity action whose partition function 
leads to the product of a particular form of 15-$j$ symbol which 
corresponds to a 4-simplex. 
The action is explicitly constructed by lattice $B$ field defined on 
triangles and link variables defined on dual links and is shown to 
be invariant under lattice local Lorentz transformation and Kalb-Ramond 
gauge transformation. 
We explicitly show that the partition 
function is Pachner move invariant and thus topological.  
The action includes the vanishing holonomy constraint which
can be interpreted as a gauge fixing condition. 
This formulation of lattice $BF$ theory 
can be generalized into arbitrary dimensions. 

\end{abstract}

\vspace{5 cm}

\begin{tabular}{lcl}
\small
 PACS code &:& 04.60.Nc, 11.15.Ha \\
 Keywords &:&  Lattice gravity, BF theory, Quantum gravity
\end{tabular}

\end{titlepage}

\setcounter{equation}{0}

\section{Introduction}

In the mid 70's nobody expected that it was possible to evaluate the hadron 
masses by any means in the near future. 
The later developments of lattice QCD made it possible and gave us an idea 
that non-perturbative formulation of gauge theory on the lattice is 
crucial for the realistic numerical analyses of the phenomena based on the 
field theory like QCD. 
As for the lepton and quark masses we don't have any possible formulation 
at this moment. 
It would, however, be not unreasonable to expect that gravity theory might 
play a crucial role in the realistic unified model leading to the lepton 
and quark spectrum calculations. 
On the other hand it is quite natural to expect that some means
to evaluate phenomenological numbers numerically is obviously needed. 
We believe that a lattice theory will play again a crucial role in the 
formulation. 
This line of thinking naturally encourages us to find a formulation of 
gravity theory on the lattice. 

In 2-dimensional gravity the dynamical triangulation of lattice 
theory was very successful analytically%
\cite{2dgravity-anlitic}\cite{2dgravity-anlitic2} and 
numerically\cite{KKSW}\cite{2dgravity-numerical} to evaluate
the fractal dimensions\cite{KKMW}. 
The matrix model for the dynamical triangulation of random surface
and the Liouville theory are the analytic means on one hand 
while the recursive sampling
method\cite{KKSW}\cite{2dgravity-numerical}\cite{Migdal} and the 
flip flop moves by Monte-Carlo are the numerical means 
on the other in 2-dimensional quantum gravity.
In general there is no local action to describe
the 2-dimensional quantum gravity.

In 3 dimensions Einstein gravity has been successfully formulated 
by Chern-Simons action as a gauge theory of $ISO(2,1)$ group\cite{Witten}.
After Turaev and Viro\cite{Turaev-Viro} proposed the possibility 
based on the analyses of $q$-deformed version of Ponzano-Regge model 
that the Ponzano-Regge\cite{Ponzano-Regge} model formulated by 6-$j$ 
symbols may be the lattice version of Chern-Simons gravity, there are 
several authors to try to prove the connection of these two 
formulations\cite{Ooguri-Sasakura}\cite{PR-related}.
The proofs of the connection has been, however, indirect. 
In our recent paper we have shown by explicit calculations that the 
partition function of the lattice 
version of Chern-Simons action indeed leads to the Ponzano-Regge 
model\cite{KNS}.

In 4 dimensions Ooguri pointed out the possibility that the partition 
function of $BF$ theory\cite{Horowitz} leads
to the partition function constructed from 
15-$j$ symbols\cite{Ooguri}.
This is quite parallel to the 3-dimensional case where the Chern-Simons 
gravity theory is associated with the Ponzano-Regge model. 
This 4-dimensional proposal triggered a new way of
defining 4-dimensional quantum gravity theory by spin foam\cite{Baez}
and a new type 
of 4-dimensional topological invariant
by the quantum $q$-deformed formulation%
\cite{Crane-Yetter}\cite{Crane-K-Yetter}. 

There are plausible arguments that the partition function
of the continuum $BF$ theory and 15-$j$ symbols are related,
but there is no explicit derivation to connect the two models.
Here in this paper we first introduce the lattice version of $BF$ theory 
and explicitly show that the partition function of the lattice $BF$ theory 
leads to the product of 15-$j$ symbols. 
We then show explicitly that the particular combination of 15-$j$ product 
is Pachner move invariant and thus the partition function is independent 
how the 4-dimensional simplicial manifold is divided by the Pachner moves 
and is thus topological. 
Then the continuum limit of the lattice $BF$ theory can be taken 
analytically and leads to the continuum $BF$ action
since the partition function is division independent 
of the 4-dimensional simplicial manifold. 

In formulating lattice gravity theory, 
we intend to couple the formulation of the lattice QCD
and Regge calculus\cite{Regge}.
It has already been proposed that 
the link variable of lattice QCD can be used to generate
the curvature on the simplexes where Regge calculus claims
as the location of gravitational
curvature\cite{Kawamoto-Nielsen}\cite{Dadda}.

It is known that the $BF$ theory is not equivalent to the Einstein 
gravity but if the 2-form $B$ field is related to the vierbein as 
$B=*(e\wedge e)$, then the $BF$ action leads to the Einstein-Hilbert 
action\cite{Plebanski}. 
It is also known that the metric can be written by $B$ field directly
if some constraints are fulfilled.
In this case the Einstein gravity will be formulated by 2-form 
field $B$ and equations of motion. 
In this sense the $BF$ theory is closely related to
the gravity theory but in the limited sense. 
The lattice version of $BF$ theory is, therefore, a very good starting point 
to the realistic 4-dimensional lattice gravity theory. 

One of the authors (N.K.) and Watabiki discovered the formulation to 
generalize the standard Chern-Simons action
into arbitrary dimensions by employing 
all degrees of form as gauge fields and parameters\cite{GCS}. 
It was shown that the topological gravity and the topological
conformal gravity can be formulated by 2- and 4-dimensional
generalized Chern-Simons action\cite{GCSG}.
We show concrete expressions of the generalized Chern-Simons 
actions in 2, 3 and 4 dimensions where we omit fermionic degrees 
for simplicity, 
\begin{eqnarray*}
S_2 &=& - \int \mbox{Tr}\{\phi(d\omega + \omega^2) + \phi^2 B\}, \\
S_3 &=& - \int \mbox{Str} \left\{
			   \frac{1}{2}\omega d\omega + \frac{1}{3}\omega^3 -
              \phi(d B + [\omega,B]) + \phi^2 \Omega \right\}, \\
S_4 &=& - \int \mbox{Tr}\{ B(d\omega + \omega^2) + \phi(d \Omega + 
               \{\omega, \Omega\}) + 
             \phi B^2 + \phi^2 H \}, 
\end{eqnarray*} 
which are invariant under the following gauge transformations: 
\begin{eqnarray}
\delta \phi &=& [\phi,v],                      \nonumber \\
\delta \omega &=& dv + [\omega,v] - \{\phi,u\},    \nonumber \\
\delta B &=& du + \{\omega,u\} + [B,v] + [\phi,b],      \nonumber \\
\delta \Omega &=& db + [\omega,b] + [\Omega,v] - \{B,u\} + 
\{\phi,U\},\nonumber \\
\delta H &=& -dU-\{\omega,U\}+\{\Omega,u\}+[H,v]+[B,b]+[\phi,V],  
\nonumber
\end{eqnarray} 
where $[~,~]$ and $\{~,~\}$ are, respectively, commutator and
anti-commutator, and 
$\phi$, $\omega$, $B$, $\Omega$, $H$ and
$v$, $u$, $b$, $U$, $V$ are, respectively, 
0-, 1-, 2-, 3- and 4-form gauge fields and parameters. 
Here odd forms of gauge fields and even forms of gauge parameters
are ordinary even Lie algebra valued fields
while the rest of gauge fields and parameters
are odd super Lie algebra valued fields\cite{GCS}.
Since all these generalized Chern-Simons actions are formulated by 
form degrees, the fields of forms are very naturally accommodated on the 
simplicial manifold. 

It is important for us to recognize that 3-dimensional Chern-Simons action and 
4-dimensional $BF$ action are the leading terms of the 
above generalized Chern-Simons actions of $S_3$ and $S_4$. 
In the previous paper and present paper we focus on the
formulation of the leading terms of 
the 3- and 4-dimensional generalized Chern-Simons action as the lattice 
gravity theory on a simplicial manifold.
It is very natural to generalize the lattice formulation of this paper 
to include all the form degrees of the generalized Chern-Simons 
action on the lattice. 
We believe that the matter field will be accommodated via this generalized 
gauge theory formulation which might lead to the unified theory including 
gravity on the simplicial manifold\cite{Nishi}. 
In this sense the $BF$ action on the lattice is the good starting formulation 
for the generalized gauge theory as well. 

In section 2 we first summarize the continuum formulation of $BF$ theory 
then introduce a lattice version of $BF$ theory. 
Since the lattice $BF$ theory is a gauge theory, we discuss the gauge 
invariance of the theory on the lattice in section 3. 
We then integrate out the $B$ fields and the link variables and 
then explicitly show that the partition function of the lattice $BF$ action 
leads to the product of 15-$j$ symbols in section 4. 
We then prove the Pachner move invariance of our partition function 
by graphical method explicitly in section 5. 
We summarize the result with several discussions in the final section.

\setcounter{equation}{0}
\renewcommand{\theequation}{\arabic {section}.\arabic{equation}}

\section{Formulation of the lattice $BF$ theory}

\subsection{Continuum $BF$ Theory}

We first summarize characteristics of the continuum 4-dimensional $BF$ 
theory\cite{Horowitz}.
The action of $BF$ theory is given by
\begin{equation}
 S_{BF} = \int_M \langle B, F \rangle,
 \label{BF action 1}
\end{equation}
where
$M$ is 4-dimensional manifold.
We take our gauge group any Lie group $G$ whose Lie algebra ${\cal L_{G}}$ 
is equipped with an invariant nondegenerate bilinear form 
$\langle~, ~\rangle$ for the pair $B$ and $F$. 
Here $F$ is curvature 2-form constructed from Lie algebra ${\cal L_{G}}$ 
valued 1-form $\omega$, and $B$ is dual Lie algebra ${\cal L_G^*}$
(= ${\cal L_G}$ in the present case) valued generic 2-form.

It has been noticed since quite sometime that the $BF$ action leads 
to the Palatini type of Einstein-Hilbert action
if we take the Lie group $G$ to be local Lorentz group
and identify $B = *(e \wedge e)$ with $*$ as Hodge dual operation
to the local Lorentz suffices.
In particular the self dual component of the 2-form $B$ plays a 
crucial role in formulating the Einstein gravity
by the $SL(2,C)$ spinor representation\cite{Plebanski}.

Using the fact that the 4-dimensional Euclidean version of
local Lorentz group is a direct product of chiral $SU(2)$ groups:  
$SO(4) \simeq SU(2)_L \times SU(2)_R$, 
we can factorize the $BF$ action with local Lorentz group into 
the chiral parts,
\begin{equation}
 S_{BF} = \int_M B_{a'b'} F^{a'b'} 
  = \int_M B^+_a F^+_a + \int_M B^-_a F^-_a,
 \label{BF action 1-2}
\end{equation}
where $F^{\pm}_a \equiv \pm F^{\pm}_{0a}$,
$F^{\pm}_{a'b'} \equiv F_{a'b'}
\pm \frac{1}{2} \epsilon_{a'b'c'd'} F^{c'd'}$,
and similar for $B^{\pm}_a$ and $B^{\pm}_{a'b'}$.
The suffices, $a',b',..$ are local Lorentz suffices 
while $a$ is the chiral $SU(2)$ suffix.
In this paper we only consider the one chiral, say left handed, 
counterpart of the action.
We may consider that the right-handed part is nothing but
the copy of the left-handed part, at least in the continuum limit. 
On the lattice there would be a possibility to accommodate the 
same copy of chiral counterpart which we will discuss later. 
If we take the left-chiral part of 
the action, we can formulate the lattice $SU(2)$ $BF$ theory by naive 
extension from 3-dimensional lattice Chern-Simons gravity
into 4 dimensions.

In formulating left-chiral $BF$ model, we can expand both $B$ and $F$ 
by the Pauli matrices, $\sigma^a$ $(a = 1,2,3)$ since $SU(2)$ has  
self-dual Lie algebra, 
\begin{equation}
 B = \frac{1}{2} ~B^a \sigma^a, \quad F = \frac{1}{2}~F^a \sigma^a.
\end{equation}
Hereafter we omit the suffix $+$ to denote the left chirality.
Then the action of $SU(2)$ $BF$ theory leads to
\begin{equation}
 S_{BF} = \int_M \tr B F = \frac{1}{2} \int_M B^a F^a,
  \label{BF action 2}
\end{equation}
where the trace is taken for the Pauli matrices.
It is worth to mention that the factor 2 difference between
the chiral part of $BF$ action in (\ref{BF action 1-2})
and the above action (\ref{BF action 2}) is due to the fact
that the chiral decomposition of the fundamental representation 
for $SO(4)$ generator is reducible into the direct sum of
two Pauli matrices.

There are two independent gauge transformations in this theory.
One of them is the local Lorentz transformation,
\begin{equation}
 \delta_{\tau} \omega^a = D \tau ^a, \quad
 \delta_{\tau} B^a = [B,~\tau]^a,
 \label{gauge_trans1}
\end{equation}
where $\tau$ is 0-form gauge parameter.
There is another independent gauge transformation 
\begin{equation}
 \delta_{u} \omega^a = 0, ~~
 \delta_{u} B^a = D u^a,
 \label{gauge_trans2}
\end{equation}
where $u^a$ is 1-form gauge parameter
according to the 2-form nature of $B^a$.
This transformation may be called Kalb-Ramond symmetry 
transformation\cite{Kalb-Ramond} 
and corresponds to the diffeomorphism transformation of dreibein $e$ 
in 3 dimensions.
This transformation is on-shell reducible.
In fact the following gauge change of the gauge parameter $u$ itself
by 0-form gauge parameter $v^a$,
\begin{equation}
 \delta_{v} u^a = D v ^a,
\label{one-form-gauge-tr}
\end{equation}
leads to the vanishing contribution of the gauge change
\begin{equation}
 \delta_{v} (\delta_{u} B^a)
  = D \delta_{v} u^a
  = D D v^a = [F,~ v]^a = 0,
\label{continuum-reducibility}
\end{equation}
where we have used the equation of motion $F=0$.

\subsection{Lattice $BF$ Gravity Action} \label{action}

We intend to formulate a gravity version of lattice gauge theory 
where the product of link variables $U$ along the boundary of 
a square plaquette leads to the exponential of curvature. 
Thus we may identify the curvature locating at the center of 
the plaquette and then the trace leads to the Yang-Mills action 
in the leading order of the lattice constant in the lattice QCD. 
On the other hand Regge calculus tells us that the curvature is
located at sites in 2 dimensions, links in 3 dimensions, 
triangles in 4 dimensions and so on. 
It is thus very natural to formulate the gravity version of 
lattice gauge theory in terms of the link variable $U$ and 
take a product of $U$ variables surrounding the simplexes suggested 
by Regge calculus\cite{Kawamoto-Nielsen}\cite{Dadda}. 

In fact in 3 dimensions, we have successfully formulated 
lattice Chern-Simons gravity based on the correspondence between the 
lattice gauge theory and Regge calculus\cite{KNS}.
The link variable $U(\tilde{l}) = e^{\omega(\tilde{l})}$ is 
defined on the dual link $\tilde{l}$ which is located on the 
boundary of dual plaquette $\tilde{P}$. 
Since the dual of the dual plaquette $\tilde{P}$ is original link $l$ 
which intersect $\tilde{P}$ in 3 dimensions, the curvature is located 
in the center of the dual plaquette or equivalently on the original link 
and thus consistent with the Regge calculus. 
We needed to introduce vanishing holonomy constraint which enforces 
the parallelism of the dreibein $e^a$ and the lattice curvature 
$F^a \equiv \frac{1}{2} \epsilon^{abc} F^{bc}$ to obtain gravity theory.
It turns out that the partition function formulated in this way 
exactly coincides with that of the Ponzano-Regge model.
Then the action leads to the continuum Chern-Simons gravity 
since the Ponzano-Regge partition function is Alexander move invariant 
and thus invariant under the division of the simplicial manifold
and then the naive continuum limit can be taken.

\begin{figure}[b]
\begin{center}
 \begin{minipage}{\textwidth}
  \epsfxsize=.4\textwidth \epsfbox{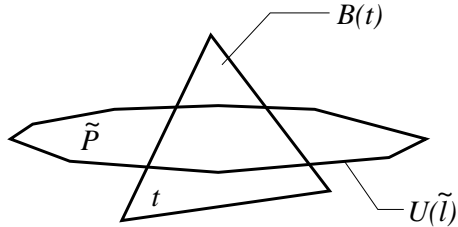}
 \end{minipage} 
\end{center}
\caption{$\tilde{P}$ is the dual plaquette dual to the
 triangle $t$ associated to the 2-form $B(t)$. Dual link variable
 $U = e^{i \omega}$ is located on the dual link
 which constructs the boundary of the dual plaquette $\tilde{P}$.}
\label{fig:BF action}
\end{figure}

In this paper we extend the formulation of
3-dimensional lattice gravity into 4 dimensions. 
We formulate a lattice version of $SU(2)$ $BF$ theory.
We consider a piecewise linear 4-dimensional simplicial manifold.
According to the Regge calculus, curvature is located at a triangle $t$.
There are in general several 4-simplexes which have the 
triangle $t$ in common. 
A dual link $\tilde{l}$ is defined as the line connecting the 
center of neighbouring 4-simplexes and is located on the 
boundary of the dual plaquette $\tilde{P}$. 
We introduce dual link variables $U(\tilde{l}) = e^{i \omega(\tilde{l})}$ 
on the dual link $\tilde{l}$
\footnote{In \cite{KNS} we have used anti-hermitian generators of $SO(3)$
and thus denoted the link variable as $U=e^{\omega}$ 
while here we use hermitian generators of $SU(2)$
and thus denote the link variable as $U=e^{i \omega}$.}.
Then the product of the dual link variables $U(\tilde{l})$ along 
the boundary of the dual plaquette $\tilde{P}$ leads to a 
curvature located on the center of the dual plaquette $\tilde{P}$ 
which coincides with the center of the triangle $t$ in 4 dimensions. 
We then locate the 2-form field $B$ on the original triangle $t$. 
See Fig.\ref{fig:BF action}.

We define the ``curvature'' $F(t)$ located on the triangle $t$
by the following equation:
\begin{equation}
 \plaq \equiv e^{i F(t)}.
\end{equation}
The leading term in $F$ with respect to the lattice unit is the
ordinary curvature $d\omega + \omega^2$ similar in structure to the 
ordinary lattice gauge 
theory, but here we define the lattice ``curvature'' $F$ in such a way that 
it contains all orders of terms of lattice unit.

Using these variables, we consider the following
lattice version of the $BF$ action:
\begin{equation}
 S_{LBF}
  = \sum_{t} \tr \left( -i B(t) \Bigl[ \ln \plaq \Bigr] \right)
  = \frac{1}{2} \sum_{t} B^a(t) F^a(t).
\label{LBF} 
\end{equation}

In formulating continuum Chern-Simons gravity in 3 dimensions, we need to 
impose a torsion free condition as an equation of motion at the 
classical level. 
The torsion free nature is lost at the quantum level when we integrate out 
the dreibein and spin connection. 
In formulating lattice Chern-Simons gravity and showing the equivalence with 
the Ponzano-Regge model in 3 dimensions, we needed to introduce the 
following vanishing holonomy constraint which 
relates the dreibein and spin connection even at the quantum level:
\begin{equation}
 \Bigl[ \plaq \Bigr]^{ab} e^b = e^a \label{econstraint},
\end{equation}
where $a,b$ are $SO(3)$ suffices. 
This constraint has been interpreted as a gauge fixing condition of 
lattice gauge diffeomorphism transformation. 
Geometrically the constraint is equivalent to the parallelism of 
the dreibein and the curvature: $e^a \propto F^a$, where  
$F^a \equiv \frac{1}{2} \epsilon ^{abc} F^{bc}$. 

In contrast with the 3-dimensional case
we introduce the following constraint in 4 dimensions
\begin{equation}
 \left[ \left(\plaq\right) B 
\left(\plaq \right)^{\dagger} \right]^{\alpha \beta}
  = B^{\alpha \beta} 
\label{constraint1},
\end{equation}
where $\alpha \beta$ are suffices of $SU(2)$ spinor representation. 
Geometrically the constraint means that 
the each of spinor suffix of $B^{\alpha \beta}$ field
located at the center of a triangle can be 
parallel transported in the opposite direction
along the boundary of the dual plaquette $\tilde{P}$ and
comes back to the starting point and should coincide with
the original direction and thus the holonomy of each suffix vanishes. 
We now derive more concrete expression for the constraint
(\ref{constraint1}) which is equivalent to the following relations:
\begin{equation}
 \sum_{n=0}^{\infty} \frac{i^n}{n!}
  [F, \cdots [F,[F,~B]] \cdots ] = B.
\end{equation}
This equation can be satisfied if
\begin{equation}
 [F,~B] = i \epsilon _{abc} F^a B^b \frac{\sigma^c}{2} = 0,
\end{equation}
or equivalently $B^a \propto F^a$. 

Analogous to the arguments in 3 dimensions,
the constraint thus leads to 
the parallelism of the 2-form $B^a$ and the curvature:   
$B^a \propto F^a$. 
More precisely the constraint can be rewritten in a concrete form by taking 
into account the parallel and antiparallel nature of $B^a$ and $F^a$
\begin{equation}
\frac{B^3}{|B|} 
\left[ \prod_{a=1}^{2}
 \delta \left( \frac{F^a}{|F|} + \frac{B^a}{|B|} \right)
+ \prod_{a=1}^{2}
\delta \left( \frac{F^a}{|F|} - \frac{B^a}{|B|} \right) \right], 
\label{constraint2}
\end{equation}
where $|B| \equiv \sqrt{B^a B^a}$ and
$|F| \equiv \sqrt{F^a F^a}$ are
the length of $B^a$ and $F^a$, respectively.
The term $B^3/|B|$ is needed to keep the rotational
invariance of the expression.

One of the important characteristic of the lattice $BF$ action 
(\ref{LBF}) is that the discreteness of the length of the 2-form field 
$|B|$ comes out as a natural consequence of the specific choice of 
the logarithmic action. 
We first note the following identity: 
\begin{equation}
 e^{i |F| I} = \hbox{cos}\left(\frac{|F|}{2}\right) 
                            + i2I \hbox{sin}\left(\frac{|F|}{2}\right),
\end{equation}
where $I=\frac{F^a}{|F|}\frac{\sigma^a}{2}$
with $\sigma^a$ as Pauli matrix, then 
\begin{equation}
 e^{i 4 \pi n I} = 1, \quad n \in \bm{Z} \label{periodicity1}.
\end{equation} 

Using the above relation 
and $F^a \propto B^a$ by the constraint (\ref{constraint1}),
we find that our lattice $BF$ action $S_{LBF}$ has 
the following ambiguity:
\begin{eqnarray*}
 S_{LBF}
  &=& \sum_{t} \tr \left( -iB(t)  \ln e^{F(t)} \right)\\
  &=& \sum_{t} \tr \left( -iB(t)  \ln e^{F(t)+ i 4 \pi n I} \right)\\
  &=& S_{LBF} + \frac{1}{2}\sum_{t} 4 \pi n |B(t)|.
\label{ambiguity}
\end{eqnarray*}
This ambiguity leads to an ambiguity in the partition 
function
\begin{equation}
 Z = \int {\cal D}U {\cal D}B ~ e^{i S_{LBF}}
   = \int {\cal D}U {\cal D}B ~
   e^{ i S_{LBF} + i \sum_{t} 2 \pi n |B|}.
\end{equation}
Imposing the single valuedness of 
$e^{i S_{LBF}}$, we obtain the constraint 
that $\sum_{t} |B(t)| $ should be integer,
or equivalently $|B(t)|$ should be integer.
We can now write down the explicit form of partition function
with constraints
\begin{equation}
 Z = \int {\cal D}U {\cal D}B ~
  \delta\left( \left (\simpplaq \right) B \left( \simpplaq \right)^{\dagger} 
  - B \right) \sum_{N}  \delta(|B| - N) 
  e^{i S_{LBF}},
\label{Z of LBF}
\end{equation}
where $N \in Z$.

It is important to notice that the length of the 2-form $|B(t)|$ 
is proportional to
the area of triangle $t$ on which the 2-form field $B^a(t)$ is 
defined. It is then interesting to note that the ``area of the triangle'' 
is discretized due to the logarithmic form of our lattice $BF$ action. 
It has already been pointed out by Rovelli and Smolin that the 
area of triangle in 4-dimensional gravity is discretized
in the square root of angular momentum square $\sqrt{J(J+1)}$ 
by investigating an area operator\cite{Rovelli-Smolin}.
The origin of the discreteness of the area of triangle in our formulation 
is quite different from that of Rovelli and Smolin.

\setcounter{equation}{0}
\renewcommand{\theequation}{\arabic {section}.\arabic{equation}}

\section{Gauge Invariance on the Lattice}

The gauge transformations of the continuum $BF$ theory have been given 
by the local Lorentz transformation (\ref{gauge_trans1})   
and the Kalb-Ramond transformation (\ref{gauge_trans2}). 
Under the local Lorentz transformation both of the $B$ field and curvature  
$F$ transform adjointly in the continuum theory
\begin{equation}
 \delta_{\tau} B^a = [B,~\tau]^a,\quad 
 \delta_{\tau} F^a = [F,~\tau]^a.
\end{equation}
We consider that the lattice versions of the local Lorentz gauge 
parameters are located on the dual sites and the center of the original 
triangles, the same point where the $B$ field is located. 
Then the dual link variable $U(\tilde{l}) = e^{i \omega(\tilde{l})}$ 
transforms under the lattice local Lorentz transformation as 
\begin{equation}
 U(\tilde{l}) \rightarrow V^{-1} U(\tilde{l}) V',
\end{equation}
where $V = \exp \left(\frac{i}{2} \tau ^a \sigma ^a\right)$
is the $SU(2)$ matrix 
on the dual site and $V'$ is the one on the neighboring site.
Then the following lattice counterpart of the local Lorentz transformation:  
\begin{equation}
 \begin{array}{ccc}
  B(t) & \rightarrow & V^{-1} B(t) V, \\
  F(t) & \rightarrow & V^{-1} F(t) V, 
 \end{array}
\end{equation}
leads to the obvious invariance of the action (\ref{BF action 2}). 

There are, however, some subtleties on the gauge invariance of the 
lattice action due to the fact that the center of the triangle 
where $B$ field is located and the dual links where dual link variables 
$U(\tilde{l})$ are located are not connected by a link variable $U$. 
We can arbitrarily choose a dual site and introduce new link variable
to bridge between the dual site and the center of the original triangle 
to identify the lattice local Lorentz transformation. 
The newly defined link variables, however, can be gauged away. 
The details of these arguments go parallel to the 3-dimensional 
Chern-Simons gravity case and can be found in \cite{KNS}.

Next we will investigate the lattice version of Kalb-Ramond transformation.
Since the invariance nature of the continuum action
under this transformation is due to the Bianchi identity,
\begin{equation}
 DF = dF + [\omega,~ F] = 0, \label{Bianchi}
\end{equation}
we need to identify Bianchi identity on the lattice.
We will consider the integrated version of the Bianchi identity,
\begin{equation}
 \int _M DF = \int_{\partial M} F + \int _M [\omega, ~F] = 0,
  \label{integrated Bianchi}
\end{equation}
where $M$ is a 3-dimensional manifold corresponding to the 
3-form nature of $DF$ in (\ref{Bianchi}).
\begin{figure}[t]
\begin{center}
 \begin{minipage}[c]{.4\textwidth}
 \epsfxsize=\textwidth \epsfbox{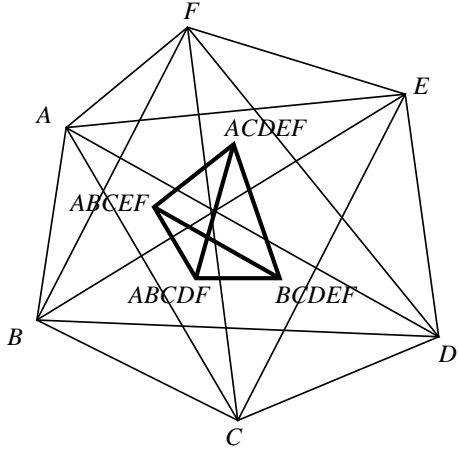}
 \end{minipage}
\end{center} 
\caption{A simple setup to show
 the integrated lattice Bianchi identity.
 $A$, $B$, $C$, $D$, $E$ and $F$ are sites and
 $ACDEF$, $ABCEF$, $ABCDF$, $BCDEF$ are the
 centers of 4-simplex.
 Thick lines are the dual links and thin lines are
 original links.}
 \label{fig:Bianchi}
\end{figure}

In 3 dimensions the geometrical structure of the manifold $M$ 
can be visually well understood since the dual cell of a 
original site associated to the 0-form gauge parameter
of the gauge diffeomorphism transformation is easily recognized.
This type of setup was just what
we needed to show the 3-dimensional lattice 
counterpart of the Bianchi identity\cite{KNS}. 

Correspondingly in 4 dimensions, we have to consider the dual cell 
of the original link $CF$ as a 3-dimensional simplicial submanifold $M$ 
since the gauge parameter of the Kalb-Ramond transformation 
(\ref{gauge_trans2}),
which we now identify as the parameter on the link $CF$, 
is 1-form in 4 dimensions.
In general the 3-dimensional cell dual to a given link may have 
complicated simplicial structure.
For simplicity we here consider the dual cell as a tetrahedron. 
The tetrahedron as the dual cell has four dual sites which correspond 
to the center of original 4-simplexes. 
We then need to introduce four original sites $A$, $B$, $D$ and $E$ 
to obtain four 4-simplexes which share the common link $CF$ as shown in 
Fig.\ref{fig:Bianchi}.
With this setup we can discuss the invariance of the lattice 
$BF$ action (\ref{LBF}) under a lattice version of Kalb-Ramond 
transformation. 

We introduce the following notations: 
$U_1 \equiv U_{ACDEF \rightarrow ABCDF}$ is a dual link variable 
on the dual link which connect the dual sites or 
equivalently the center of original links denoted as $ACDEF$ and $ABCDF$. 
The 2-form gauge field 
$B_1 \equiv \displaystyle \int_{BCF} B_{\mu\nu} dx^{\mu} dx^{\nu}$ 
is located on the center of the triangle $BCF$. 
The 1-form gauge parameter
$u_0 \equiv \displaystyle \int_{FC} u_{\mu} dx^{\mu}$ 
is located on the original link $FC$.
The 0-form gauge parameter 
$ v_{1} \equiv v (B)$ 
is located on the original site $B$. 
Here we summarize the notations:
\renewcommand{\arraystretch}{1.5}
\begin{eqnarray*}
&& \left\{
\renewcommand{\arraystretch}{1.5}
 \begin{array}{c}
  U_1 \equiv U_{ABCDF \rightarrow ACDEF},~
  U_2 \equiv U_{ABCDF \rightarrow BCDEF},~
  U_3 \equiv U_{ABCDF \rightarrow ABCEF},~ \\
  U_4 \equiv U_{BCDEF \rightarrow ABCEF},~
  U_5 \equiv U_{ABCEF \rightarrow ACDEF},~
  U_6 \equiv U_{ACDEF \rightarrow BCDEF},\\
 \end{array}
\right. \\
&& \left\{
\renewcommand{\arraystretch}{2.0}
\begin{array}{c}
 B_1 \equiv \displaystyle \int_{BCF} B_{\mu\nu} dx^{\mu} dx^{\nu},~
 B_2 \equiv \displaystyle \int_{ACF} B_{\mu\nu} dx^{\mu} dx^{\nu},~ \\
 B_3 \equiv \displaystyle \int_{CDF} B_{\mu\nu} dx^{\mu} dx^{\nu},~ 
 B_4 \equiv \displaystyle \int_{CEF} B_{\mu\nu} dx^{\mu} dx^{\nu},~ \\
\end{array}
\right. \\
&& \left\{
\renewcommand{\arraystretch}{2.0}
\begin{array}{l}
 u_0 \equiv \displaystyle \int_{FC} u_{\mu} dx^{\mu},~
 u_1 \equiv \displaystyle \int_{CB} u_{\mu} dx^{\mu},~
 u_2 \equiv \displaystyle \int_{BF} u_{\mu} dx^{\mu},~
 u_3 \equiv \displaystyle \int_{CA} u_{\mu} dx^{\mu},~
 u_4 \equiv \displaystyle \int_{AF} u_{\mu} dx^{\mu},~ \\
 u_5 \equiv \displaystyle \int_{CD} u_{\mu} dx^{\mu},~
 u_6 \equiv \displaystyle \int_{DF} u_{\mu} dx^{\mu},~
 u_7 \equiv \displaystyle \int_{CE} u_{\mu} dx^{\mu},~
 u_8 \equiv \displaystyle \int_{EF} u_{\mu} dx^{\mu},~ \\
\end{array}
\right. \\
&& \left\{
\renewcommand{\arraystretch}{1.5}
\begin{array}{c}
 v_{1} \equiv v (B),~
 v_{2} \equiv v (A),~
 v_{3} \equiv v (D),~
 v_{4} \equiv v (E),~
 v_{5} \equiv v (F),~
 v_{6} \equiv v (C).
\end{array}
\right. 
\end{eqnarray*}
We first note the following identity,
\begin{equation}
 \prod U \equiv U_2 U_4 U_3^{-1} \cdot U_3 U_5 U_1^{-1}
  \cdot U_1 U_6 U_2^{-1} \cdot U_2 U_6^{-1} U_5^{-1} U_4^{-1} U_2^{-1} = 1.
\end{equation}
Using the following definitions of curvature on the triangles,
\begin{equation}
 e^{F_1} \equiv U_2 U_4 U_3^{-1},~
 e^{F_2} \equiv U_3 U_5 U_1^{-1},~
 e^{F_3} \equiv U_1 U_6 U_2^{-1},~
 e^{F_4} \equiv U_2 U_6^{-1} U_5^{-1} U_4^{-1} U_2^{-1},~
\end{equation}
and the Baker-Hausdorff formula,
\begin{equation}
 e^A \cdot e^B = \exp\left( A+B+\frac{1}{2}[A,~B] \cdots \right),
\end{equation}
we can rewrite the above identity in the following form:
\begin{equation}
 0 = \ln \prod U
  = \ln \left(e^{F_1} \cdot e^{F_2} \cdot e^{F_3} \cdot e^{F_4}\right)
  =  \sum_{k=1}^{4} F_k + \sum_{k=1}^{4} [\Omega_k, ~ F_k].
\label{lattice Bianchi}
\end{equation}
Here we have introduced the following variables $\Omega_k$,
\begin{equation}
\renewcommand{\arraystretch}{2}
 \begin{array}{lcl}
\displaystyle
 \Omega_1 = -\frac{1}{4}(F_2+F_3+F_4)+\cdots, & ~ &
\displaystyle
 \Omega_2 = -\frac{1}{4}(-F_1+F_3+F_4)+\cdots, \\
\displaystyle
 \Omega_3 = -\frac{1}{4}(-F_1-F_2+F_4)+\cdots, & ~ &
\displaystyle
 \Omega_4 = -\frac{1}{4}(-F_1-F_2-F_3)+\cdots, 
\end{array}
\label{Omega}
\end{equation}
which can be evaluated order by order of the curvature $F$ 
by using the Baker-Hausdorff formula.
We identify (\ref{lattice Bianchi}) as the lattice version of the
Bianchi identity (\ref{Bianchi}).

Now we consider the lattice version of
the gauge transformation for the 2-form gauge field $B_k$. 
We first note that $B_1$ field is located on the center of 
the triangle $BCF$ and thus the gauge transformation of $B_1$ field 
is associated with the gauge parameters $u_0$, $u_1$ and $u_2$ 
located on the boundary $FC$, $CB$ and $BF$, respectively. 
Then the lattice version of $B_k$ transformation will be given   
as follows:   
\begin{equation}
 \delta B_k =  u_{0} + u_{2k-1} + u_{2k} 
  - [\Omega_k, ~ u_0] - [\Omega'_{2k-1}, ~ u_{2k-1}]  
  - [\Omega''_{2k}, ~ u_{2k}],
  \label{lattice KR}
\end{equation}
where $\Omega_k$ is defined in (\ref{Omega}) while $\Omega'_{2k-1}$ and 
$\Omega''_{2k}$ are the similar quantities associated with the gauge 
parameters $ u_{2k-1}$ and $u_{2k}$ defined on the links 
just as  $\Omega_k$ is associated with the gauge parameter $u_0$ on the 
link $FC$.
Then the transformation of the action leads to
\begin{eqnarray*}
 \delta S_{LBF}
 &=& \sum_{k=1}^{4} \tr \left(\delta B_k \cdot F_k \right) + \cdots \\
 &=& \sum_{k=1}^{4} \tr
  \left( \left(  u_0 + u_{2k-1} + u_{2k} 
   - [\Omega_k, ~ u_0] - [\Omega'_{2k-1}, ~ u_{2k-1}] 
   -[\Omega''_{2k}, ~ u_{2k}]\right) F_k \right)
   + \cdots \\
 &=& \tr \left(
  u_0 \sum_{k=1}^4 \left( F_k + [\Omega_k, ~F_k]\right)
  \right)
  + \cdots \\
 &=& 0,
\end{eqnarray*}
where in the last expression we have only picked up
the terms associated with $u_0$. 
We can identify this transformation as 
the Kalb-Ramond transformation on the lattice.
Thus the lattice $BF$ action is invariant
under the transformation (\ref{lattice KR}) due to the lattice version of 
Bianchi identity (\ref{lattice Bianchi}).

In 4 dimensions the Kalb-Ramond transformation is on-shell reducible.
We now show how the reducibility of the 2-form gauge transformation is 
realized on the lattice.
Due to the equation of motion $F=0$,
we obtain $\Omega = 0$ since $\Omega$ is a polynomial of $F$
as we can see in the definition (\ref{Omega}).
Hence the transformation of 
$B$ in (\ref{lattice KR}) can be reduced as follows:
\begin{equation}
 \delta B_k = u_{2k-1} + u_{2k} + u_{0},
\end{equation}
which corresponds to the continuum gauge transformation 
(\ref{gauge_trans2}).
Considering the continuum reducibility relation, we identify the 
transformation of 1-form gauge parameter $u$ defined on a link 
as the difference of 0-form gauge parameter defined on a site,
\begin{equation}
 \delta u (AB) = v (B) - v (A), 
\end{equation}
more explicitly, 
\begin{equation}
\begin{array}{ccc}
 \delta u_{2k} &=& v_5 - v_k ,\\
 \delta u_{2k-1} &=& v_k - v_6, \\
 \delta u_0 &=& v_6 - v_5.
\end{array}
\end{equation}
Then the continuum reducibility relation (\ref{continuum-reducibility}) 
leads to 
\begin{equation}
 \delta (\delta B_k)
=  (v_k - v_6) + (v_5 - v_k) + (v_6 - v_5) 
=0, 
\end{equation}
on the lattice.
We can consider that this result is corresponding to
the on-shell reducibility of the Kalb-Ramond transformation
on the lattice.

We now point out that the constraint (\ref{constraint1}) or equivalently 
(\ref{constraint2}) 
breaks the lattice version of Kalb-Ramond transformation while the 
lattice $BF$ action itself is invariant under the transformation 
as shown in the above.
The lattice $B$ field is transformed but the lattice curvature is unchanged 
under the Kalb-Ramond gauge transformation. 
Therefore we can use a part of degrees of freedom of the lattice 1-form 
gauge parameter $u_k$ to align the 2-form $B^a$ field and the curvature 
$F^a$ in accordance with the constraint. 
In fact the reducibility allows the total gauge degrees of freedom of 
the 1-form 
gauge parameter $u_\mu^a$ to be $4\times 3 - 3 = 9$, while the parallel 
or anti-parallel nature of the 2-form chiral $B^a_{\mu\nu}$ and 
$F^a_{\mu\nu}$ needs 
$6\times 3/2=9$ constraints where we have taken into account the chiral 
nature for the space-time suffix as well. 
There are thus necessary degrees of freedom to fulfill the constraint.
The 3 degrees of freedom out of 9 should be exhausted to adjust the 
discrete nature of the $B^a$ field.
We now claim that the vanishing holonomy constraint can be identified 
as the ``gauge fixing condition'' of the Kalb-Ramond gauge transformation 
on the lattice. 

\setcounter{equation}{0}
\renewcommand{\theequation}{\arabic {section}.\arabic{equation}}

\section{Calculation of the Partition Function}

In our previous paper we have explicitly shown that 3-dimensional 
version of lattice Chern-Simons action with the constraint of vanishing 
holonomy along dual links leads to the Ponzano-Regge model formulated by 
6-$j$ symbol\cite{KNS}. 
Analogously to the 3-dimensional case we explicitly show that the 
lattice version of 
$BF$ action (\ref{LBF}) with the vanishing holonomy constraint 
(\ref{constraint1}) leads to a 
topological gravity model formulated by 15-$j$ symbol in 4 dimensions.

The partition function with the constraint (\ref{constraint1}) 
or equivalently (\ref{constraint2}) and the 
discreteness of $|B|$ taken into account is given by 
\begin{eqnarray}
 Z &=& \int {\cal D}U  \prod_{t} Z_{t}, \\ 
 Z_{t} &=& \int d^3 B ~\frac{B^3}{|B|}
  \left[ \prod_{a=1}^{2}
   \delta \left( \frac{F^a}{|F|} + \frac{B^a}{|B|} \right)
  + \prod_{a=1}^{2}
  \delta \left( \frac{F^a}{|F|} - \frac{B^a}{|B|} \right)
  \right] \nonumber \\ 
 && \hspace*{1cm} \times
  \sum_{J=0}^{\infty} \delta \left( |B| - 2J \right)
  e^{\frac{i}{2} B^a F^a}  \quad \left( J \in \frac{\bm{Z}}{2} \right), 
\end{eqnarray}
where $Z_{t}$ is the part of partition function
associated with a triangle $t$.
We have introduced the discrete nature of $|B|$  
by using the following relation:
\begin{equation}
  \int^{|B_f|}_{|B_i|} d|B| ~=~ 
  \int^{|B_f|}_{|B_i|}
  \sum_{J=0}^{\infty} \delta \left( |B| - 2J \right) d|B|, 
\end{equation}
where $J \in \bm{Z}/2$ and thus $|B_f|$ and $|B_i|$ 
should be integer.

We can now evaluate the $B$ integral of $Z_t$ straightforwardly 
thanks to the delta functions and obtain 
\begin{equation}
Z_{t}= \sum_{J} 8 J^2 \cos(J|F|). 
\label{B-integration}
\end{equation}
At this stage it is important to recognize that arbitrary change of the 
normalization constant for the lattice $BF$ action (\ref{LBF}) 
gives the same result for (\ref{B-integration}).
This can be understood by the following arguments: 
Suppose we take the action of the form 
$S_{LBF} = \frac{\alpha}{2} \sum_{t} B^a(t) F^a(t)$, 
we need to take $\alpha|B(t)|=2J$ ($J \in \bm{Z}/2$), 
based on the similar arguments as in section 2. 
Taking into account the vanishing holonomy constraint (\ref{constraint1}) 
or equivalently (\ref{constraint2}), the action 
essentially leads to the form $S_{LBF} = \sum_{t} J |F(t)|$ 
and thus to (\ref{B-integration}).

Using the following formula for the character $\chi_J$
of the spin-$J$ representation of $SU(2)$,
\begin{equation}
 \chi _{J} (e^{i \theta^a J_a})
  = \chi _{J} (|\theta|)
  = \frac{\sin\left( (2J+1) \frac{|\theta|}{2} \right)}
  {\sin \left(\frac{|\theta|}{2} \right)},
\end{equation}
where $|\theta|$ is the length of $\theta^a$,
we find 
\begin{eqnarray}
 \chi _{J}(|F|) - \chi _{J-1}(|F|)
 &=& 2 \cos(J|F|). 
\end{eqnarray}
Hence we can naively calculate the triangle partition function,
\begin{eqnarray*}
 Z_{t}
  &=& \sum_{J=1}^{\infty} 4 J^2 (\chi _{J} - \chi _{J-1}) \\
  &=& -4 \sum_{J=0}^{\infty} (2J+1) \chi _J.
\end{eqnarray*}
This calculation is not precise, because the summation is not convergent.
We can, however, justify the above calculation by introducing the 
heat kernel regularization.
The details of the regularization procedure can be found in our previous 
paper\cite{KNS}.

Integrating out $B$ field and introducing heat kernel regularization 
and dividing the unimportant constant factor, 
we obtain the total partition function 
\begin{equation}
 Z = \int {\cal D}U \prod_{t} \sum_{J=0}^{\infty} \left(2J+1\right)
  \chi_{J}\Bigl(\plaq \Bigr) ~ e^{-J(J+1) \tau}, 
\label{z_after_dbintegration}
\end{equation}
where a regularization parameter $\tau$ is introduced.
The parameter $\tau$ will be sent to zero ($\tau \rightarrow 0$) 
at the end of calculation.

It is interesting to note that the formulation has close similarity with 
the 3-dimensional case.  
In 3 dimensions the angular momentum $J$ is associated to the original 
link while it is associated to a triangle in 4 dimensions. 
This has natural correspondence with Regge calculus in the sense that 
the discretized angular momentum $J$ is localized at the simplexes where 
the gravitational curvature is localized.

We now carry out ${\cal D}U$ integration of this partition function. 
Thanks to the character of the partition function, ${\cal D}U$
integration is straightforward. 
We first note the following relation specific to the character: 
\begin{equation}
   \chi_{J}\Bigl(\plaq \Bigr) ~ =~ \sum_{\{k_i\}}
     \prod^m_{i=1}D^{J}_{k_i k_{i+1}}(U_i) \quad\quad (k_{m+1}=k_1), 
\label{Dfunction}
\end{equation}
where $D^{J}_{k_i k_{i+1}}(U_i)$ is the $D$-function, 
a spin $J$ matrix representation 
for a $SU(2)$ group element $U_i \equiv U(\tilde{l}_i)$ located on the 
dual link $\tilde{l}_i$. 
Here we assume that there are $m$ dual link variables 
$U_i$ along the boundary of dual plaquette $\tilde{P}$. 

\begin{figure}[h]
\begin{center}
  \begin{minipage}[c]{.5\textwidth}
  \epsfxsize=\textwidth \epsfbox{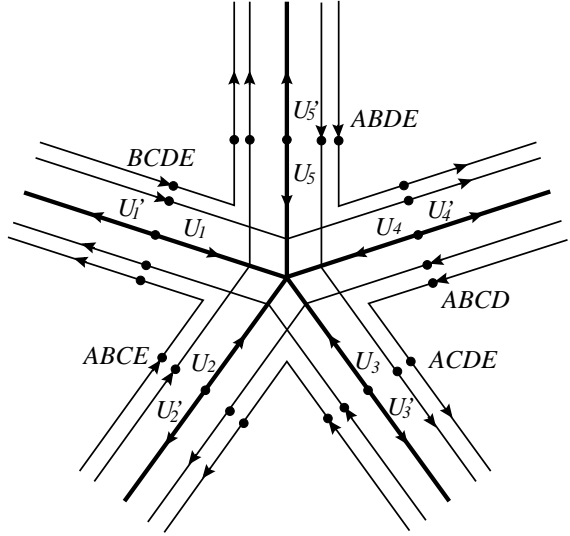}
 \end{minipage}
\end{center}
\caption{
Graphical presentation of geometrical structure of a 
4-simplex $ABCDE$
}
 \label{fig:U-integration}
\end{figure}

We then clear out the geometrical structure of a 4-simplex. 
A 4-simplex can be constructed by five sites 
for which we name $A,B,C,D$ and $E$. 
We may denote the 4-simplex itself composed of these sites as $ABCDE$, 
which has five tetrahedra $BCDE, ACDE,ABDE,ABCE$ and $ABCD$ on the 
boundary. 
Each tetrahedron, such as $BCDE$, has boundary triangles $CDE, BDE, BCE$ 
and $BCD$ and so on. In this way we can specify all the simplexes included 
in the 4-simplex $ABCDE$. 
We can graphically denotes the geometrical structure of the 
4-simplex $ABCDE$. 
See Fig.\ref{fig:U-integration}.  

We now clarify the locations of the $U_i$ and $D$-functions 
on the 4-simplexes. 
The center of the figure in Fig.\ref{fig:U-integration} 
is corresponding to that of
the 4-simplex $ABCDE$ which has five neighboring 4-simplexes and is connected 
to the 4-simplexes by five dual links denoted by the thick lines.
The five points on the five thick lines 
correspond to the center of the five boundary tetrahedra, 
$BCDE$, $ACDE$, $ABDE$, $ABCE$ and $ABCD$, respectively. 
Arrows on the dual links indicate 
the direction of the dual link variables $U_i$, $U'_i$ and 
the thin lines with arrows indicate the direction of the 
product of the $D$-functions. 
As we can see in (\ref{Dfunction}) the product of the $D$-functions 
goes around a original triangle associated with the representation $J$. 
For example the boundary triangle between the tetrahedra $ABCD$ and 
$ABCE$ is $ABC$ and thus there is a thin line going from $ABCD$ to 
$ABCE$. 
Thus there are $10$ thin lines corresponding to the number of surrounding 
triangles. 
It should be noted that there are two dual link variables $U_i$ and $U'_i$ 
each of which belongs to the neighboring 4-simplexes separated by the 
boundary tetrahedron and is located 
on the same dual link $\tilde{l}_i$. 

Before getting into the details we briefly figure out 
how 15-$j$ symbols appear after the $dU$ integration. 
As we have shown in Fig.\ref{fig:U-integration}, there are five boundary 
tetrahedra in a 4-simplex.
Each tetrahedron has four triangles and thus there are four thin lines 
carrying $D$-functions associated to a $dU$ integration.
Since each $dU$ integration for the product of four $D$-functions
reproduces four 3-$j$ symbols, we obtain twenty 3-$j$ symbols 
for each 4-simplex. 
Then the ten out of the twenty 3-$j$ symbols lead to
a 15-$j$ symbols and the rest of the ten 3-$j$ symbols together with those 
from the neighbouring 4-simplex reproduce 
a trivial factor $2J+1$ attached to each tetrahedron.

We now explicitly show how the 15-$j$ symbols appear after the $dU$ 
integration. 
As a concrete example we focus on the $dU_1 dU'_1$ integration 
on the boundary tetrahedron $BCDE$ of the 4-simplex $ABCDE$. 
There appear eight $D$-functions in $dU_1 dU'_1$ integration, 
\begin{eqnarray*}
 I_{BCDE}
&=& \sum_{\{k_i\}} \int dU_1 dU'_1 ~
  D^{J_1}_{m'_1 k_1}(U^{\prime \dagger}_1)
  D^{J_1}_{k_1 m_1}(U_1) 
  D^{J_2}_{m_2 k_2}(U^{\dagger}_1)
  D^{J_2}_{k_2 m'_2}(U'_1) \\
&& \times D^{J_3}_{m'_3 k_3}(U^{\prime \dagger}_1)
  D^{J_3}_{k_3 m_3}(U_1)
  D^{J_4}_{m_4 k_4}(U^{\dagger}_1)
  D^{J_4}_{k_4 m'_4}(U'_1) 
\end{eqnarray*}

We note that the integration of four $D$-functions leads to 
the product of four 3-$j$ symbols\cite{Formula},
\begin{eqnarray}
 \int dU \prod_{i=1}^{4} D^{J_i}_{m_i n_i}(U) 
&=& \sum_{J,m,n} (2J+1) 
 \threej{J_1}{J_2}{J}{m_1}{m_2}{m} (-)^{J+m}
 \threej{J_3}{J_4}{J}{m_3}{m_4}{-m} \nonumber \\ 
&& \hspace*{4em} \times \threej{J_1}{J_2}{J}{n_1}{n_2}{n} (-)^{J+n}
 \threej{J_3}{J_4}{J}{n_3}{n_4}{-n}. 
\label{4D int}
\end{eqnarray}
Using the above formula and the following relation: 
\begin{equation}
 D^J_{mn} (U^{\dagger}) =
 D^{J *}_{nm} (U) = (-)^{n-m} D^J_{-n-m} (U), 
  \label{D conjugate}
\end{equation}
we can straightforwardly carry out $dU_1 dU'_1$ integration, 
\begin{eqnarray*}
 I_{BCDE}
  &=& \sum_{I,L,k,l,i}
  (-)^{k_1-m'_1} (-)^{k_2-m_2} (-)^{k_3-m'_3} (-)^{k_4-m_4}
  (2L+1)(2I+1) \\
 &\times& 
  \threej{J_1}{J_3}{L}{m_1}{m_3}{l_1}
  (-)^{L-l_1} \threej{L}{J_2}{J_4}{-l_1}{-m_2}{-m_4} \\
 &\times& 
  \threej{J_1}{J_3}{L}{k_1}{k_3}{l_2}
  (-)^{L-l_2} \threej{L}{J_2}{J_4}{-l_2}{-k_2}{-k_4} \\
 &\times& 
  \threej{J_2}{J_4}{I}{k_2}{k_4}{i_2}
  (-)^{I-i_2} \threej{I}{J_1}{J_3}{-i_2}{-k_1}{-k_3} \\
 &\times& 
  \threej{J_2}{J_4}{I}{m'_2}{m'_4}{i_1}
  (-)^{I-i_1} \threej{I}{J_1}{J_3}{-i_1}{-m'_1}{-m'_3}.
\end{eqnarray*}
We can evaluate $k_i$ summations by using
the following formula:
\begin{equation}
 \sum_{m_1 m_2}
  \threej{J_1}{J_2}{J}{m_1}{m_2}{m}
  \threej{J_1}{J_2}{J'}{m_1}{m_2}{m'}
  = \frac{1}{2J+1} ~ \delta_{JJ'} ~ \delta_{mm'},
\end{equation}
and then obtain 
\begin{eqnarray}
 I_{BCDE} = \sum_{L,l,i} (2L+1)
  &\threej{J_1}{J_3}{L}{m_1}{m_3}{l_1}& 
  \threej{L}{J_2}{J_4}{-l_1}{-m_2}{-m_4} 
  (-)^{L-l_1} (-)^{J_2-m_2} (-)^{J_4-m_4}  \nonumber \\
 \times
  &\threej{J_2}{J_4}{L}{m'_2}{m'_4}{i_1}& 
  \threej{L}{J_1}{J_3}{-i_1}{-m'_1}{-m'_3}
  (-)^{L-i_1} (-)^{J_1-m'_1} (-)^{J_3-m'_3}. \nonumber \\
 && \label{I_BCDE}
\end{eqnarray}
The factor $2L+1$ can be understood to be associated with
the tetrahedron $BCDE$.
Two 3-$j$ symbols carrying suffix $m_i$ 
which is assigned to be related to the center of 4-simplex $ABCDE$, 
will be used to construct the 15-$j$ symbol
associated with 4-simplex $ABCDE$, 
while the other two 3-$j$ symbols carrying the suffix $m'_i$,
will be used to construct the 15-$j$ symbol
associated with the neighboring 4-simplex of $ABCDE$
which share the common tetrahedron $BCDE$.

Hereafter we introduce a graphical method to explain the 
manipulation of formulae and the derivation leading to the 
generalized ``15-$j$ symbol''. 
The essence of this method is to represent the 3-$j$ symbol graphically 
by trivalent vertex (3-vertex) with a sign factor.
The sign is related to the correspondence between the ordering of 
column in the 3-$j$ symbol and right- or left-handed ordering of 
the angular momentum $J_i$ on the trivalent graph. 
We choose to define the following particular ordering 
and the sign factor: 
\begin{equation}
\threej{J_1}{J_2}{J_3}{m_1}{m_2}{m_3} = 
\begin{minipage}[c]{.2\textwidth}
 \epsfxsize=\textwidth \epsfbox{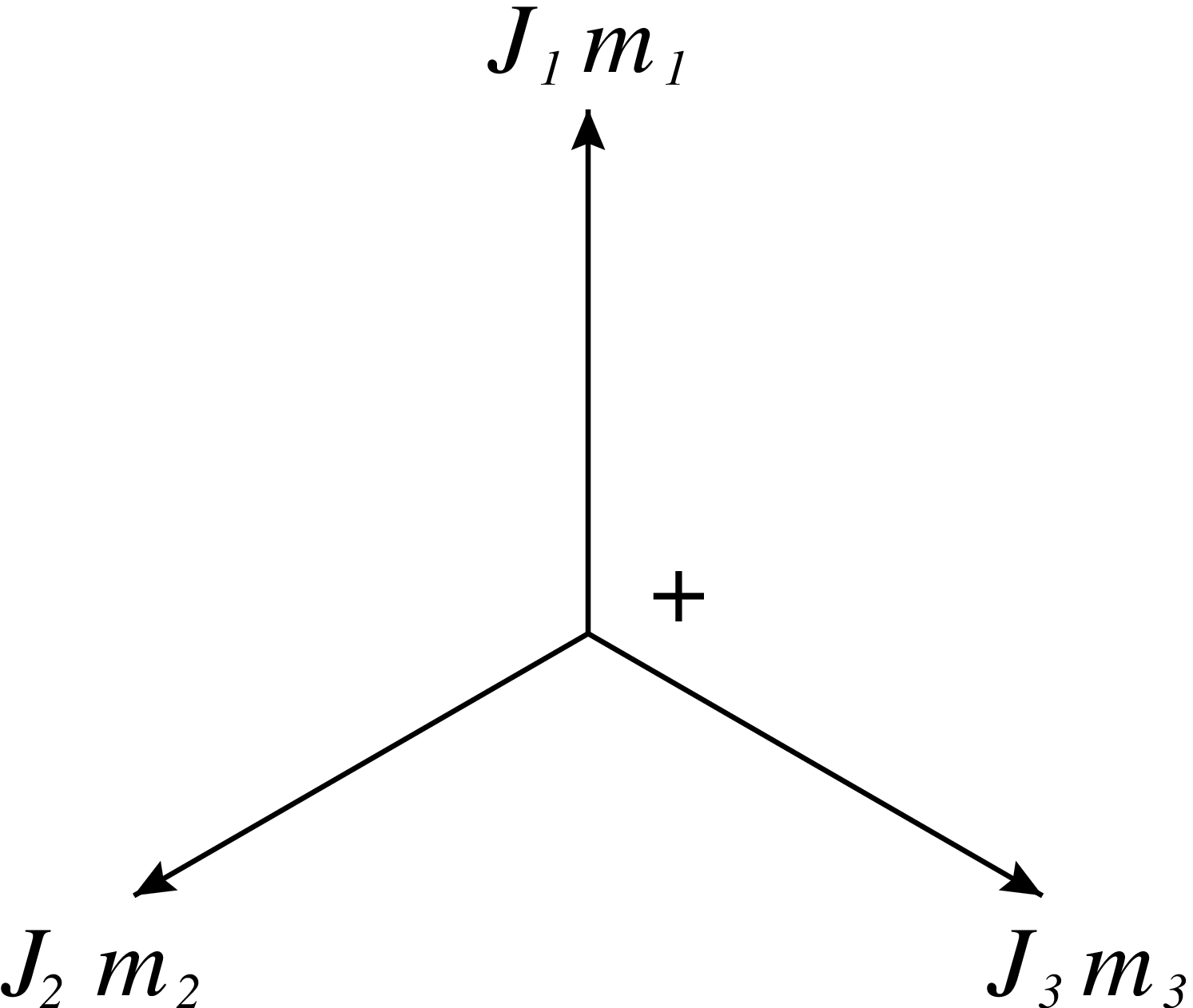}
\end{minipage}
=
\begin{minipage}[c]{.2\textwidth}
 \epsfxsize=\textwidth \epsfbox{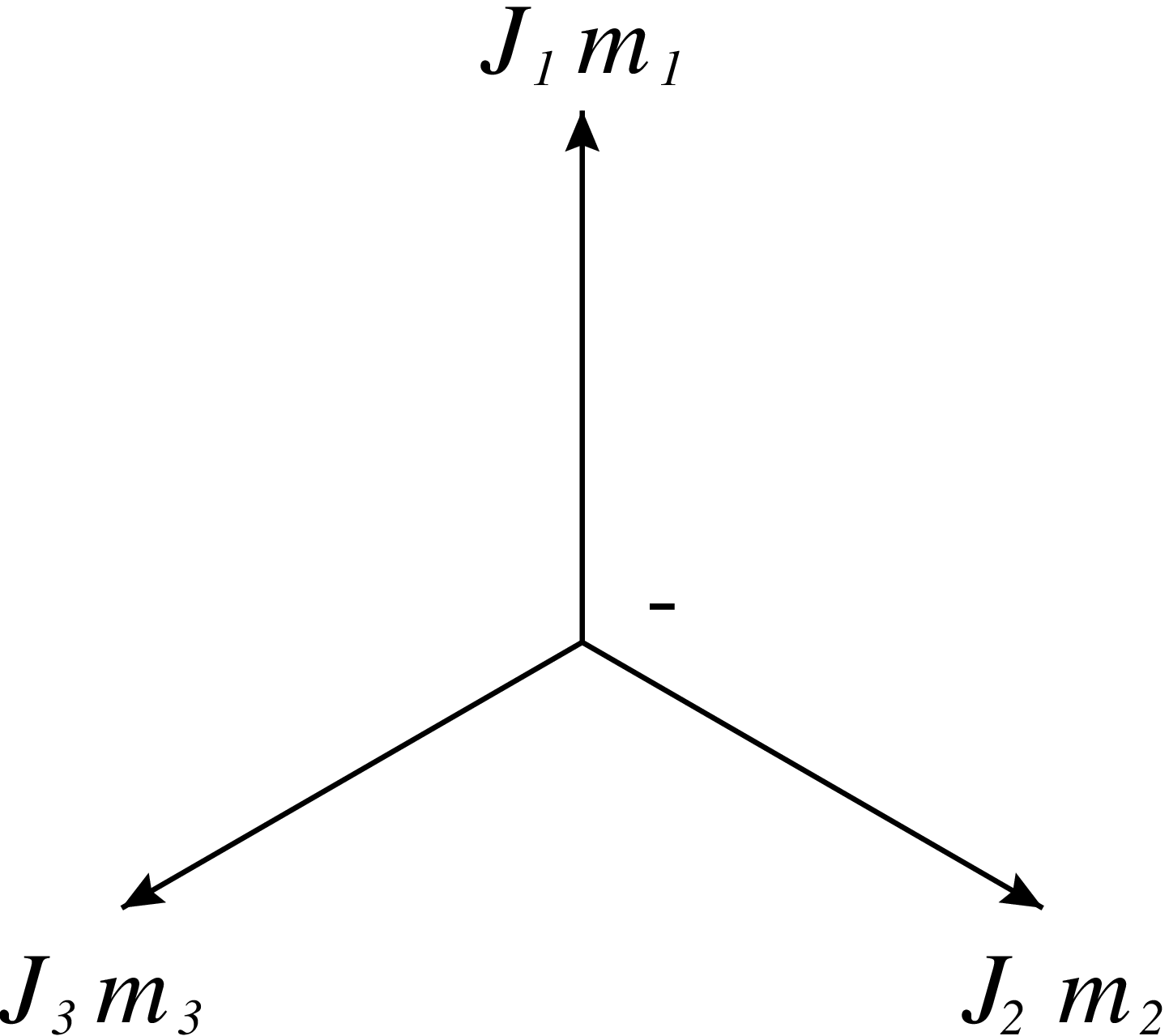}
\end{minipage}.
\end{equation}
We then define the reversal of the arrows as
$$
\begin{minipage}[c]{.2\textwidth}
 \epsfxsize=\textwidth \epsfbox{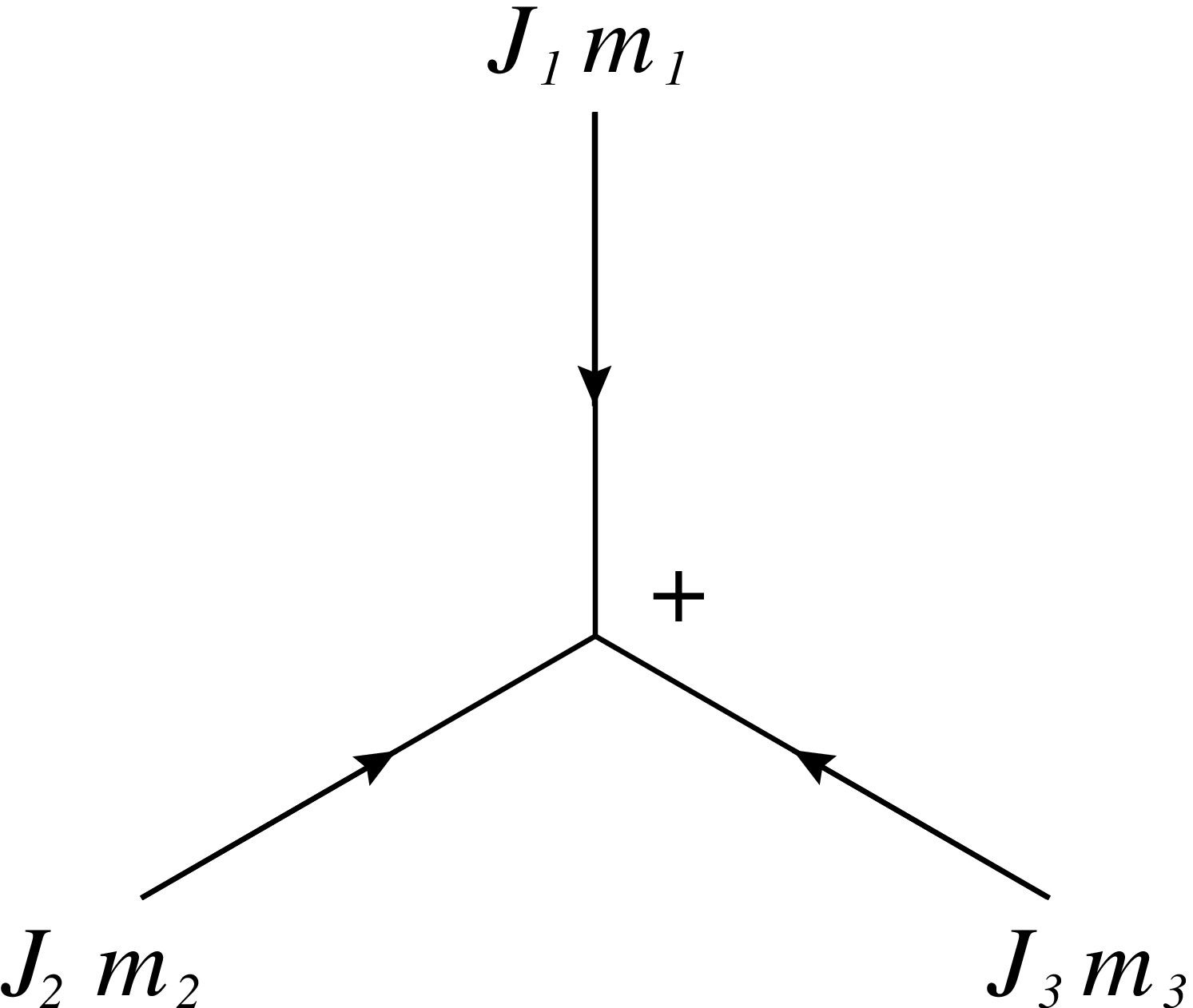}
\end{minipage}
= 
\sum_{m'_1, m'_2, m'_3}
  g^{J_1}_{m_1,m'_1} g^{J_2}_{m_2,m'_2} g^{J_3}_{m_3,m'_3}
 \threej{J_1}{J_2}{J_3}{m'_1}{m'_2}{m'_3}, 
$$
where $g^J_{m,m'} = \delta_{m,-m'} (-1)^{J-m}$ is
the $SU(2)$ invariant metric. 
The following identity
$$
 \threej{J_1}{J_2}{J_3}{m_1}{m_2}{m_3}
 =
\sum_{m'_1, m'_2, m'_3}
  g^{J_1}_{m_1,m'_1} g^{J_2}_{m_2,m'_2} g^{J_3}_{m_3,m'_3}
 \threej{J_1}{J_2}{J_3}{m'_1}{m'_2}{m'_3}
$$
can be graphically represented by
$$
 \begin{minipage}{.2\textwidth}
  \epsfxsize=\textwidth \epsfbox{def_3j-1.eps}
 \end{minipage}
=
 \begin{minipage}{.2\textwidth}
  \epsfxsize=\textwidth \epsfbox{def_3j-3.eps}
 \end{minipage}.
$$

In this way the direction of the arrow is related to the consistent choice of 
$SU(2)$ invariant metric. 
For example we can connect two 3-$j$ symbols by a metric 
\begin{eqnarray*}
 \sum_{m,m',m'_3,m'_4}
 \threej{J_1}{J_2}{J}{m_1}{m_2}{m}
  g^J_{m,m'} g^{J_3}_{m_3,m'_3} g^{J_4}_{m_4,m'_4}
 \threej{J}{J_3}{J_4}{m'}{m'_3}{m'_4}, 
\end{eqnarray*}
which is graphically represented by 
$$
\sum_{m}
 \begin{minipage}{.3\textwidth}
  \epsfxsize=\textwidth \epsfbox{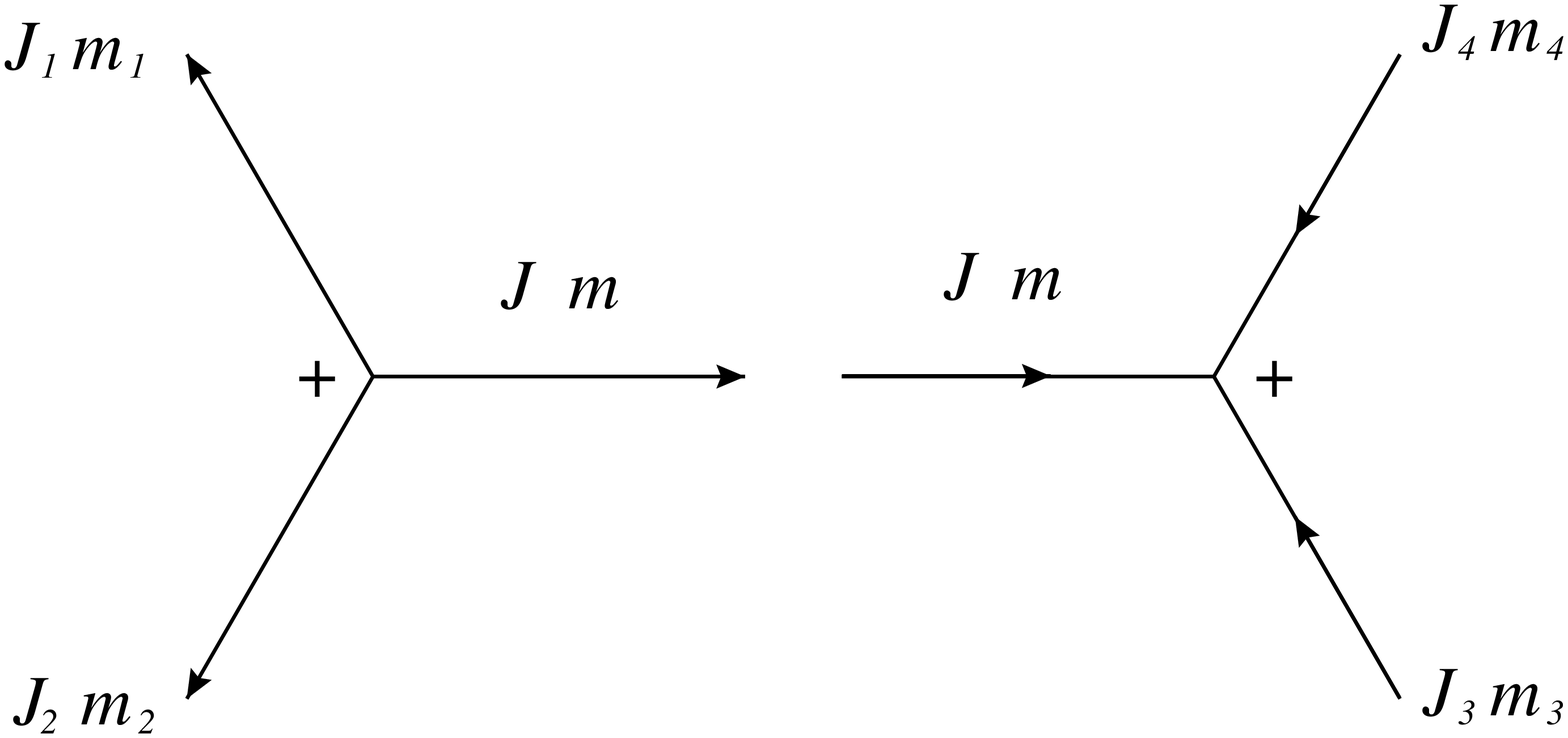}
 \end{minipage} 
=
 \begin{minipage}{.3\textwidth}
  \epsfxsize=\textwidth \epsfbox{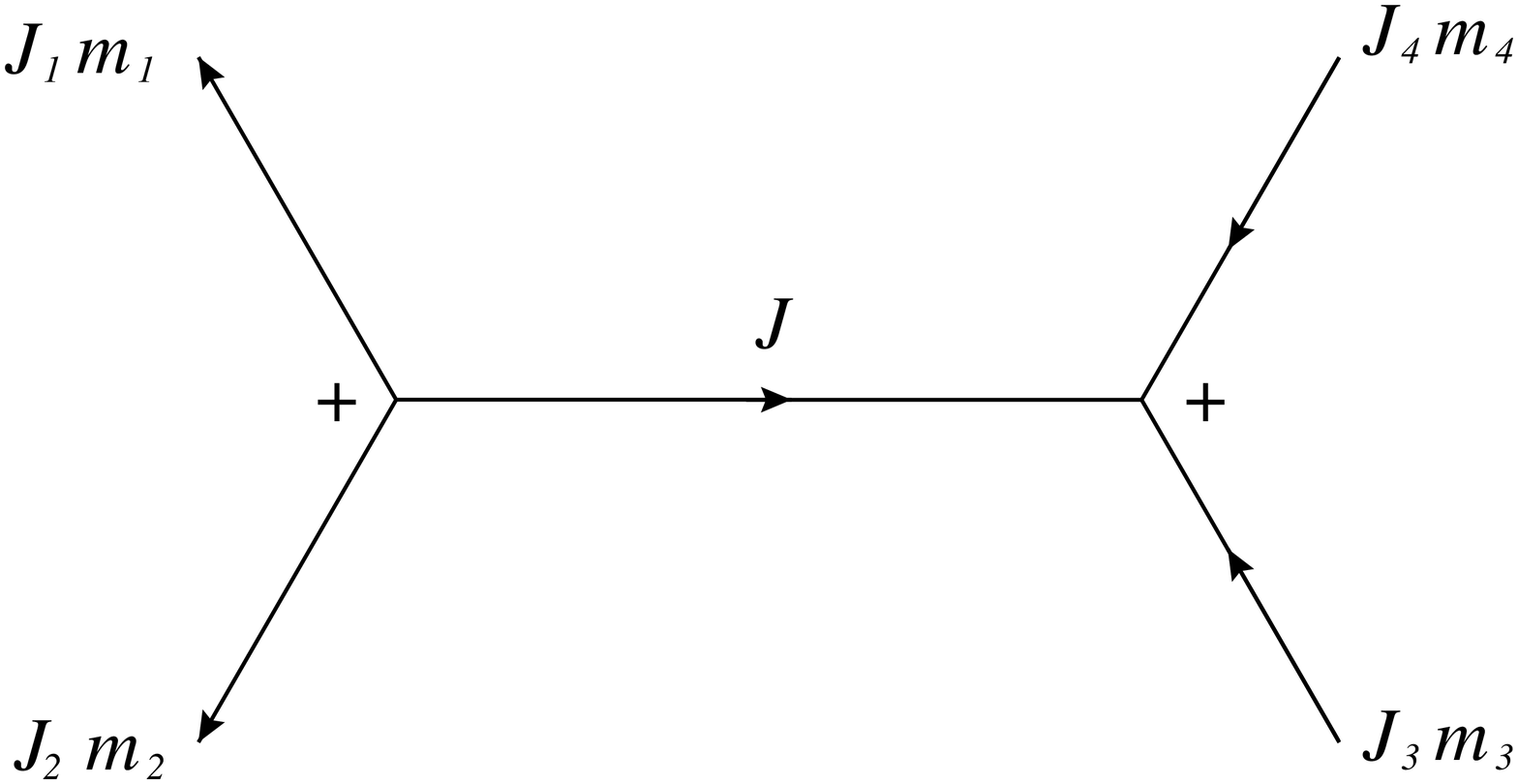}
 \end{minipage}.
$$ 

In general any $3nj$-symbols ($n=2,3,4,5,..$) can be graphically
represented by the closed trivalent graph. 
For example 6-$j$ symbol can be decomposed into
four 3-$j$ symbols by the formula 
\begin{eqnarray*}
 \sixj{J_1}{J_2}{J_3}{J_4}{J_5}{J_6} 
&=& \sum_{\hbox{\scriptsize all $m_i$}}
 (-1)^{\sum_i (J_i - m_i)}
 \threej{J_1}{J_2}{J_3}{-m_1}{-m_2}{-m_3}\nonumber \\
&& \hspace{-1cm} \times 
 \threej{J_1}{J_5}{J_6}{m_1}{-m_5}{m_6} 
 \threej{J_4}{J_2}{J_6}{m_4}{m_2}{-m_6}
 \threej{J_4}{J_5}{J_3}{-m_4}{m_5}{m_3},
\end{eqnarray*}
where the factor $(-1)^{\sum_i (J_i - m_i)}$ can be understood as a sign 
factor coming from the invariant metric.
The graphical presentation of the above 6-$j$ symbol can be given 
$$
 \begin{minipage}{.25\textwidth}
  \epsfxsize=\textwidth \epsfbox{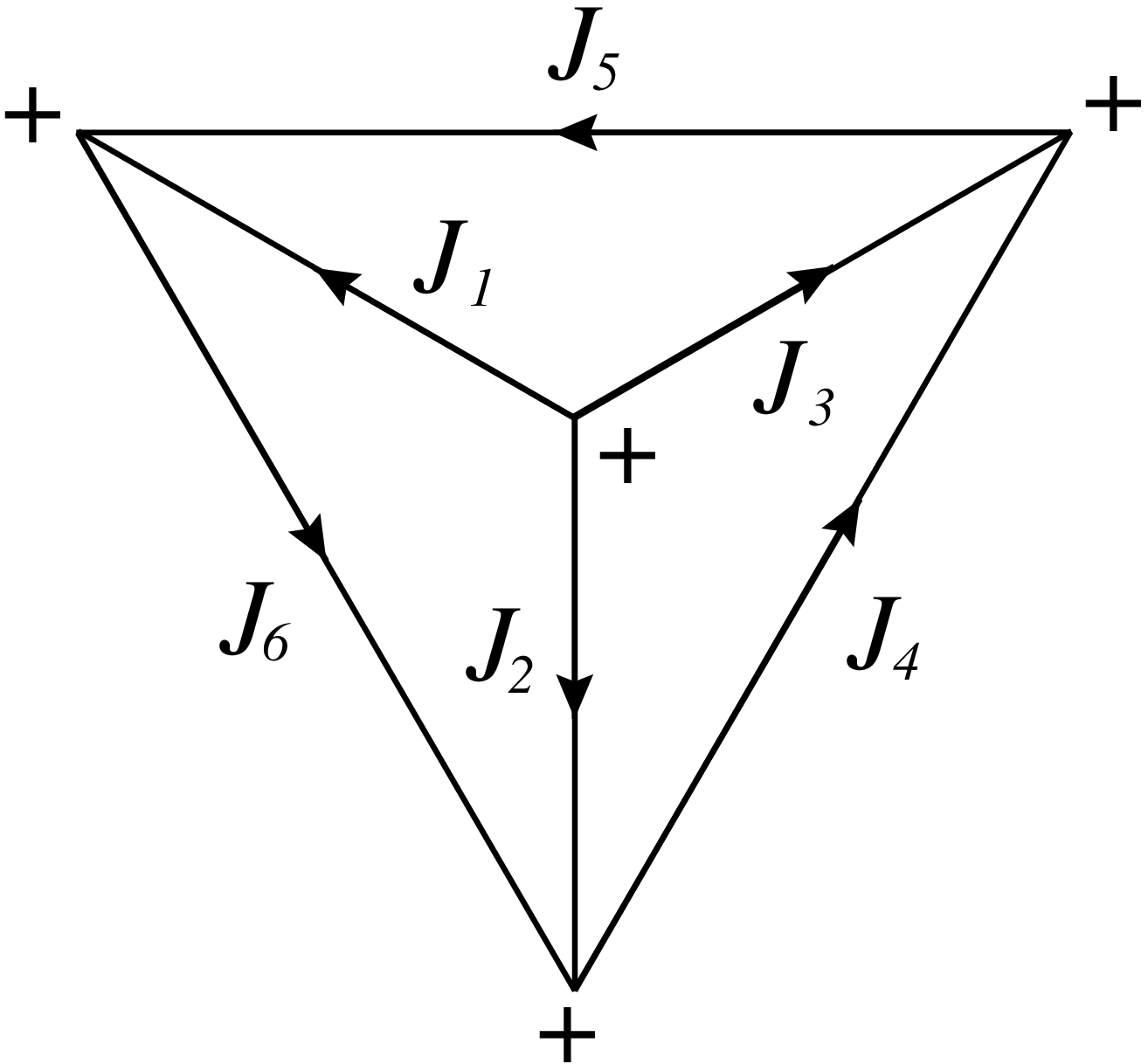}
 \end{minipage}
=
 \begin{minipage}{.25\textwidth}
  \epsfxsize=\textwidth \epsfbox{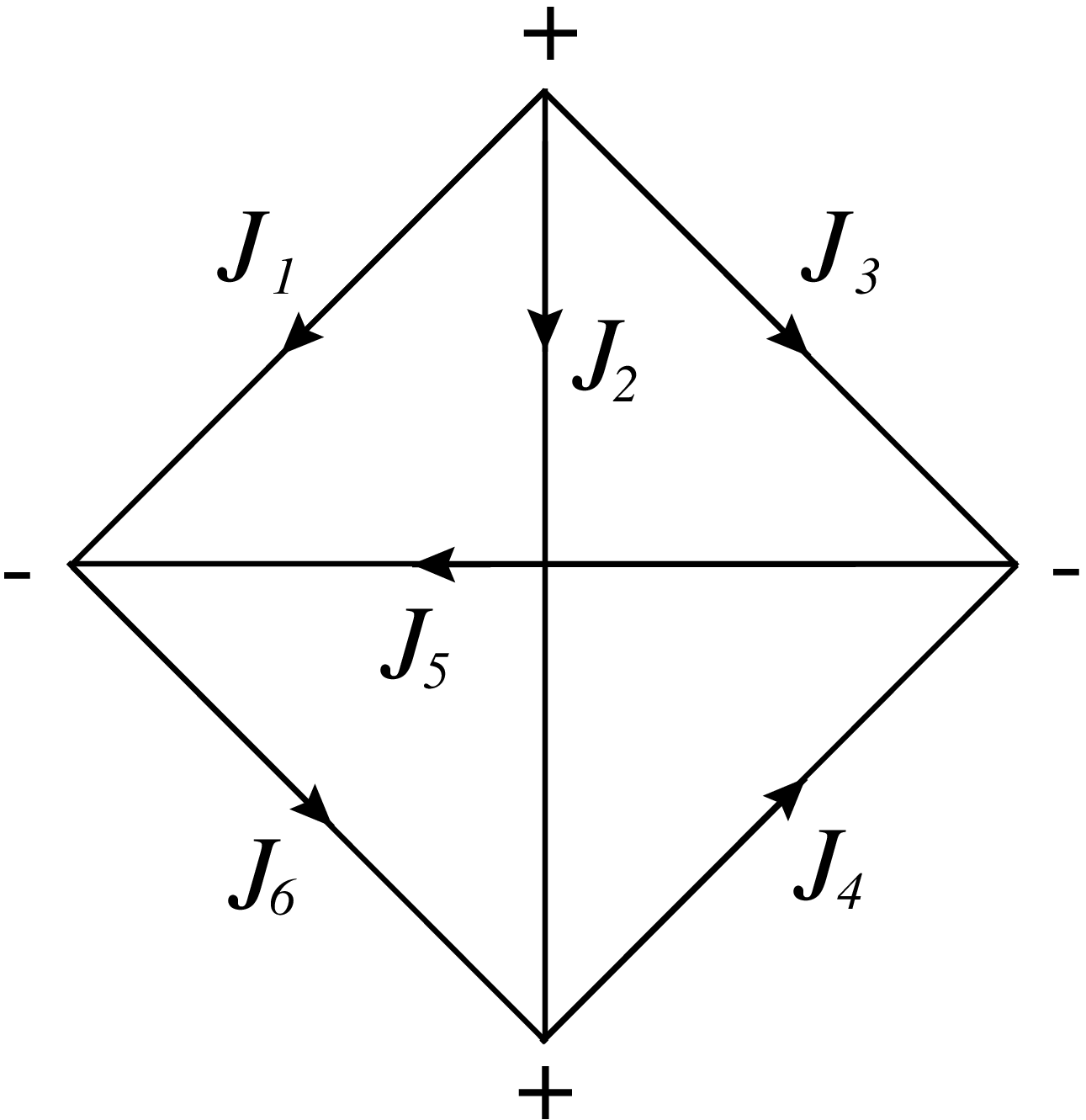}
 \end{minipage},
$$
where the rules for the signs on the vertices and the arrows on the lines 
can be understood. 
Analogously for more complicated 15-$j$ symbols, we need ten 3-$j$ symbols 
and can be represented by closed trivalent graph with ten vertices 
with sign factors. 

Using the graphical method,
we can represent the formula (\ref{I_BCDE}) as 
\begin{equation}
 I_{BCDE} = \sum_{L} (2L+1)
  \begin{minipage}[c]{.3\textwidth}
   \epsfxsize=\textwidth \epsfbox{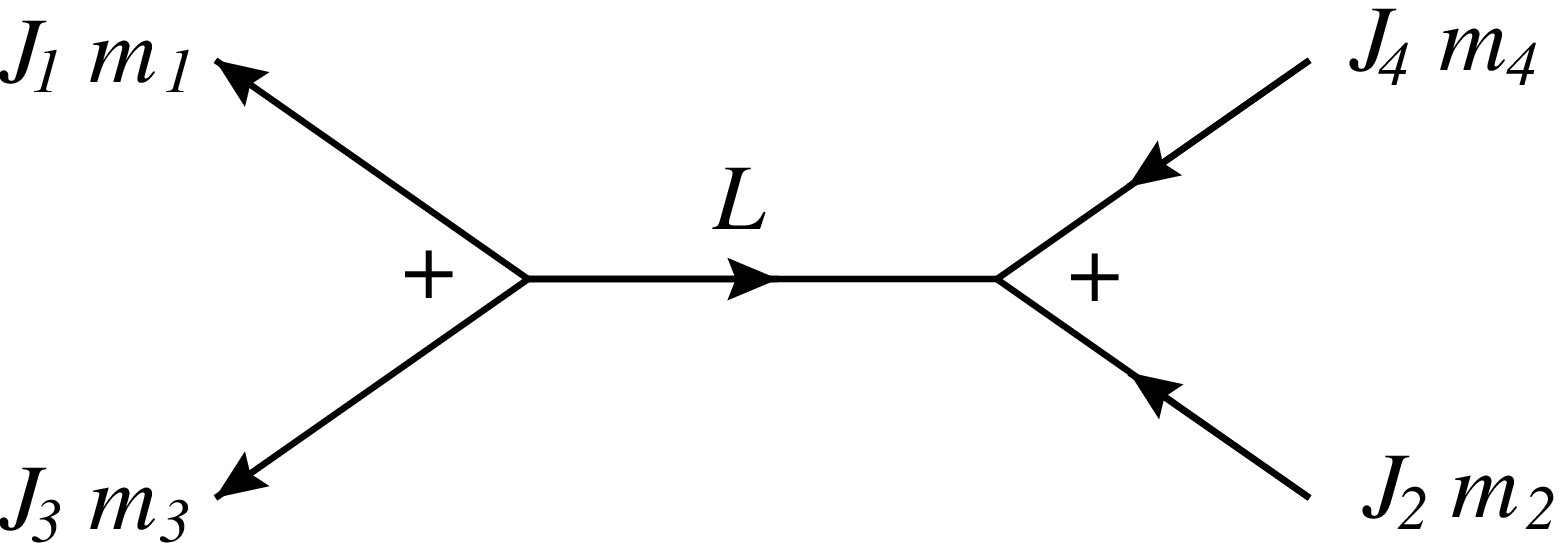}
  \end{minipage}
  \hspace*{1em}
  \begin{minipage}[c]{.3\textwidth}
   \epsfxsize=\textwidth \epsfbox{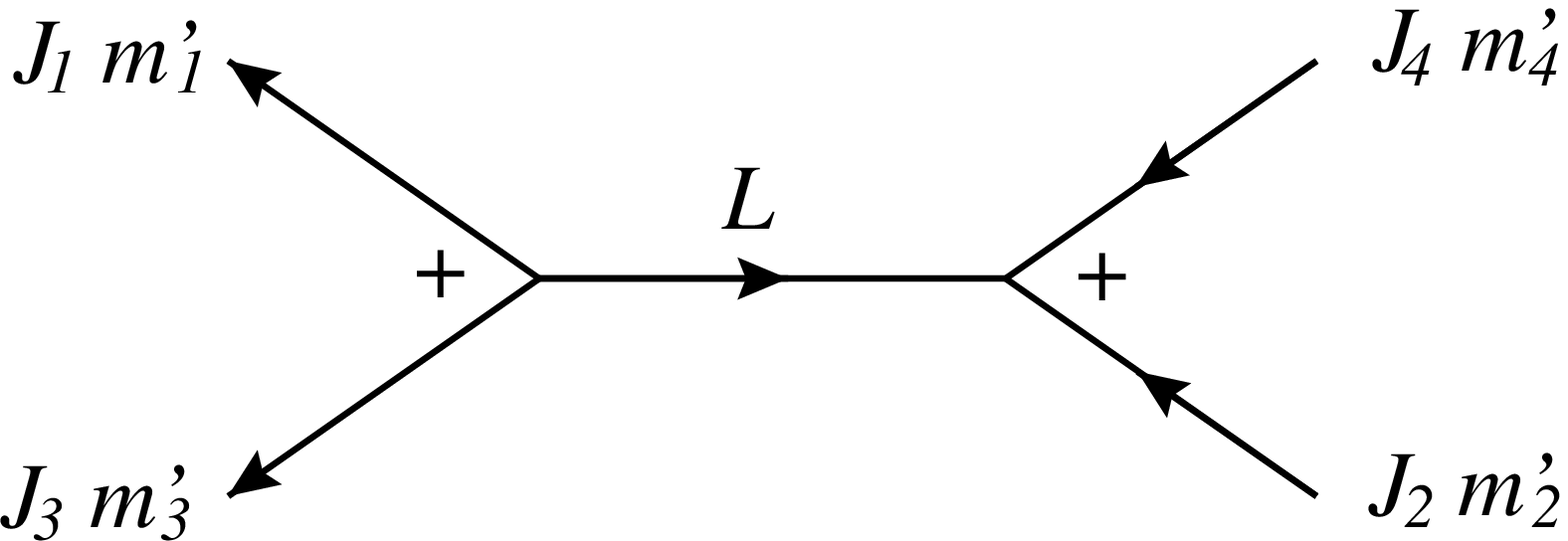}
  \end{minipage}.
\label{I_BCDE1}
\end{equation}
In accordance with (\ref{I_BCDE}) the factor $(2L+1)$ and the summation 
of the magnitude of the angular momentum $L$ are explicitly written 
while the summation of the third component $l_1,i_1$ are not written explicitly 
and should be understood in the graphical representation of the internal line.
It should, however, be noted that the above presentation is not the 
unique expression. 
There are three other equivalent expressions; 
\begin{equation}
 \renewcommand{\arraystretch}{5}
 \begin{array}{lcr}
 I_{BCDE}
  &=&
  \displaystyle
  \sum_{L}(2L+1)
  \begin{minipage}[c]{.3\textwidth}
   \epsfxsize=\textwidth \epsfbox{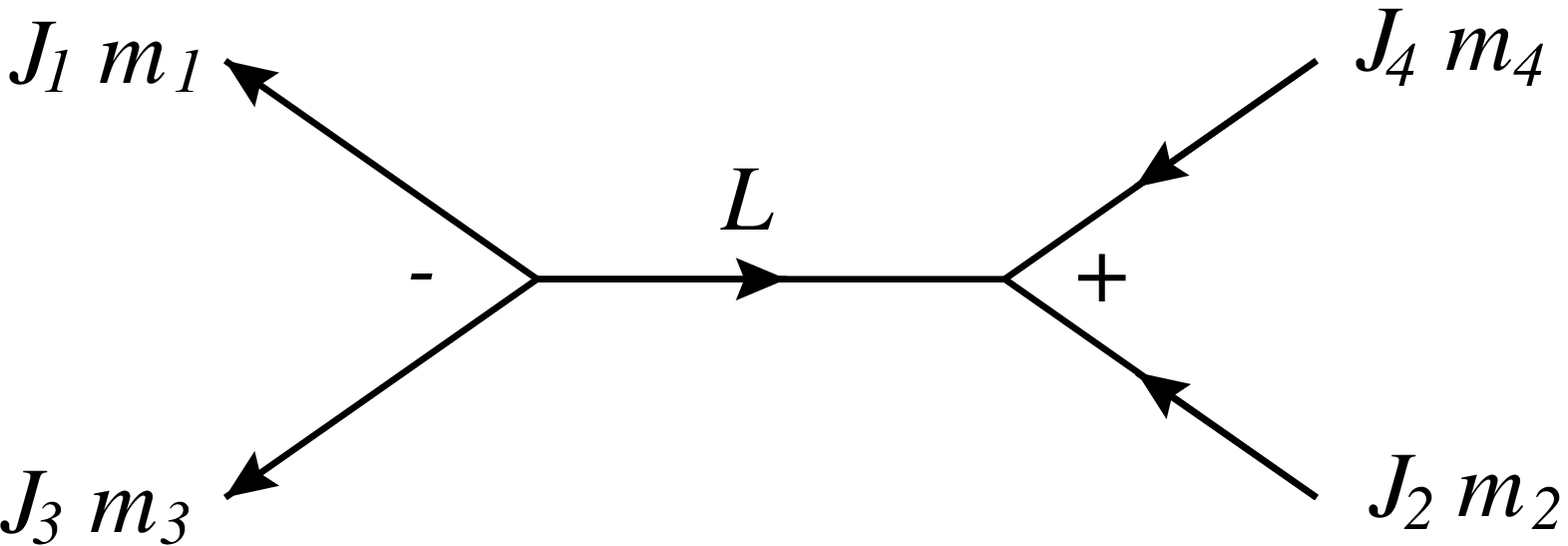}
  \end{minipage}
  \hspace*{1em}
  \begin{minipage}[c]{.3\textwidth}
   \epsfxsize=\textwidth \epsfbox{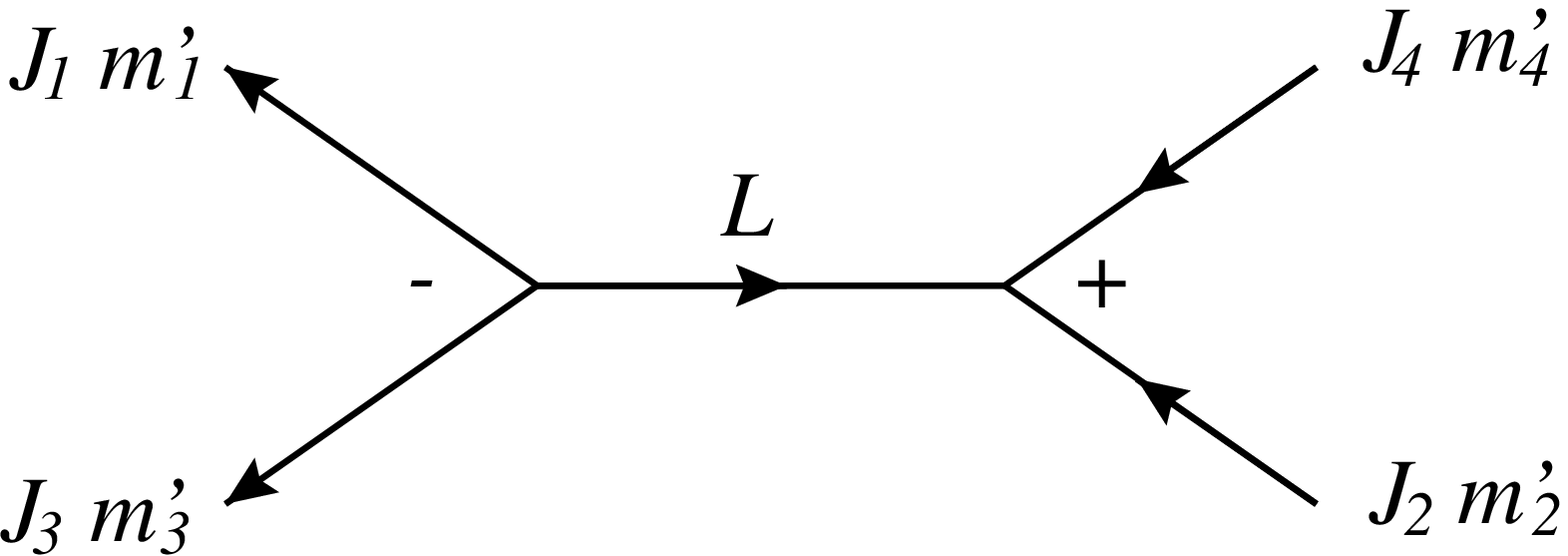}
  \end{minipage} \\
  &=&
  \displaystyle
  \sum_{L}(2L+1)
  \begin{minipage}[c]{.3\textwidth}
   \epsfxsize=\textwidth \epsfbox{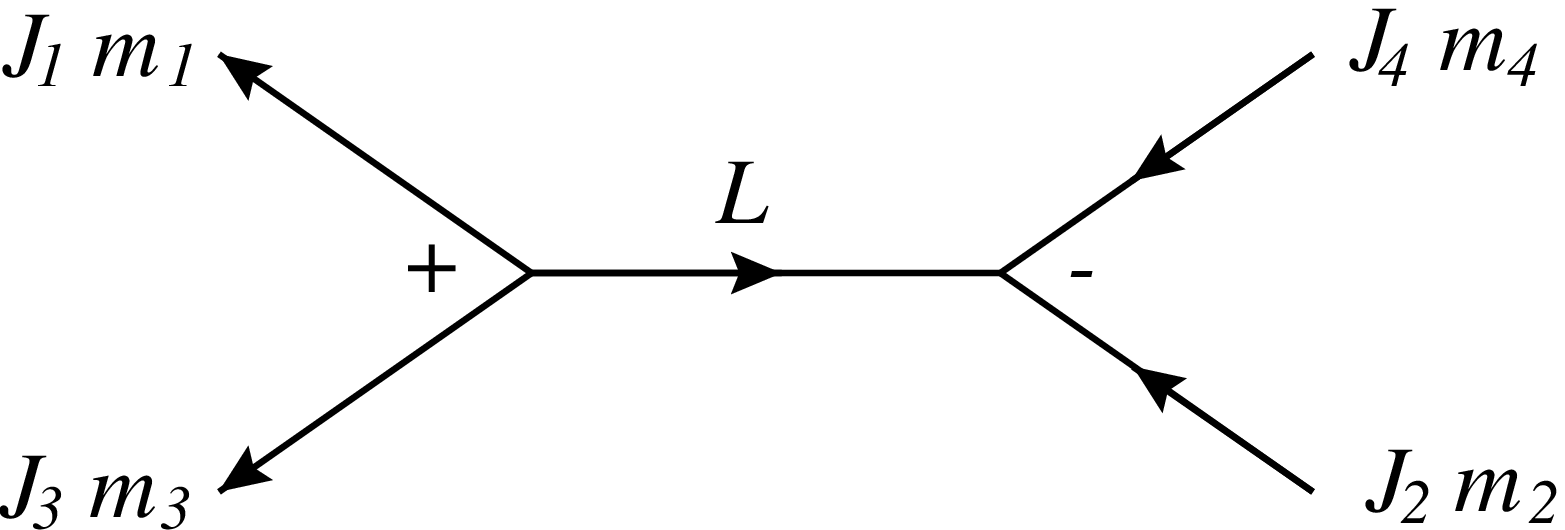}
  \end{minipage}
  \hspace*{1em}
  \begin{minipage}[c]{.3\textwidth}
   \epsfxsize=\textwidth \epsfbox{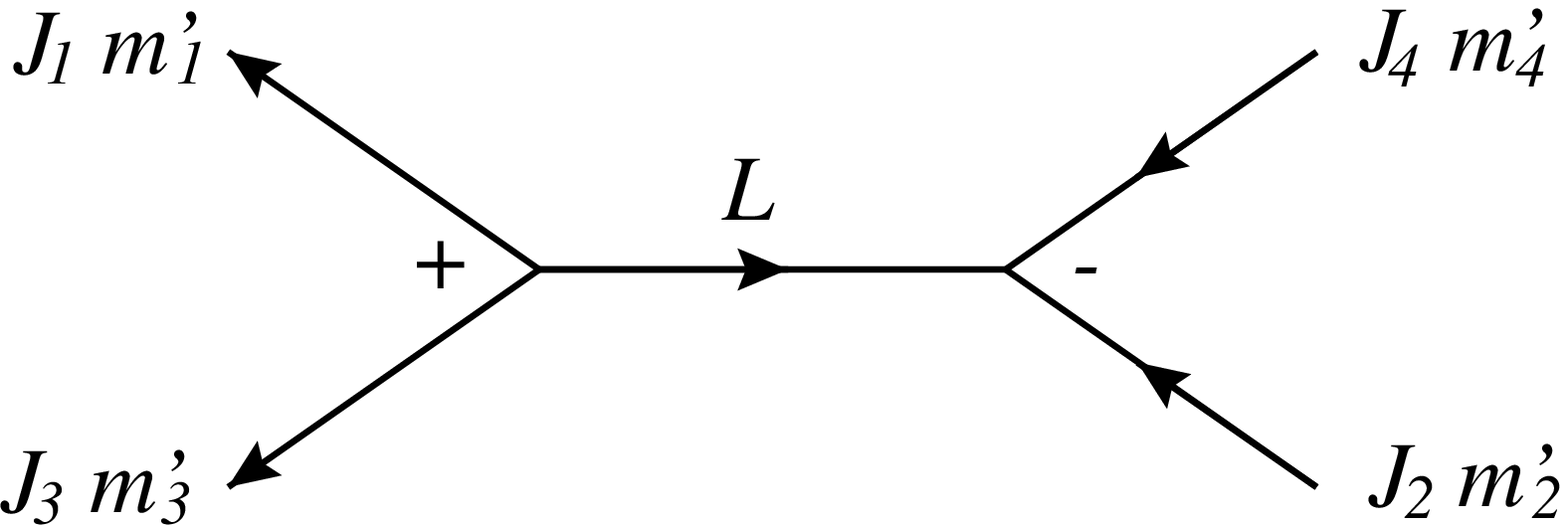}
  \end{minipage} \label{sign_ambiguity}\\
  &=&
  \displaystyle
  \sum_{L}(2L+1)
  \begin{minipage}[c]{.3\textwidth}
   \epsfxsize=\textwidth \epsfbox{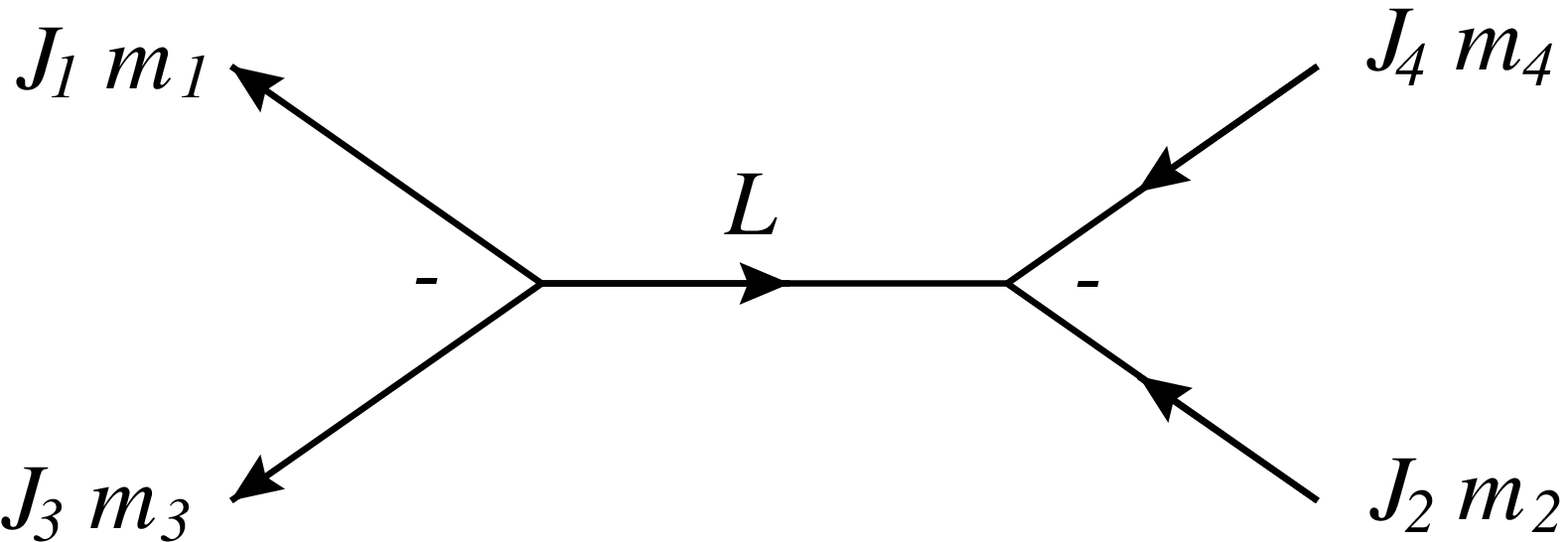}
  \end{minipage} 
  \hspace*{1em}
  \begin{minipage}[c]{.3\textwidth}
   \epsfxsize=\textwidth \epsfbox{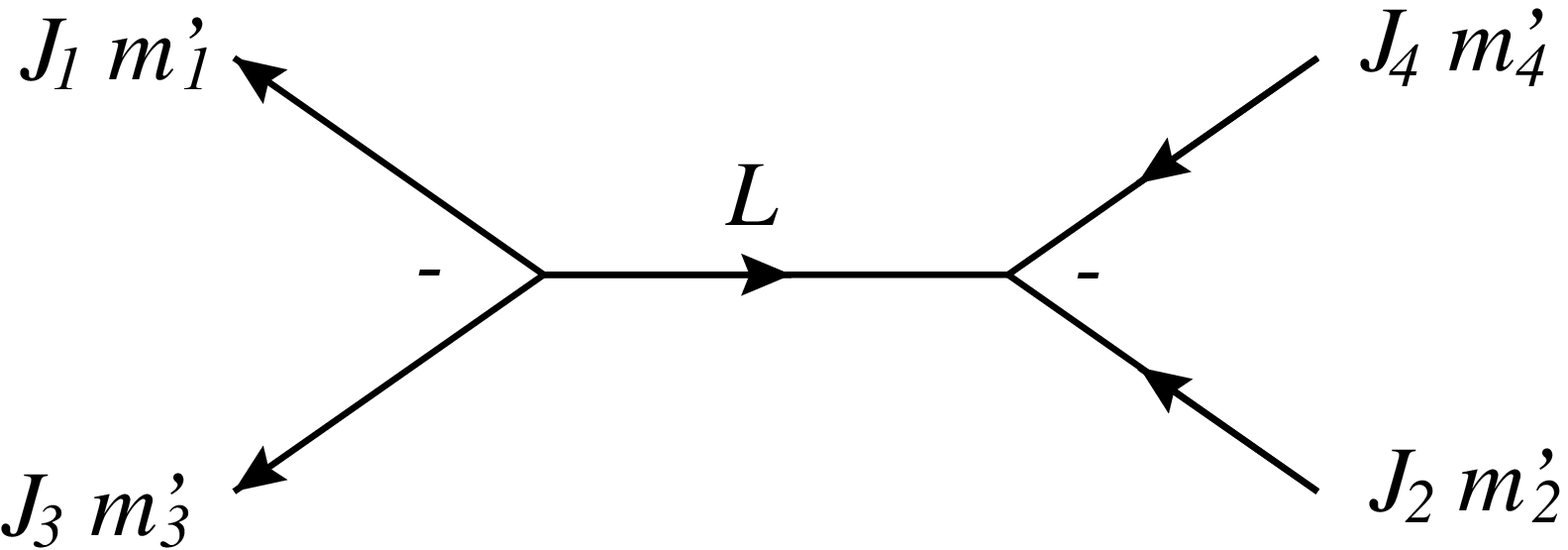}
  \end{minipage}.
\end{array}
\label{I_BCDE2}
\end{equation}
It is important to recognize that the corresponding sign factors are the 
same in each pair of the four internal line diagrams. 
In the above particular example we have chosen the dual link integration 
$dU_1dU'_1$ via the boundary tetrahedron $BCDE$ and obtained the integrated 
result $I_{BCDE}$. 
We need to carry out four other dual link integrations; 
$dU_2dU'_2$, $dU_3dU'_3$, $dU_4dU'_4$ and $dU_5dU'_5$ to perform the 
full dual link integration of the 4-simplex $ABCDE$. 
They can be carried out in the same way as $dU_1dU'_1$ integration.  

Using the formulation explained in the above, 
we can evaluate the $dU_i$ integration graphically,
\begin{eqnarray}
&& \int \prod_{i=1}^{5} dU_i dU'_i 
  \begin{minipage}[c]{.5\textwidth}
   \epsfxsize=\textwidth \epsfbox{U-int2.eps}
  \end{minipage} \label{dU_int_start} \\
&=& \prod_{i=1}^{5} \sum_{J_i} (2J_i + 1)
  \begin{minipage}[c]{.6\textwidth}
   \epsfxsize=\textwidth \epsfbox{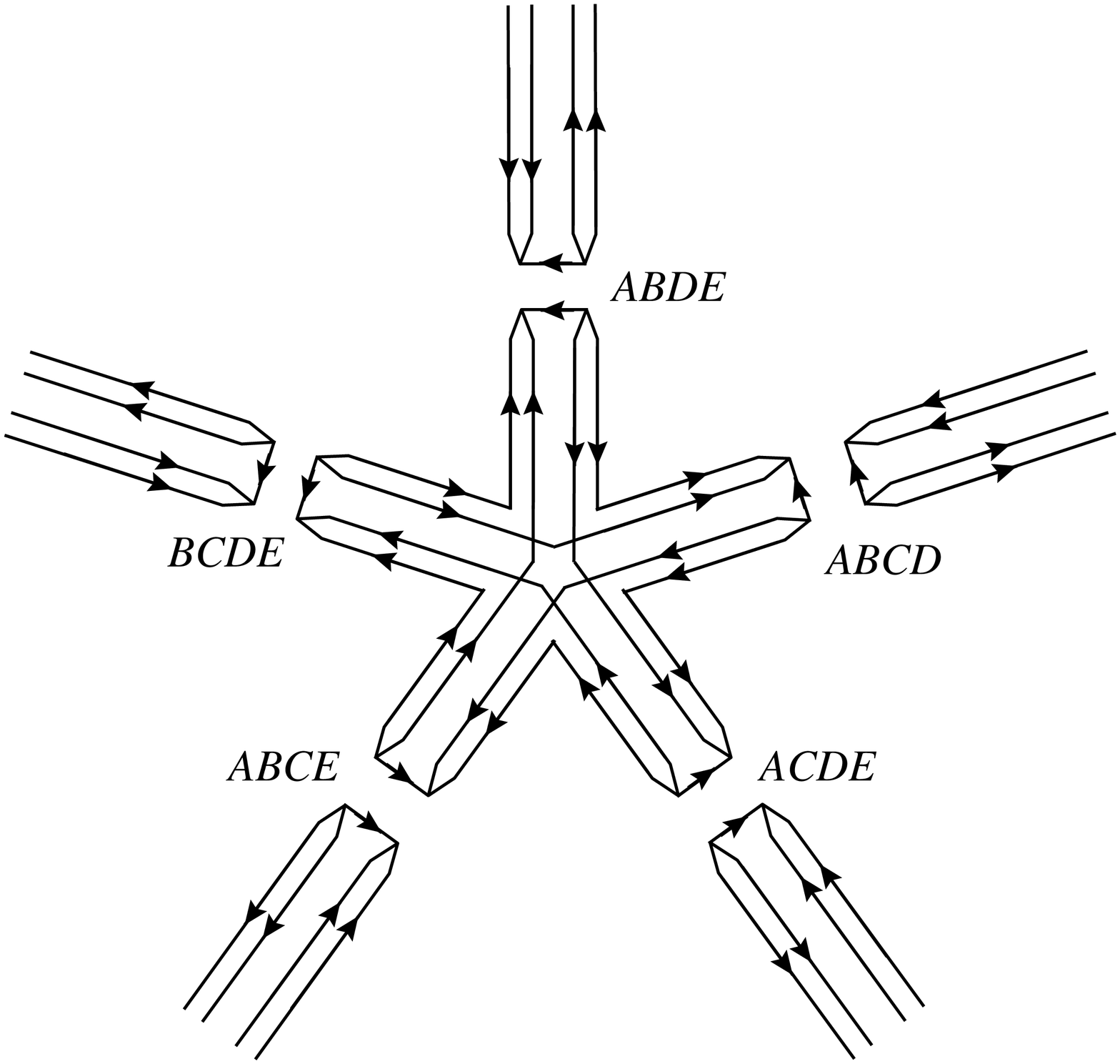}
  \end{minipage}, \label{dU_int_end}
\end{eqnarray}
where $J_1 = J_{BCDE}$, $J_2 = J_{ABCE}$, $J_3 = J_{ACDE}$,
$J_4 = J_{ABCD}$ and $J_5 = J_{ABDE}$.
Here $dU_i dU'_i$ integration graphically represented by (\ref{I_BCDE1}) 
has cut down the thin lines and newly generated a connected 
line representing the boundary tetrahedron.
The closed trivalent graph with ten vertices in the center can be 
rewritten in topologically equivalent ways, 
\begin{eqnarray}
&&  \begin{minipage}[c]{.35\textwidth}
   \epsfxsize=\textwidth \epsfbox{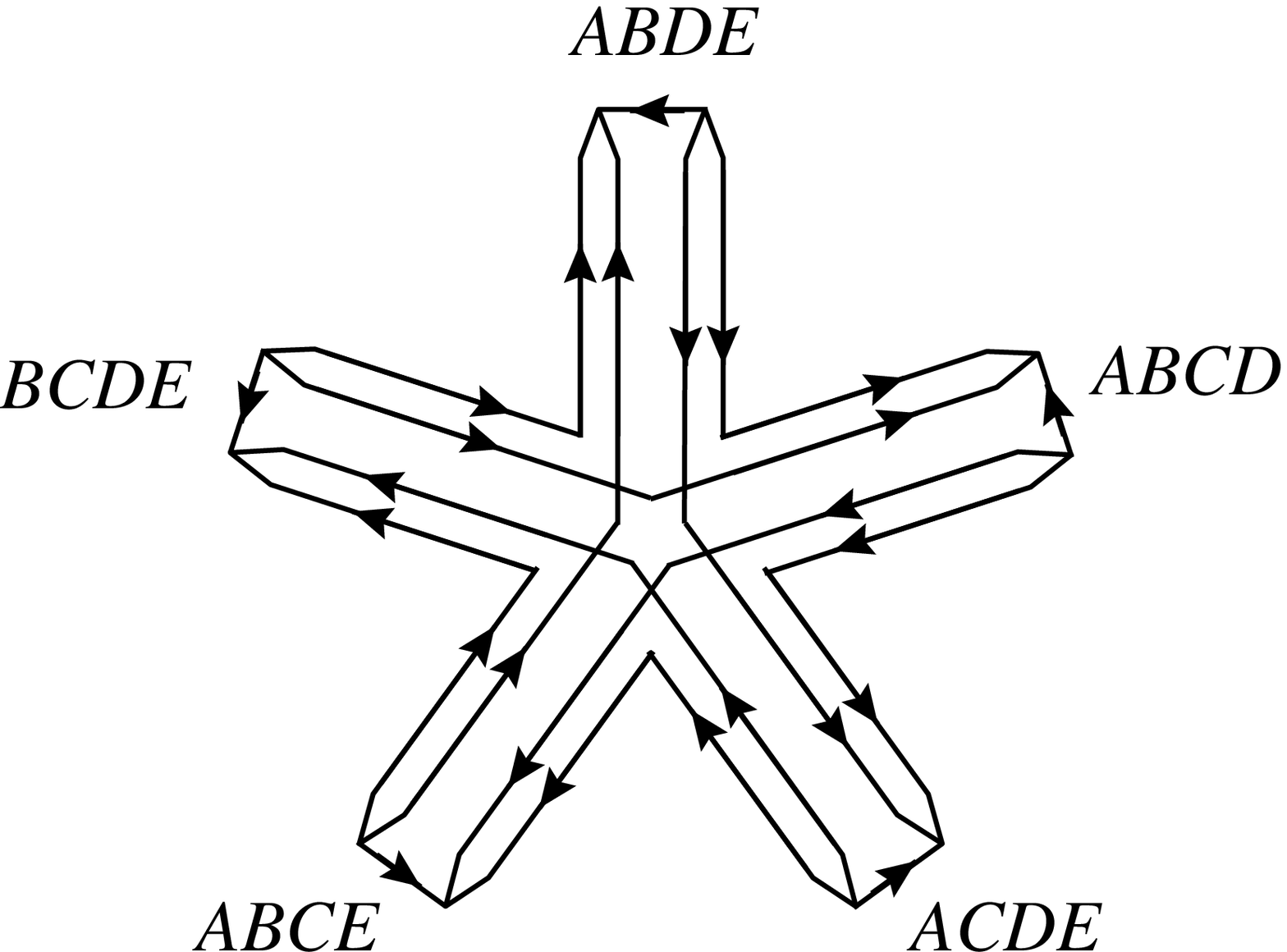}
  \end{minipage} \nonumber \\
&=&
  \begin{minipage}[c]{.35\textwidth}
   \epsfxsize=\textwidth \epsfbox{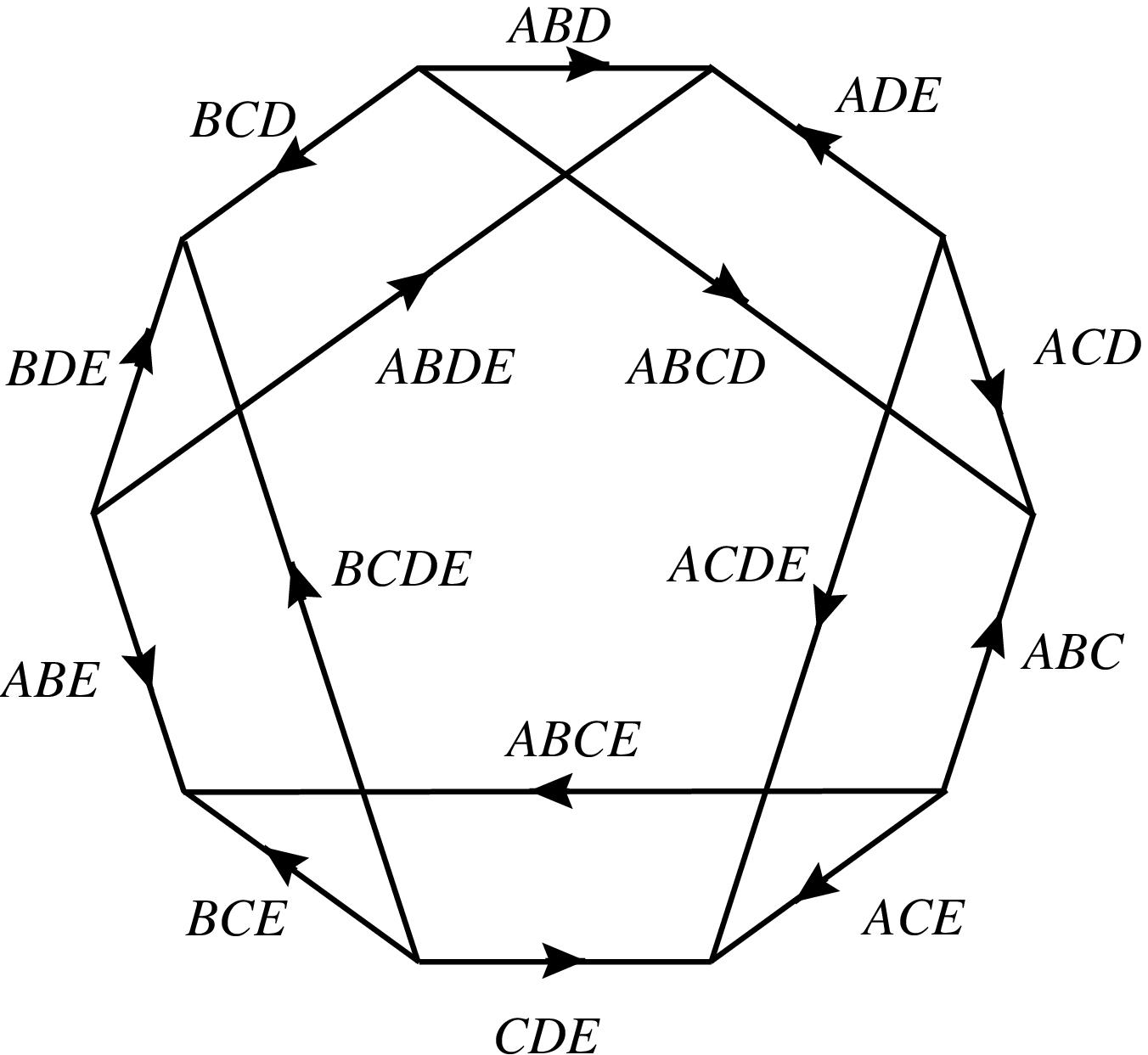}
  \end{minipage}
=
  \begin{minipage}[c]{.35\textwidth}
   \epsfxsize=\textwidth \epsfbox{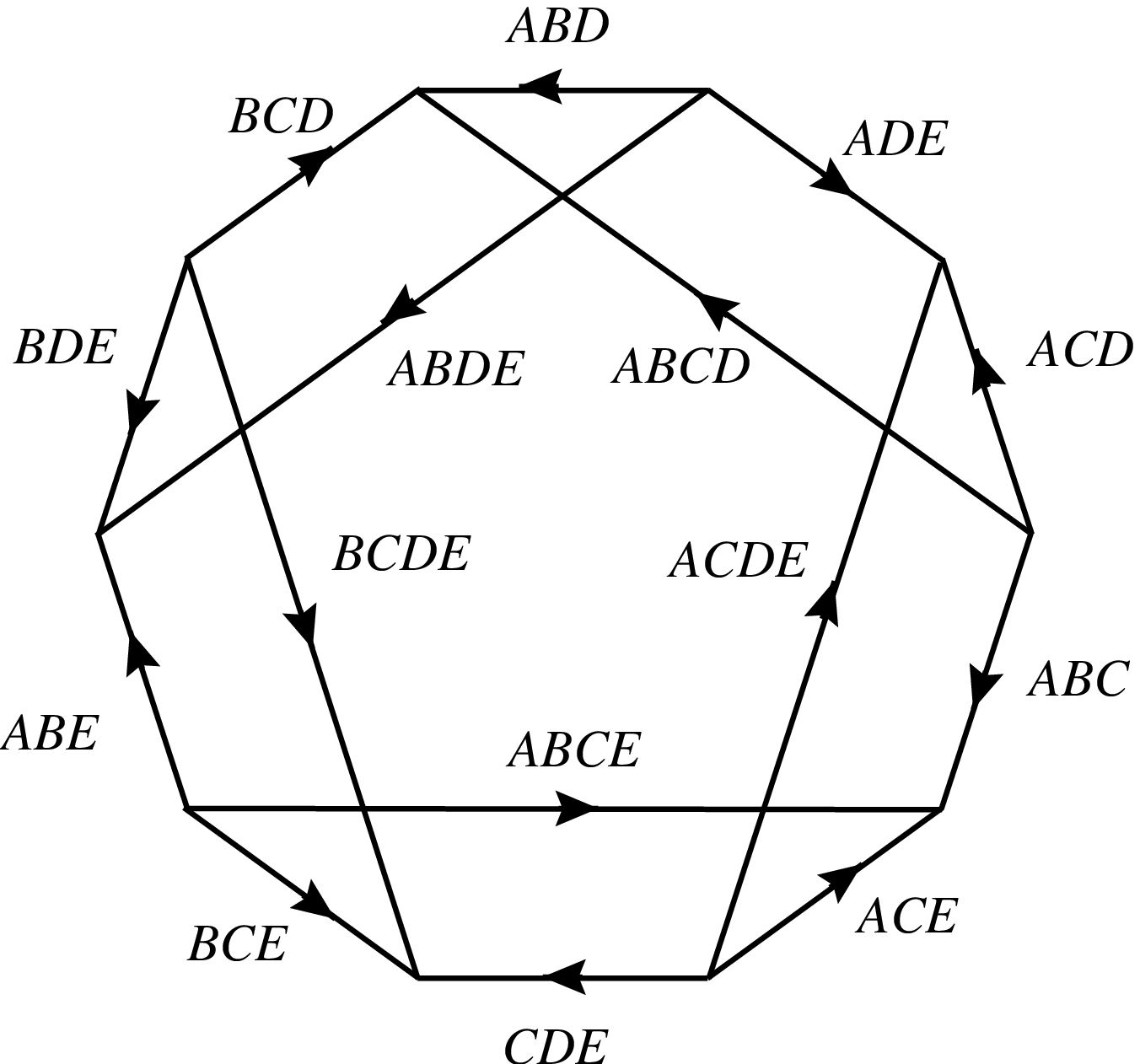}
  \end{minipage}.
\label{def of 15j}
\end{eqnarray}
The topological equivalence of the 1st and 2nd graphs are obvious while 
the 2nd graph equals to the 3rd graph by the following formula:
\begin{eqnarray}
&& \sum_{m_3} 
  \threej{J_1}{J_2}{J_3}{m_1}{m_2}{m_3}
  \threej{J_3}{J_4}{J_5}{-m_3}{-m_4}{-m_5}
  (-)^{\sum_{i=3}^{5} (J_i - m_i)} \nonumber \\
&=&
 \sum_{m_3} 
  \threej{J_1}{J_2}{J_3}{-m_1}{-m_2}{-m_3}
  \threej{J_3}{J_4}{J_5}{m_3}{m_4}{m_5}
  (-)^{\sum_{i=1}^{3} (J_i - m_i)},
\end{eqnarray}
which can be graphically represented by 
\[
  \begin{minipage}[c]{.4\textwidth}
   \epsfxsize=\textwidth \epsfbox{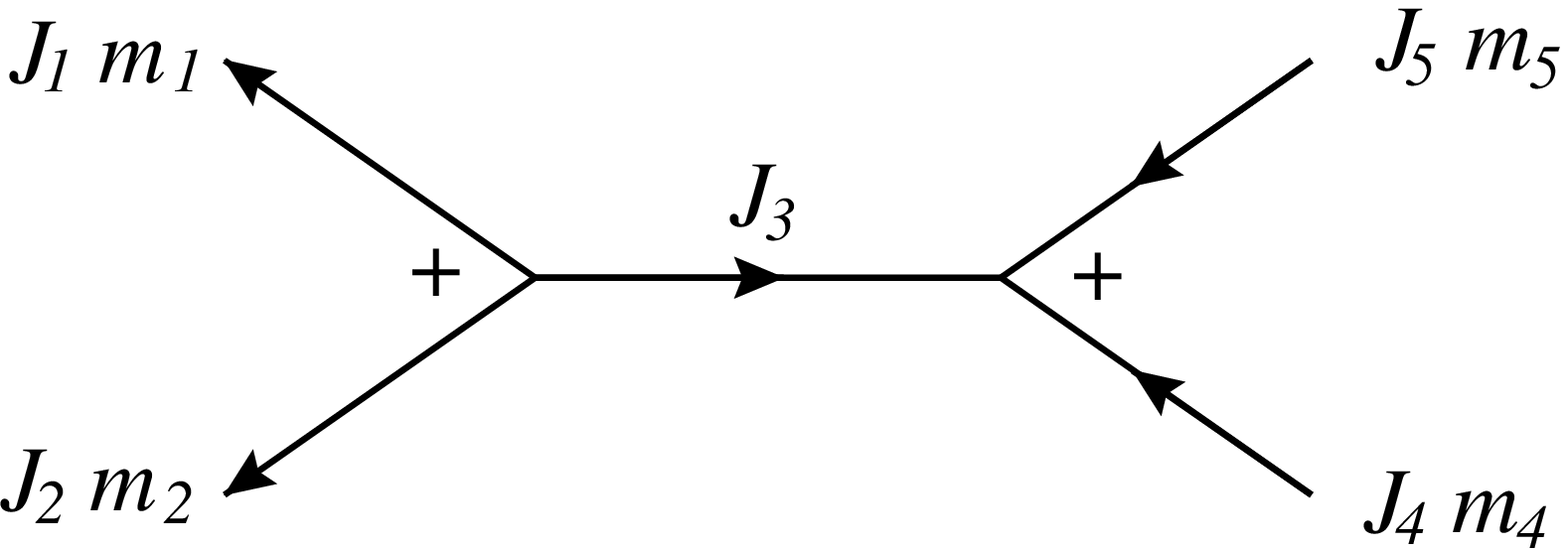}
  \end{minipage}
  = 
  \begin{minipage}[c]{.4\textwidth}
   \epsfxsize=\textwidth \epsfbox{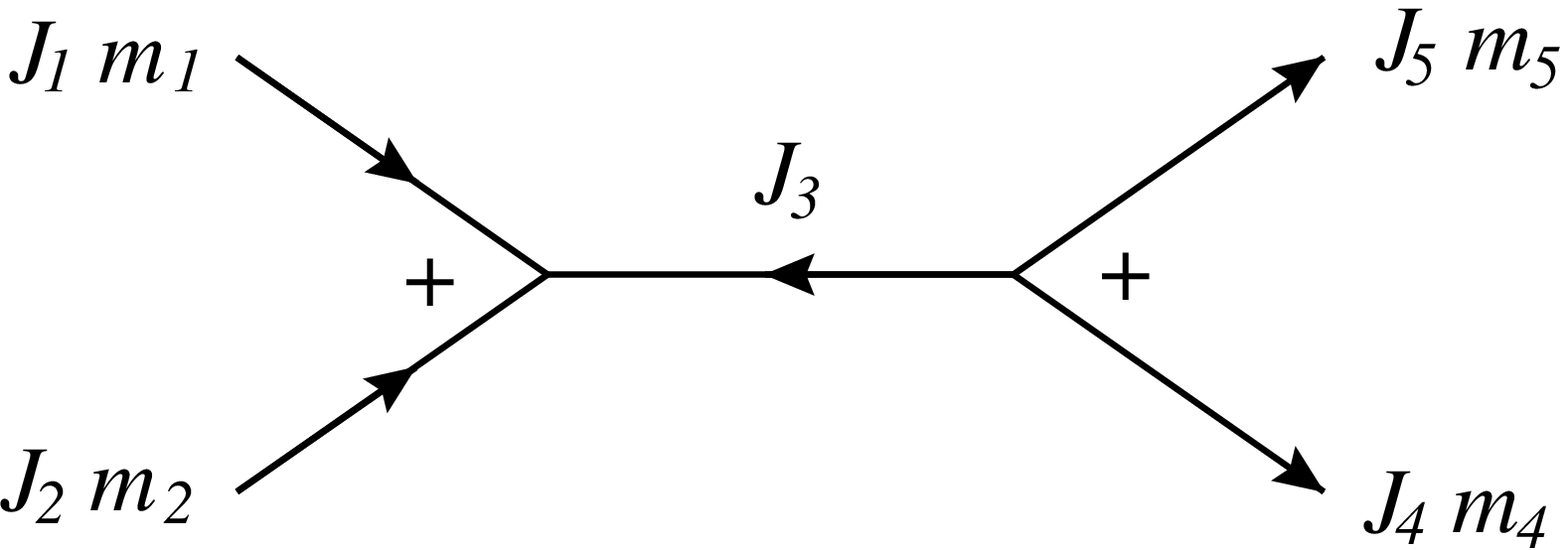}
  \end{minipage}.
\]

The decuplet graph (\ref{def of 15j}) is
the definition of our generalized 15-$j$ symbol.
There are ten peripheral lines and five internal lines.
The internal lines correspond to the tetrahedra which originally 
come from the one of the internal lines of the corresponding graph 
in (\ref{I_BCDE1}). 
The peripheral lines correspond to the common triangle between the 
neighboring tetrahedra of internal lines. 
For example the peripheral line $BDE$ is connected with two internal 
lines $ABDE$ and $BCDE$ and thus corresponds to
the common boundary of the two tetrahedra.   

We can now obtain the full expression of our partition function 
in terms of generalized 15-$j$ symbols after we carry out 
$dB$ integration for all the triangles and $dU_i$ integration for 
all the dual links whose dual are the boundary tetrahedra of each 
4-simplex, 
\begin{equation}
 Z_{LBF} = \sum_{\{J_i\}}\prod_{\hbox{\scriptsize triangle}} (2J+1)
           \prod_{\hbox{\scriptsize tetrahedron}} (2J+1)
           \prod_{\hbox{\scriptsize 4-simplex}} \{ \hbox{15-$j$} \},
\label{z_15j}
\end{equation}
where we have written the 15-$j$ symbol symbolically.
The factor $(2J+1)$ attached to a triangle is originated from the 
same factor as in (\ref{z_after_dbintegration}) appeared after $dB$ 
integration on a triangle.
The second factor $(2J+1)$ attached to a tetrahedron is originated 
from the same factor as in the graph (\ref{dU_int_end}) appeared after 
$dU$ integration for each tetrahedron (dual to dual link). 
Here we have not yet given the explicit form of the generalized 
15-$j$ symbol with the following reasons. 
As we have already pointed out in the graphical representation 
(\ref{I_BCDE1}) and 
(\ref{sign_ambiguity}), there are sign ambiguity for the 3-$j$ representation 
in the $dU$ integration, which is reflected to the definition of our 
generalized 15-$j$ symbol. 
In the next section we determine this sign ambiguity by imposing the 
topological invariance on the partition function $Z_{LBF}$ expressed 
by the 15-$j$ symbols in (\ref{z_15j}). 
The topological invariance of the partition function on the 4-dimensional 
simplicial manifold can be assured if the partition function is Pachner 
move invariant, which we will explain in the next section.

We conclude this section with a few comments.
In the Ooguri's symbolic presentation of the partition function 
there appear 6-$j$ symbols in addition to 15-$j$ symbols 
while we don't have this 6-$j$ symbols in the partition function. 
Our definition of the generalized 15-$j$ symbol
is almost the same as that of Crane-Yetter\cite{Crane-Yetter} 
except for the definition of sign factors. 

\setcounter{equation}{0}
\renewcommand{\theequation}{\arabic {section}.\arabic{equation}}

\section{Pachner Move Invariance of the Partition Function}

In 3-dimensional lattice gravity formulations, it was shown that 
the lattice Chern-Simons gravity leads to the Ponzano-Regge model 
constructed from 6-$j$ symbol\cite{KNS} which has close correspondence 
with 3-simplex, tetrahedron. 
The crucial point of the Ponzano-Regge model is that the particular 
product of 6-$j$ symbols of the model is Alexander move (2-3 move and 
1-4 move or equivalently 3-dimensional Pachner move) invariant%
\cite{Turaev-Viro}\cite{Ooguri-Sasakura}\cite{PR-related}.
Since the Ponzano-Regge model is independent how the 3-dimensional 
simplicial manifold is divided by the moves and thus the model is 
topological.
Then in the 3-dimensional case the continuum limit can be trivially 
taken and therefore the lattice Chern-Simons gravity action leads to 
the continuum $ISO(3)$ Chern-Simons gravity action. 
 
In 4 dimensions it is expected that the story proceeds quite parallel 
as in the 3-dimensional case. 
Our lattice $BF$ gravity action has led to the partition function 
constructed from 15-$j$ symbol which has close correspondence with 
4-simplex. 

As we have already mentioned that any kinds of 15-$j$ symbols can be 
represented by the closed trivalent graph which has ten vertices. 
Different types of 15-$j$ symbols are essentially distinguished 
by the topology of the graph. 
For example there are five kinds of standard 15-$j$ symbols 
which can be decomposed into the product of
6-$j$ and 9-$j$ symbols\cite{Formula}. 
Our generalized 15-$j$ symbol graphically given in (\ref{def of 15j}) 
is different from the standard 15-$j$ symbol and thus we have called the 
15-$j$ symbol of (\ref{def of 15j}) as ``generalized 15-$j$ symbol''.  

In the graphical representation of the generalized 15-$j$ symbol, we 
have not yet specified the sign factors on the trivalent vertices. 
Here in this section we show that those sign factors are determined 
by imposing the 4-dimensional Pachner move invariance to the product 
of the partition function (\ref{z_15j}). 

\begin{figure}[h]
\begin{tabular}{ccc}
 \begin{minipage}[c]{.25\textwidth}
  \epsfxsize=\textwidth \epsfbox{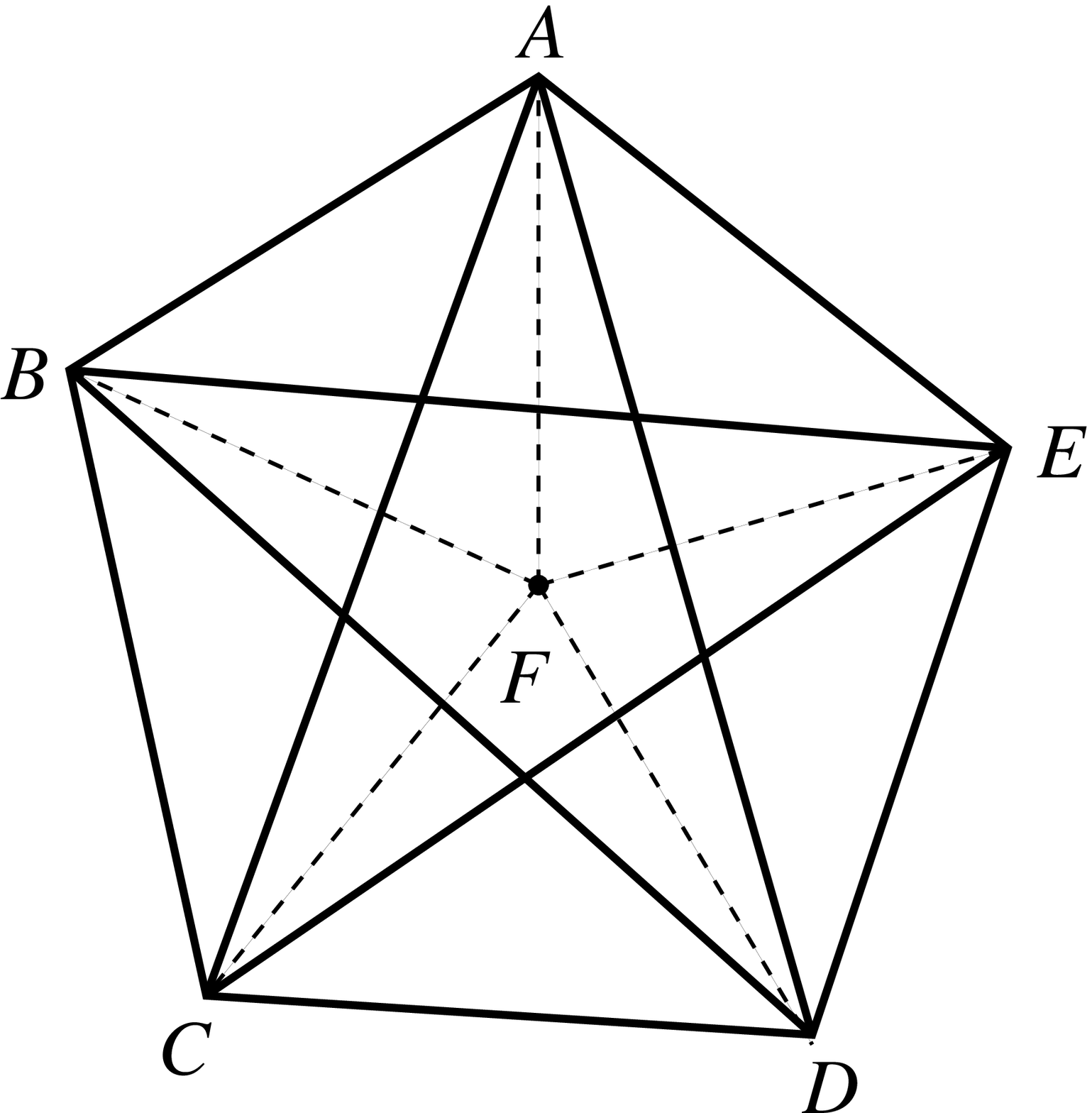}
 \end{minipage}  &
 \begin{minipage}[c]{.35\textwidth}
 \epsfxsize=\textwidth \epsfbox{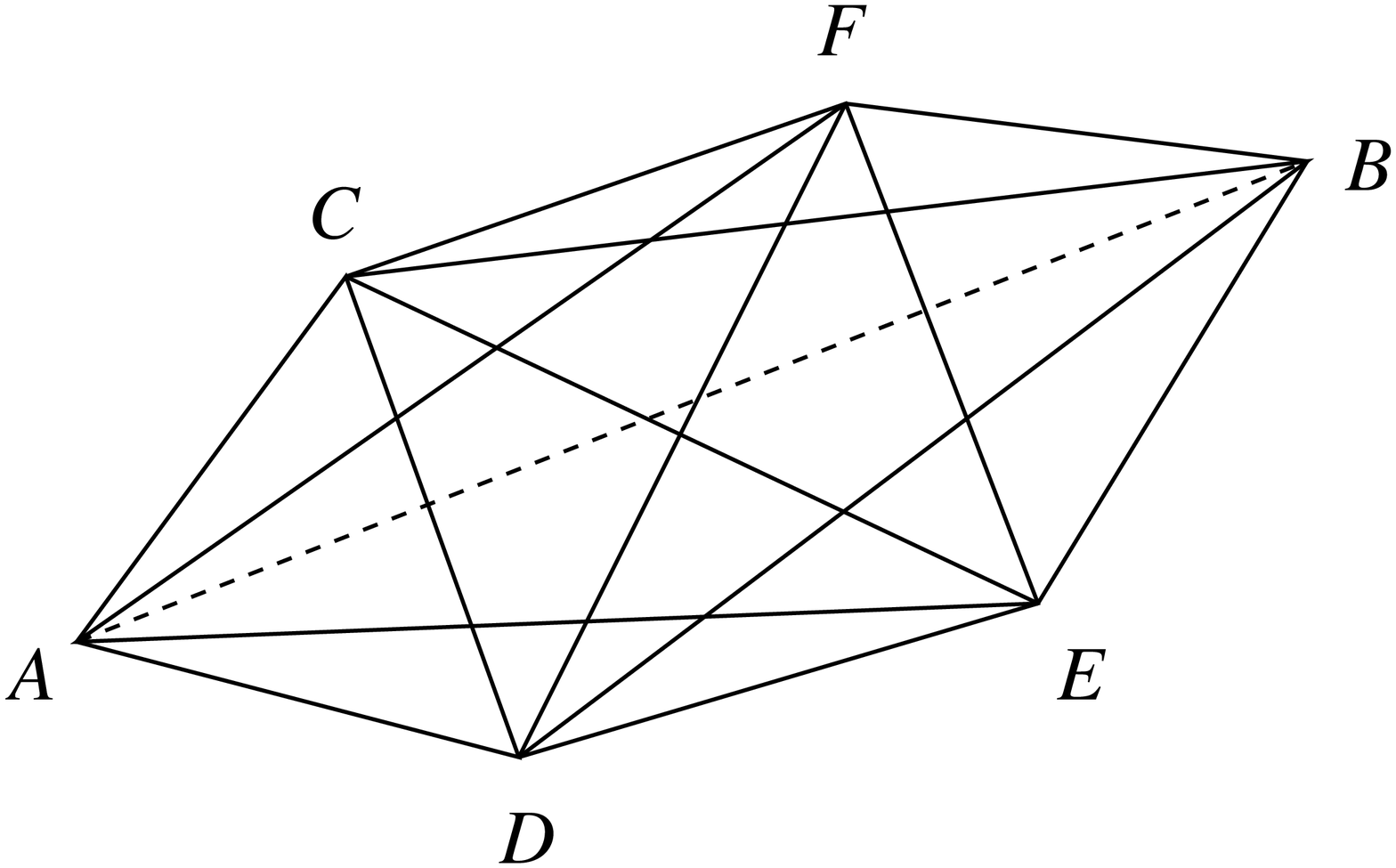}
 \end{minipage}  &
 \begin{minipage}[c]{.3\textwidth}
  \epsfxsize=\textwidth \epsfbox{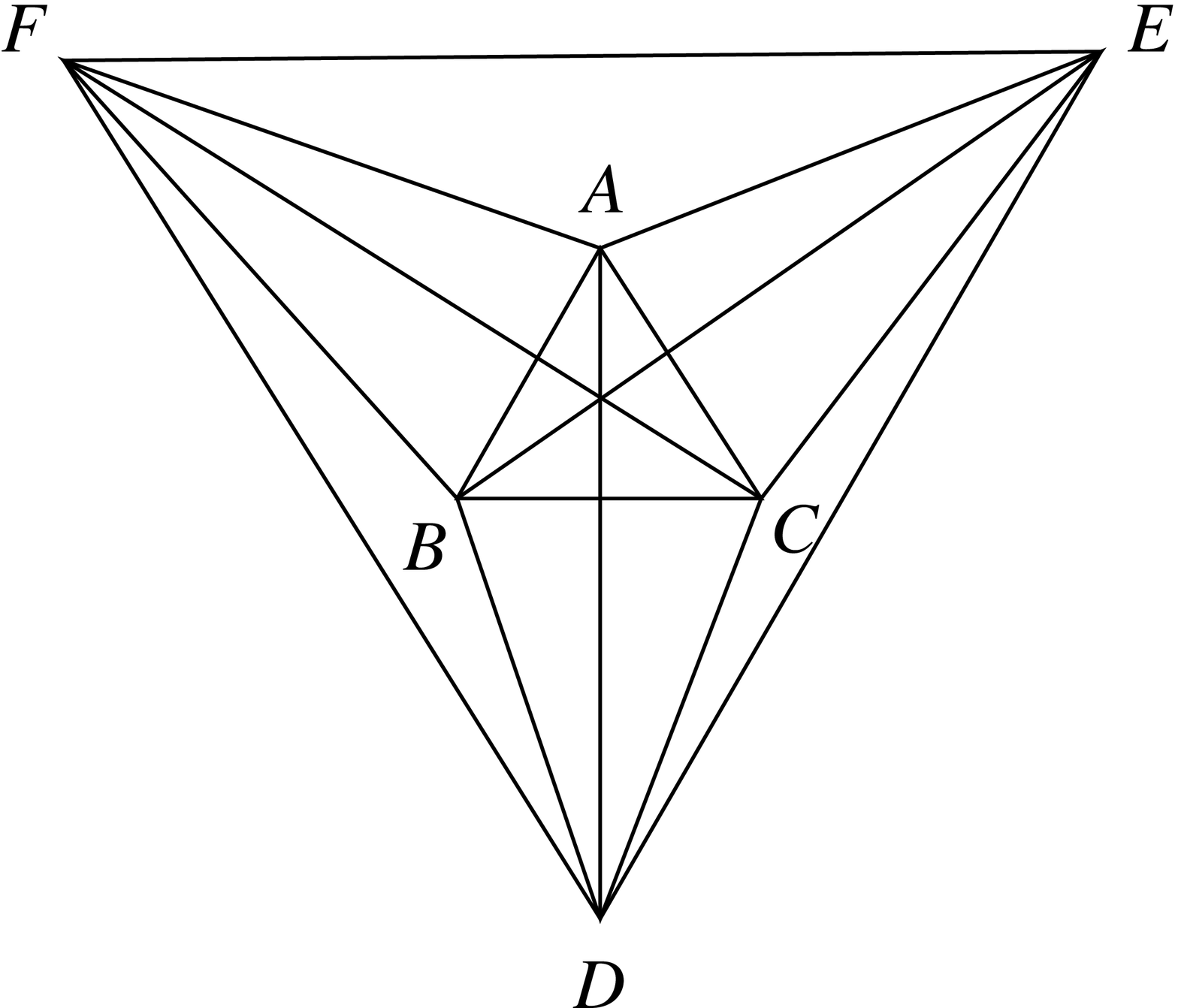}
 \end{minipage} \\
(1) & (2) & (3)
\end{tabular}
 \caption{Pachner moves in 4 dimensions. (1) 1-5 move :
 $(ABCDE) \rightarrow (BCDEF)(ACDEF)(ABDEF)(ABCEF)(ABCDF)$,
 (2) 2-4 move :
 $(ACDEF)(BCDEF) \rightarrow (ABDEF)(ABCEF)(ABCDF)(ABCDE)$ ,
 and (3) 3-3 move :
 $(BCDEF)(ACDEF)(ABDEF) \rightarrow (ABCDE)(ABCEF)(ABCDF)$.}
 \label{fig:Pachner moves}
\end{figure}

It is the known fact that any 4-dimensional simplicial manifold can be 
constructed out of 4-dimensional Pachner moves
which are composed of $n$-$m$ moves with $n+m=6$ in 4 dimensions.
Thus they include 
1-5 move, 2-4 move, 3-3 move and the inverse of the 1-5 and 2-4 
moves\cite{Ambjorn Book}. 
There are essentially three independent moves; 1-5 move, 2-4 move and 
3-3 move to reproduce an arbitrary shape of 4-dimensional simplicial 
manifold. 
Those moves are shown in Fig.\ref{fig:Pachner moves}. 
In the 1-5 move the center site $F$ in the figure is the common 
site of final five 4-simplexes. 
In the 2-4 move the dotted link $AB$ is the common link of the final 
four 4-simplexes. 
For example in 2-4 move, the initial configuration of a 4-dimensional 
simplicial manifold including two 4-simplexes $(ACDEF)$ and $(BCDEF)$ 
will be changed into another simplicial manifold including four 
4-simplexes $(ABDEF), (ABCEF), (ABCDF)$ and $(ABCDE)$ after the 
2-4 move of Pachner type.

In 3-dimensional case the Alexander move invariance of the Ponzano-Regge 
model is shown explicitly by several 
ways\cite{Turaev-Viro}\cite{Ooguri-Sasakura}\cite{PR-related}. 
Since in 4 dimensions the explicit proof of the Pachner move invariance of 
the partition function of 15-$j$ symbol type has not yet been given, 
we explicitly show the invariance here by a graphical method. 
There is a proposal of proof of Pachner move invariance 
by Crane and Yetter who used the similar
quantum 15-$j$ symbol as ours but 
didn't show explicit proof\cite{Crane-Yetter}. 
Furthermore there is some sign factor difference from ours in the 
definition of 15-$j$ symbol. 
Mathematically there are similar treatments of topologically invariant 
quantity in 4 dimensions\cite{Crane-K-Yetter}.

In order to show the Pachner move invariance of the partition function 
(\ref{z_15j}) graphically, we need three crucial formulae. 
The analytic and the corresponding graphical expressions of the formulae 
are given in the following. 
We may call an expression to be closed if all the third components of the 
angular momentum are summed up. 
Then if there are two closed expressions each of which has two 
3-$j$ symbols with the common magnitude of angular momentum, they 
can be connected into one closed expression as follows: 
\begin{eqnarray}
&&  \sum_{J} (2J+1)
   \sum_{m_i M m'_i M'}
   \threej{j_1}{j_2}{J}{m_1}{m_2}{M}
   \langle j_1 m_1 j_2 m_2 | {\cal M} | j_3 m'_3 j_4 m'_4 \rangle   
   \threej{j_3}{j_4}{J}{-m'_3}{-m'_4}{-M} \nonumber \\
&& \hspace*{8em} \times 
   \threej{j_3}{j_4}{J}{m_3}{m_4}{M'}   
   \langle j_3 m_3 j_4 m_4 | {\cal N} | j_1 m'_1 j_2 m'_2 \rangle   
   \threej{j_1}{j_2}{J}{-m'_1}{-m'_2}{-M'} \nonumber \\
&& \hspace*{8em} \times 
 (-)^{\sum_{i=1}^4 (j_i-m'_i) + (J-M) + (J-M')} \nonumber \\
&=& 
   \sum_{m_i}
   \langle j_1 m_1 j_2 m_2 | {\cal M} | j_3 m_3 j_4 m_4 \rangle
   \langle j_3 m_3 j_4 m_4 | {\cal N} | j_1 m_1 j_2 m_2 \rangle ,
\label{move_formula1}
\end{eqnarray} 
which is graphically given by 
\[
\setlength{\arraycolsep}{.5em}
\begin{array}{cccc}
 \displaystyle
 \sum_{J} (2J+1) ~ 
 \begin{minipage}[c]{.3\textwidth}
 \epsfxsize=\textwidth \epsfbox{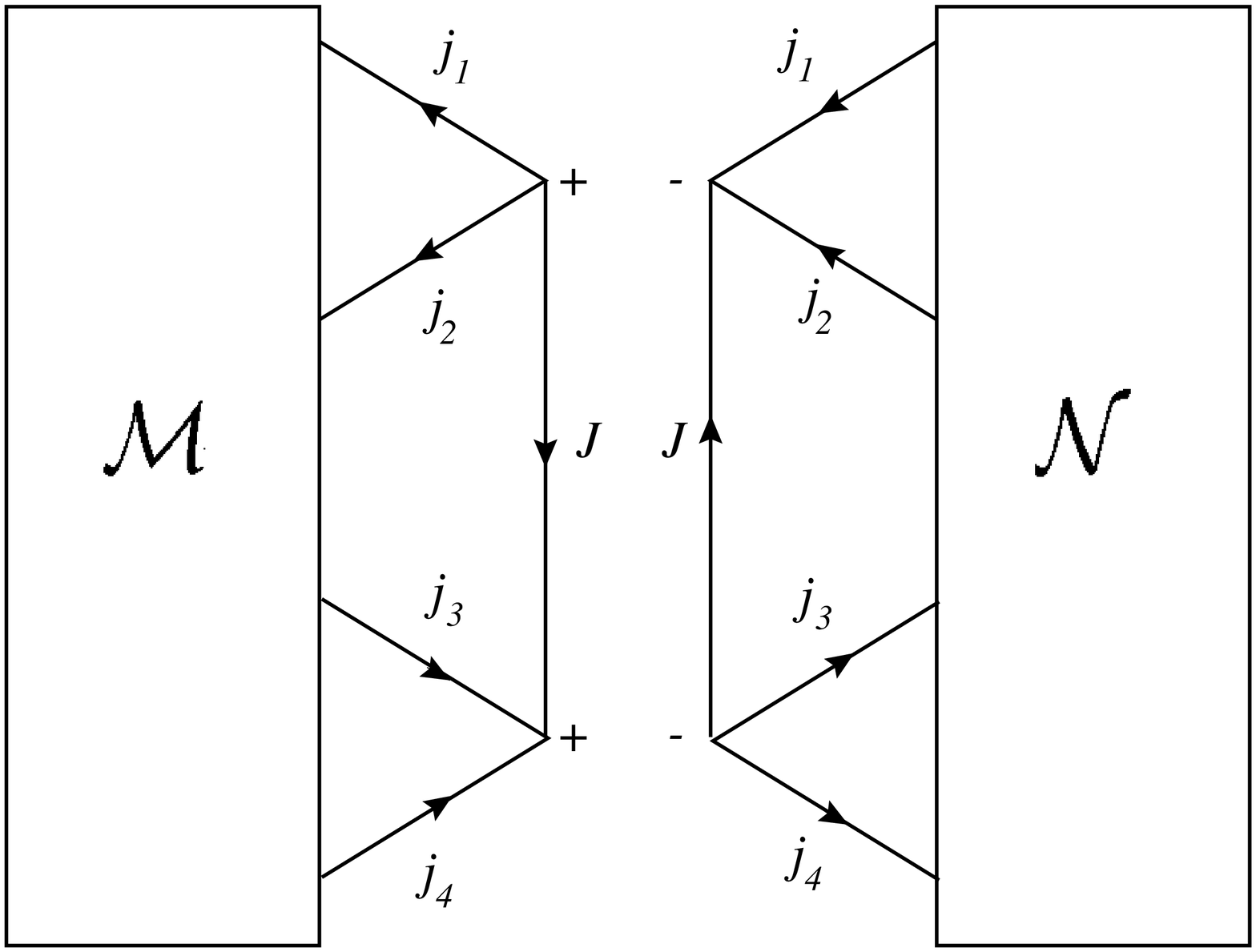}
 \end{minipage}
 & = &
 \begin{minipage}[c]{.3\textwidth}
 \epsfxsize=\textwidth \epsfbox{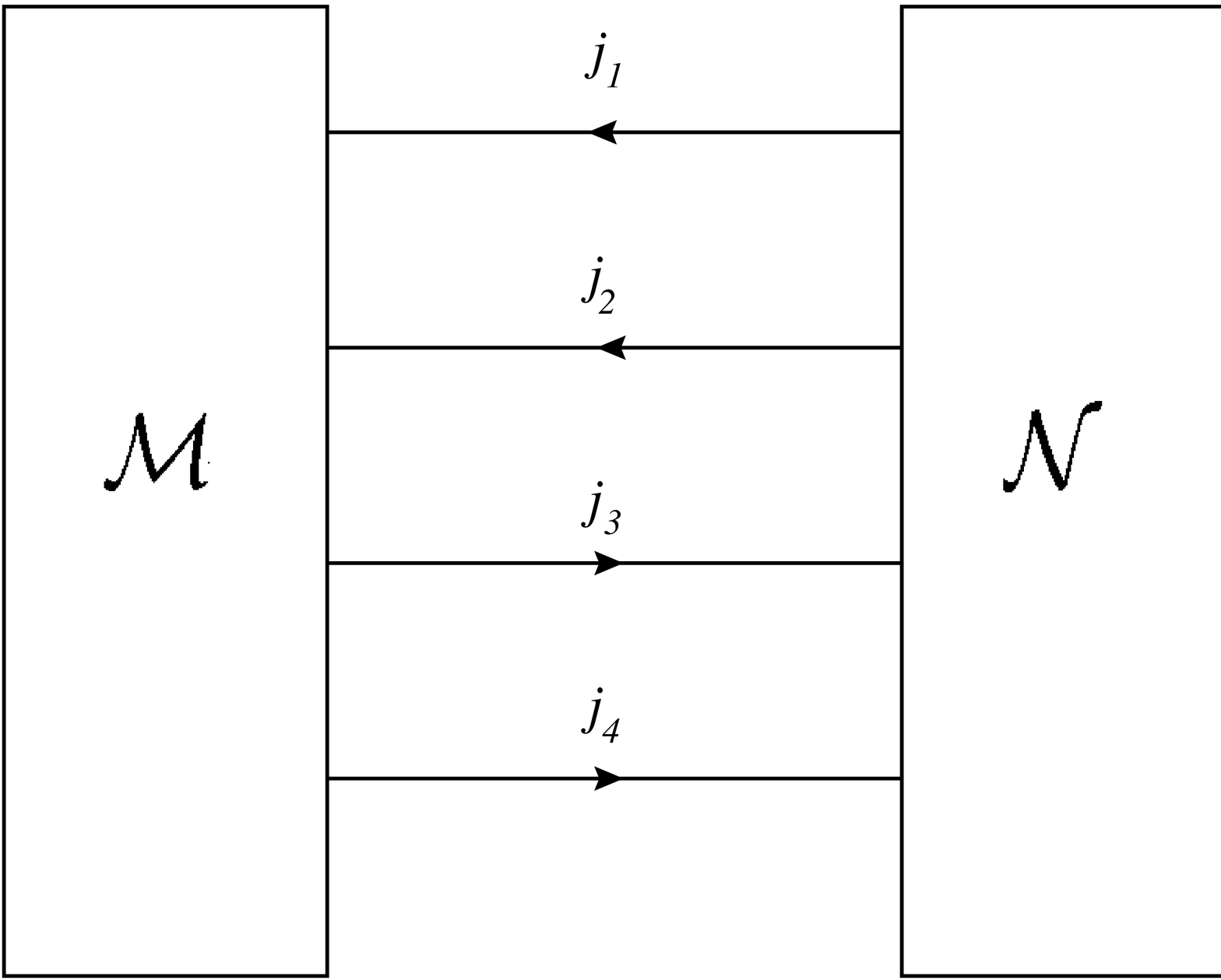}
 \end{minipage}. 
\end{array}
\]
The orthogonality of 3-$j$ symbol with a weight factor is 
\begin{equation}
 \sum_{J m'} (2J+1)
  \threej{J_1}{J_2}{J}{m_1}{m_2}{m'}
  \threej{J_1}{J_2}{J}{-m'_1}{-m'_2}{-m'}
  (-)^{\sum_{i=1}^3 (J_i-m'_i)}
 = \delta _{m_1 m'_1} \delta _{m_2 m'_2}, 
\label{move_formula2}
\end{equation}
which is graphically given by 
\[
\setlength{\arraycolsep}{.5em}
 \begin{array}{cccc}
  \displaystyle
  \sum_{J} (2J+1) &
  \begin{minipage}[c]{.3\textwidth}
   \epsfxsize=\textwidth \epsfbox{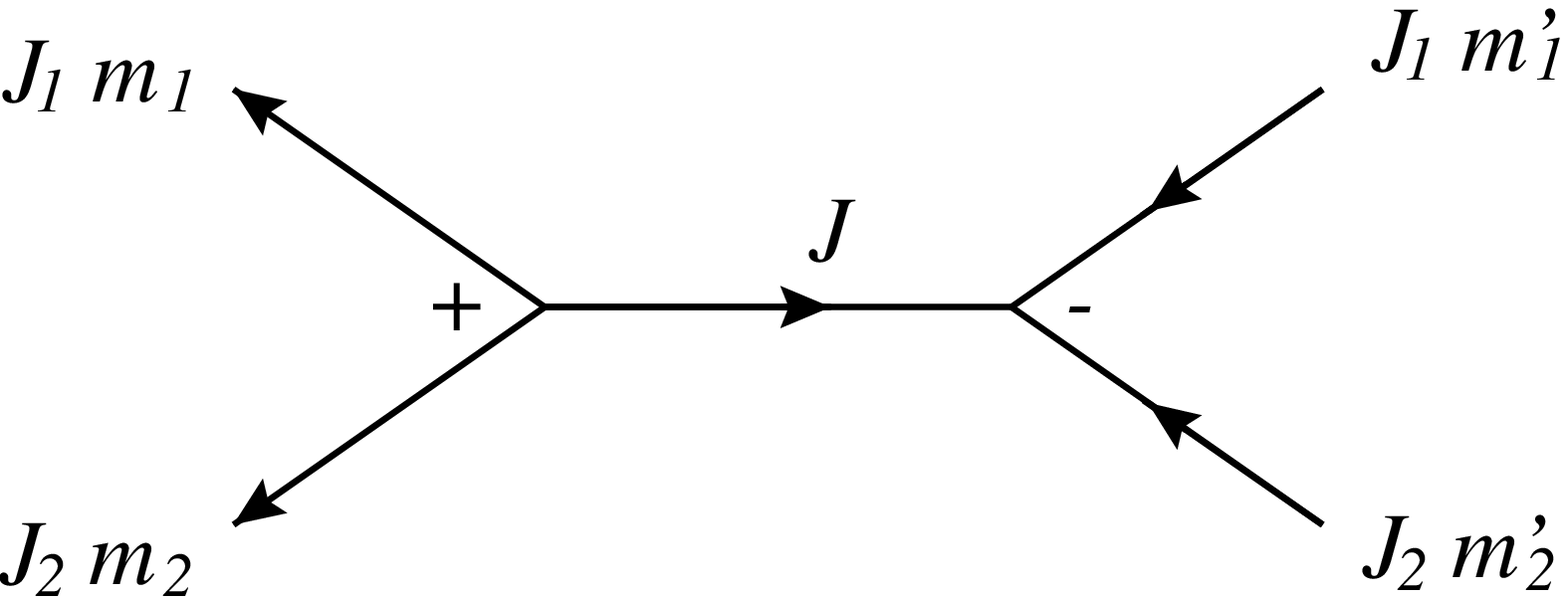}
  \end{minipage}
  & = &
  \begin{minipage}[c]{.3\textwidth}
   \epsfxsize=\textwidth \epsfbox{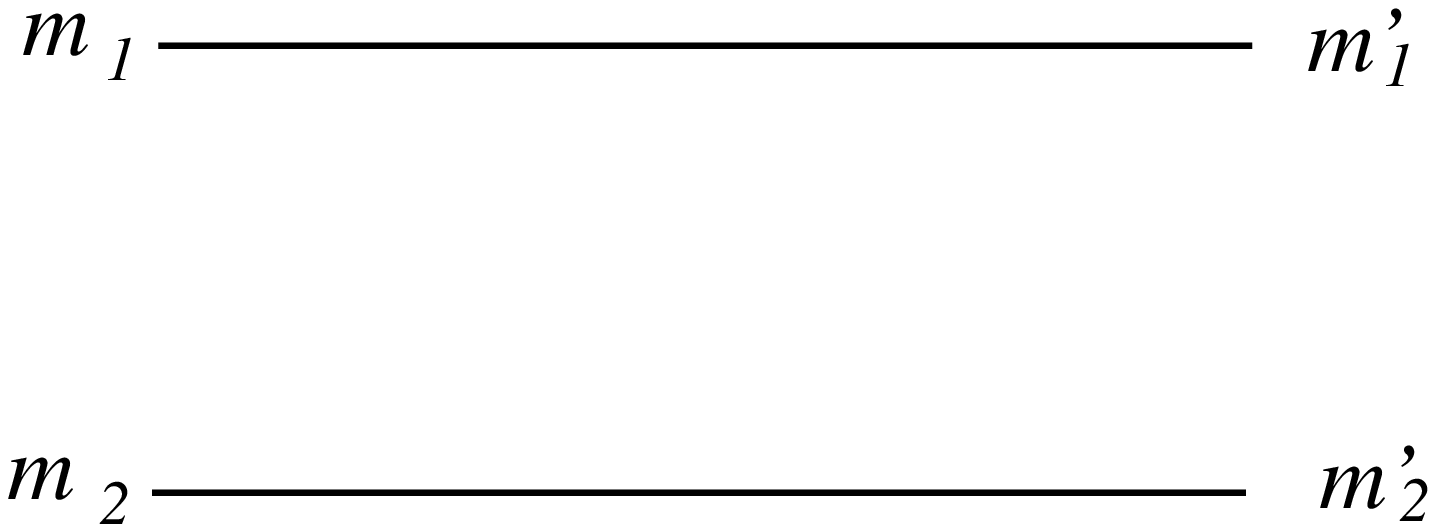}
  \end{minipage}.
  \\
 \end{array}
\]
The orthogonality of 3-$j$ symbol with two angular momentum magnitudes 
summing up leads 
\begin{eqnarray}
&& \sum_{J_2 m'_2 J_3 m'_3} (2J_2+1)(2J_3+1)
  \threej{J_1}{J_2}{J_3}{m_1}{m'_2}{m'_3}
  \threej{J_1}{J_2}{J_3}{m'_1}{m'_2}{m'_3}
  (-)^{\sum_{i=1}^3 (J_i - m'_i)} \nonumber \\
&=& \Lambda \cdot \delta _{m_1 m'_1}, 
\label{move_formula3}
\end{eqnarray}
where $\Lambda$ is the infinite constant
which needs a regularization and 
is the same factor appeared in the Ponzano-Regge model.
Here we introduce the following cut-off factor of the angular momentum 
to regularize the infinite constant 
\begin{eqnarray}
 \Lambda (\lambda)
&=& \frac{1}{2 J_1 + 1}  
\!\!\!\!\!\!\!\!\!\!
  \sum_{ 
  \mbox{ \scriptsize $
  \begin{array}{c} 
   K_2,K_3 \leq \lambda,\\ 
   |K_2 - K_3| \leq J_1 \leq K_2 + K_3
  \end{array} $}
   }
\!\!\!\!\!\!\!\!\!\!
(2 K_2 + 1)(2 K_3 + 1) \nonumber  \\
&=& \sum_{J=0}^{\lambda} (2J+1)^2    
\sim \frac{4 \lambda^3}{3} ~~ (\lambda \rightarrow \infty)
\label{Lambda}.
\end{eqnarray}
Then (\ref{move_formula3}) can be graphically given by 
\[
\setlength{\arraycolsep}{.5em}
 \begin{array}{ccc}
  \begin{minipage}[c]{.4\textwidth}
   \epsfxsize=\textwidth \epsfbox{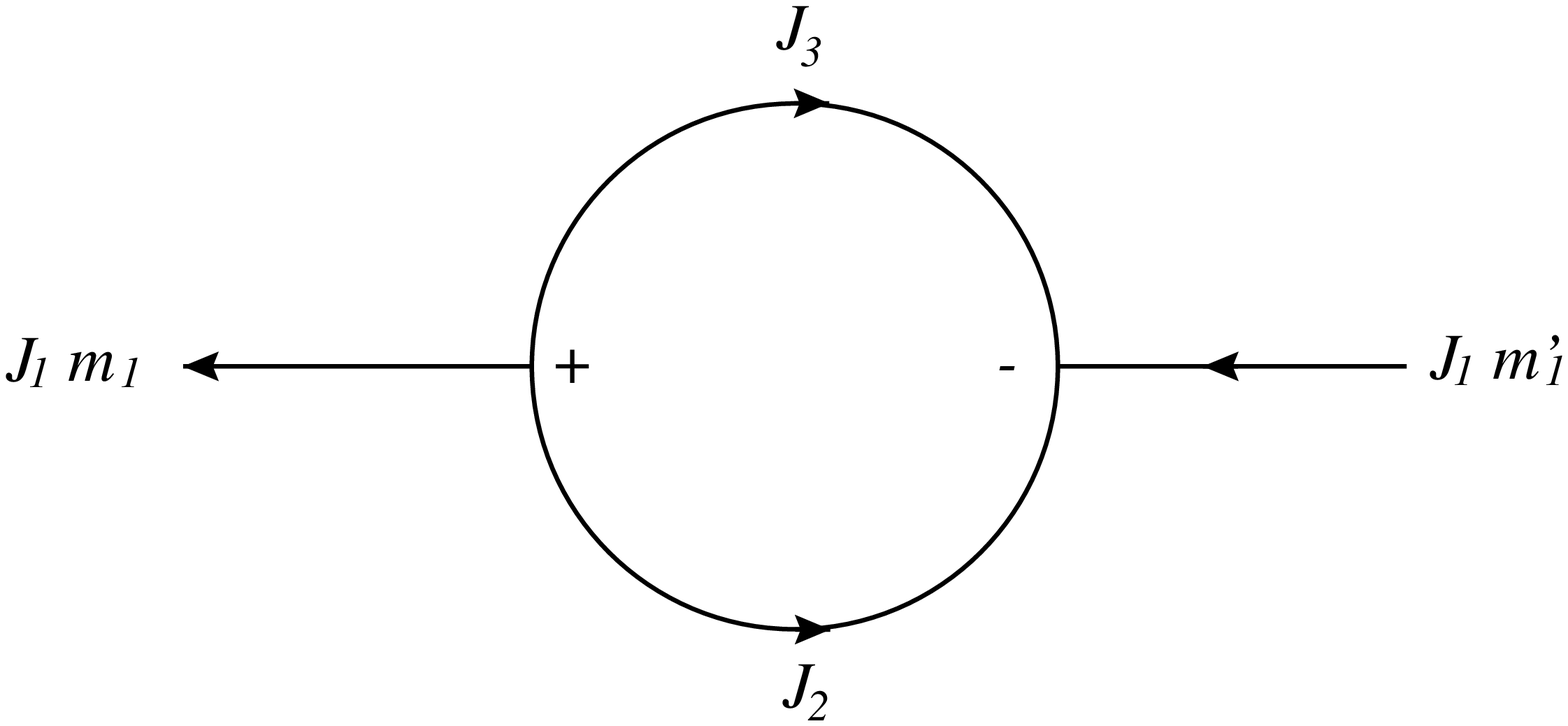}
  \end{minipage}
  & = &  \Lambda \cdot
  \begin{minipage}[c]{.4\textwidth}
   \epsfxsize=\textwidth \epsfbox{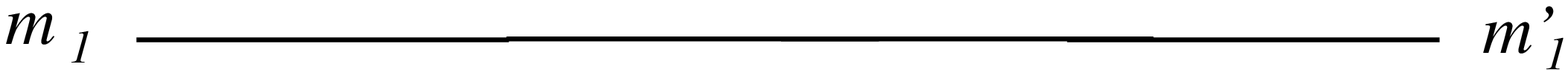}
  \end{minipage}.
  \\
 \end{array}
\]

An example of the combined graphical relation of (\ref{move_formula1}), 
(\ref{move_formula2}) and (\ref{move_formula3}) is 
\begin{eqnarray}
I &=& \sum_{J_1 J_2 J_3} \prod_{i=1}^{3}(2J_i+1) 
 \sum_{m_1 m_2 n_1 n_3} \nonumber \\
 && \times \threej{J_1}{J_3}{J_4}{m_1}{n_3}{m_4}
 \threej{J_1}{J_5}{J_2}{-m_1}{-m_5}{-m_2}
 (-)^{\sum_{i=1,2,5}(J_i-m_i)} \nonumber \\
 &&  \times \threej{J_1}{J_5}{J_2}{n_1}{n_5}{m_2} 
 \threej{J_1}{J_3}{J_4}{-n_1}{-n_3}{-n_4} (-)^{\sum_{i=1,3,4}(J_i-n_i)},
\end{eqnarray}
which is graphically given by
\begin{equation}
I = \sum_{J_1 J_2 J_3} \prod_{i=1}^{3}(2J_i+1) 
 \begin{minipage}{.4\textwidth}
  \epsfxsize=\textwidth \epsfbox{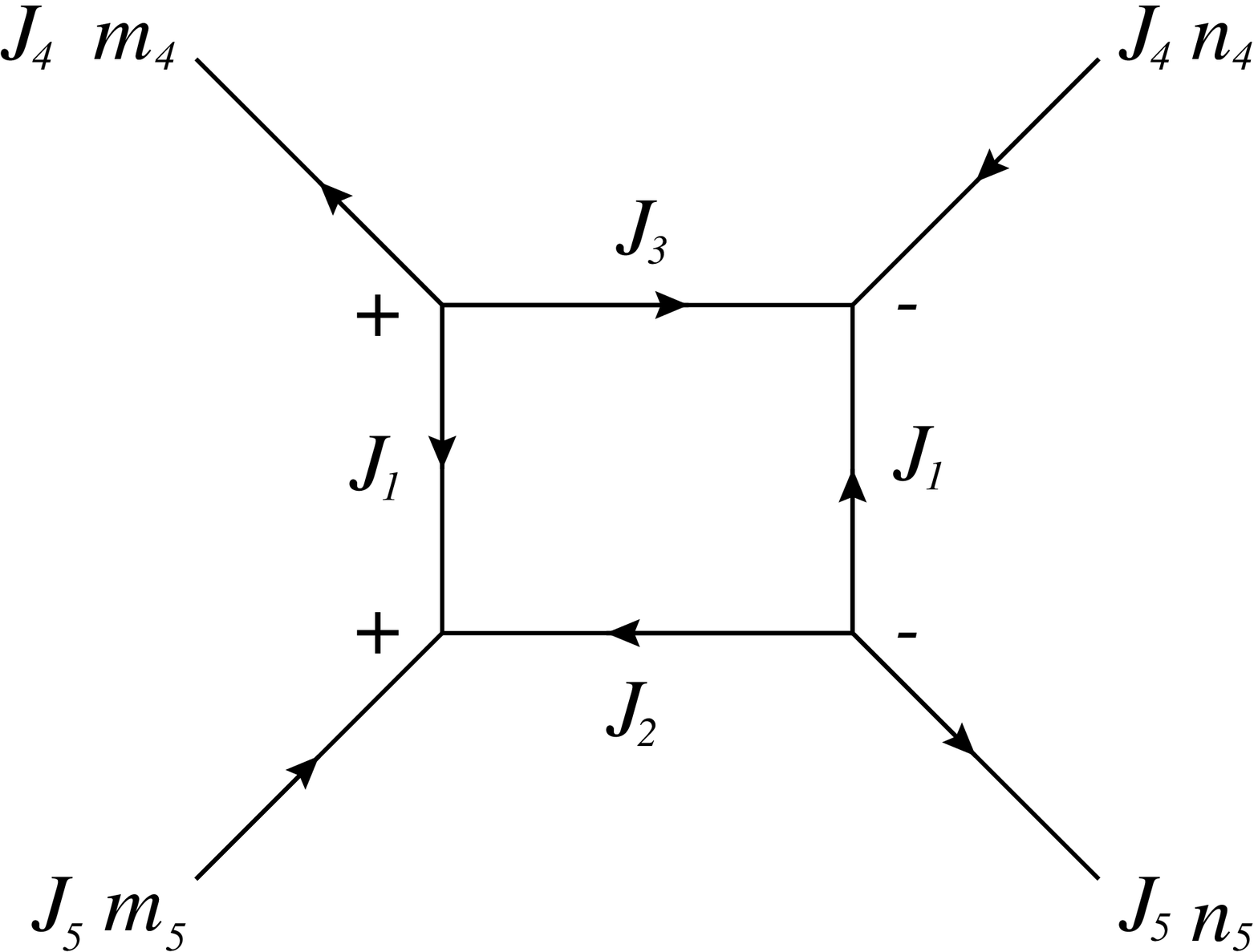}
 \end{minipage}.
\label{box1}
\end{equation}

Using the relations (\ref{move_formula2}) and (\ref{move_formula3}), 
we obtain 
\begin{eqnarray}
 I &=& \sum_{J_1 J_3} (2J_1+1) (2J_3+1) \nonumber \\
 && \times   \sum_{n_1 n_3}
 \threej{J_1}{J_3}{J_4}{n_1}{n_3}{m_4}
 \threej{J_1}{J_3}{J_4}{-n_1}{-n_3}{-n_4} (-)^{\sum_{i=1,3,4}(J_i-n_i)} 
 \cdot \delta_{m_5 n_5} \nonumber \\
 &=& \Lambda \delta_{m_4, n_4} \delta_{m_5 n_5},
\end{eqnarray}
which is graphically given by
\begin{eqnarray}
 I &=& \sum_{J_1 J_3} (2J_1+1) (2J_3+1) ~ 
 \begin{minipage}{.4\textwidth}
  \epsfxsize=\textwidth \epsfbox{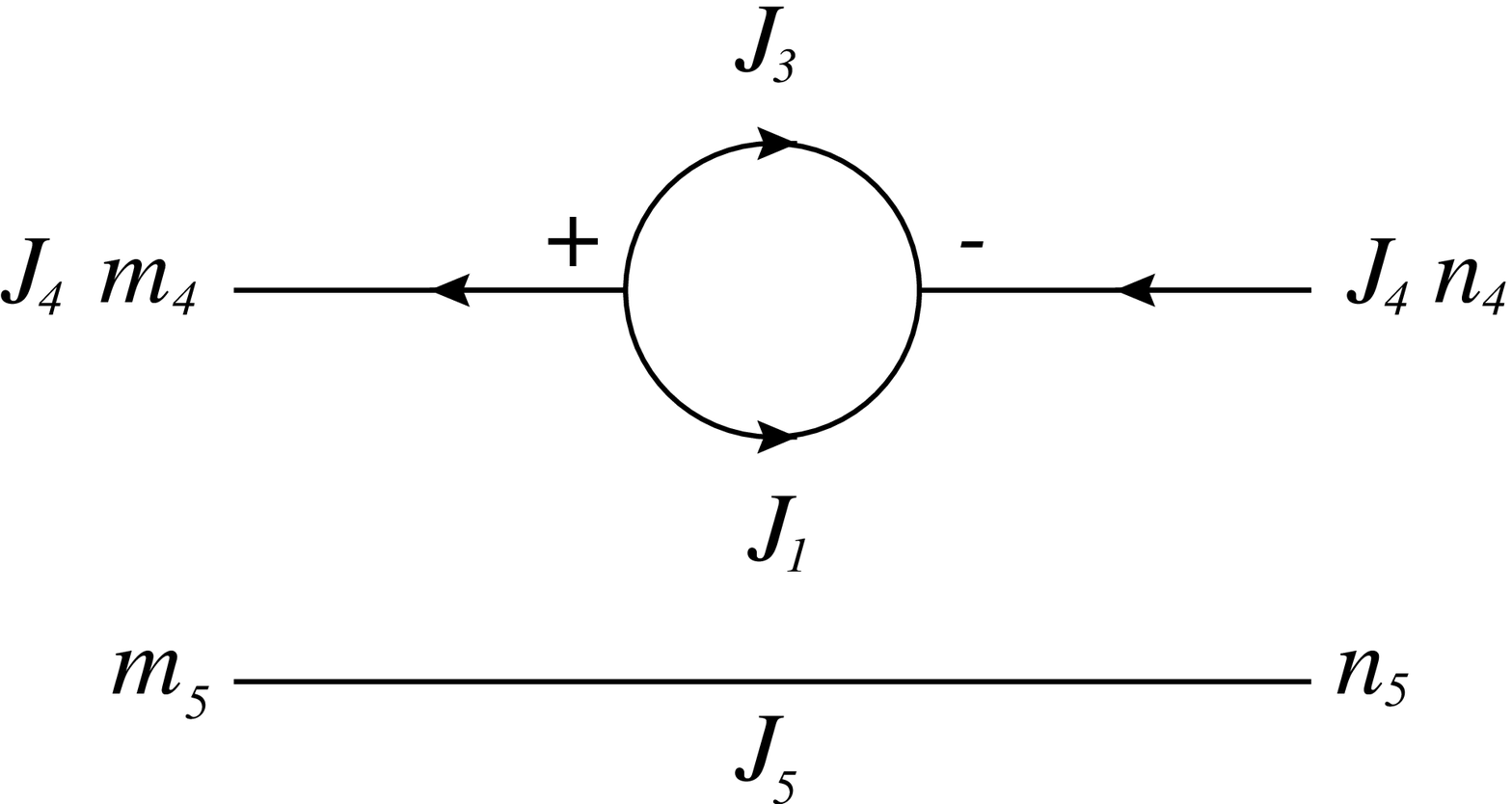}
 \end{minipage} \nonumber \\
 &=& \Lambda \cdot
 \begin{minipage}{.3\textwidth}
  \epsfxsize=\textwidth \epsfbox{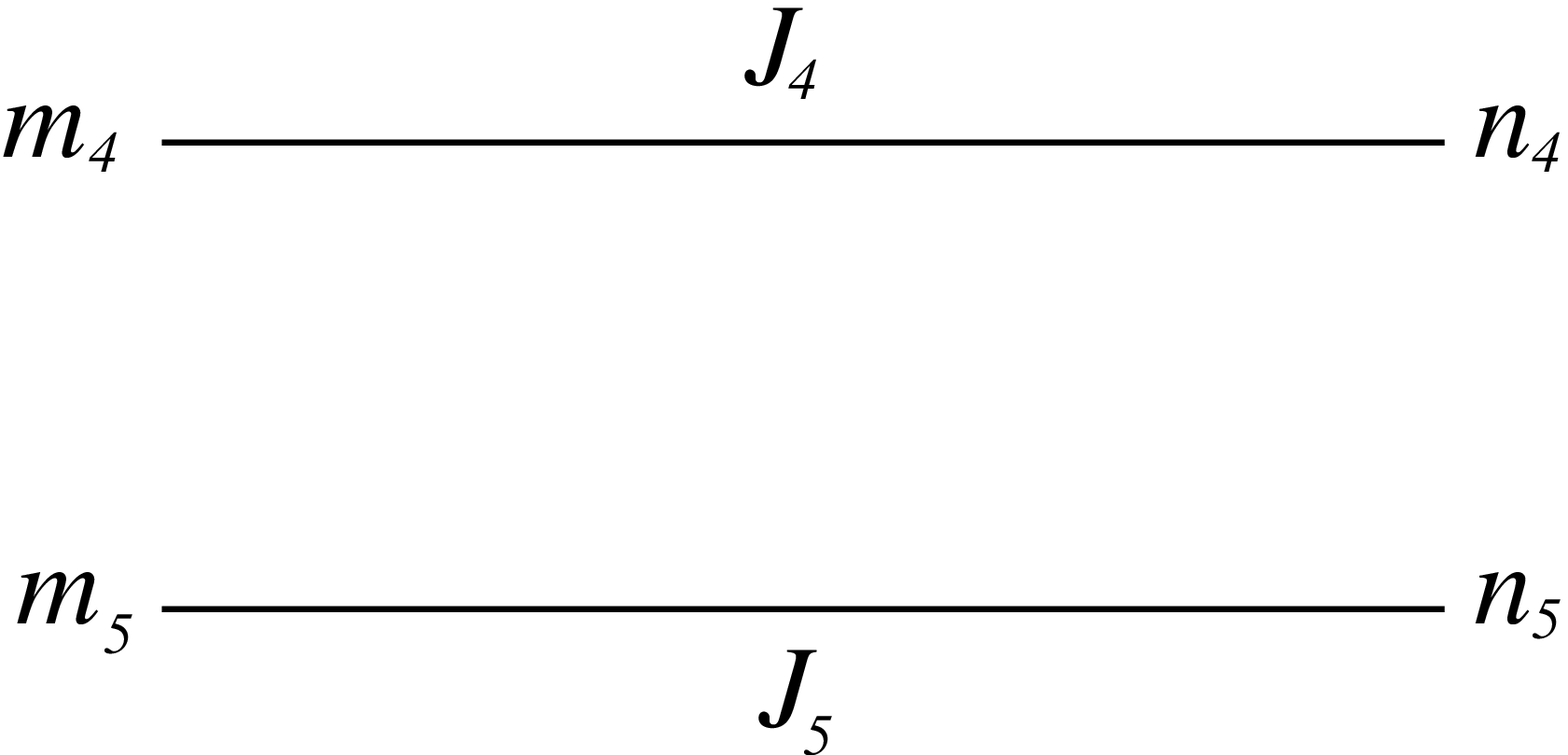}
 \end{minipage}.
\label{box2}
\end{eqnarray}

We now explicitly show the invariance of the partition function 
(\ref{z_15j}) under the 4-2 Pachner move graphically by using 
the above formulae. 
The 4-2 move is simply the inverse of 2-4 move and
includes all the necessary manipulations.
The 5-1 and 3-3 Pachner move invariance can be proved
in the similar way. 
We will abbreviate the summation of $J$ and factors $(2J+1)$
in the following graphical calculation.

We first display four decuplet diagrams corresponding to the four 
4-simplexes in particular rule. 
We first note that the boundary of the initial four 4-simplexes 
should coincide with the boundary of the final two simplexes with 
correct orientations. 
With the help of homological algebraic notations,
we obtain the following relation: 
\begin{eqnarray}
&& \partial (ABCDE) -  \partial (ABCDF)
 - \partial (ABDEF) + \partial (ABCEF) \nonumber \\
&=&  \partial (BCDEF) - \partial (ACDEF) \nonumber \\
&=& BCDE - ACDE - BDEF + ADEF   \nonumber\\
&&- BCDF + ACDF + BCEF - ACEF,
 \label{boundary_relation1}
\end{eqnarray}
where the boundary operator $\partial$ is defined by
\begin{equation}
\partial (ABCDE) = BCDE - ACDE + ABDE - ABCE +ABCD.
 \label{boundary_relation2}
\end{equation}

In the first row of the decuplet diagrams
in the following first equation,
we have shown the 
4-simplex $ABCDE$ where the internal lines corresponding to 
tetrahedra are arranged in the order starting from $BCDE$ and then 
$ACDE$ with common triangle $CDE$ and so on
as in the order of (\ref{boundary_relation2}).
In the second row of the first decuplet diagrams, we show the 
4-simplex $ABCDF$ with the opposite arrows with respect to the 
diagram $ABCDE$
reflecting the sign difference in (\ref{boundary_relation1}). 
For example the arrow of $BCDE$ for the 4-simplex $ABCDE$ in (1) is the 
opposite to the arrow of $BCDF$ for the 4-simplex $ABCDF$ in (2),
which are located on the same position of each decouplet. 
We then display the 4-simplexes $ABDEF$ and $ABCEF$ 
with the same rule in the first column of the equations. 

\newpage
\[
 \begin{array}{ccc}
  \begin{minipage}[c]{.4\textwidth}
   \epsfxsize=\textwidth \epsfbox{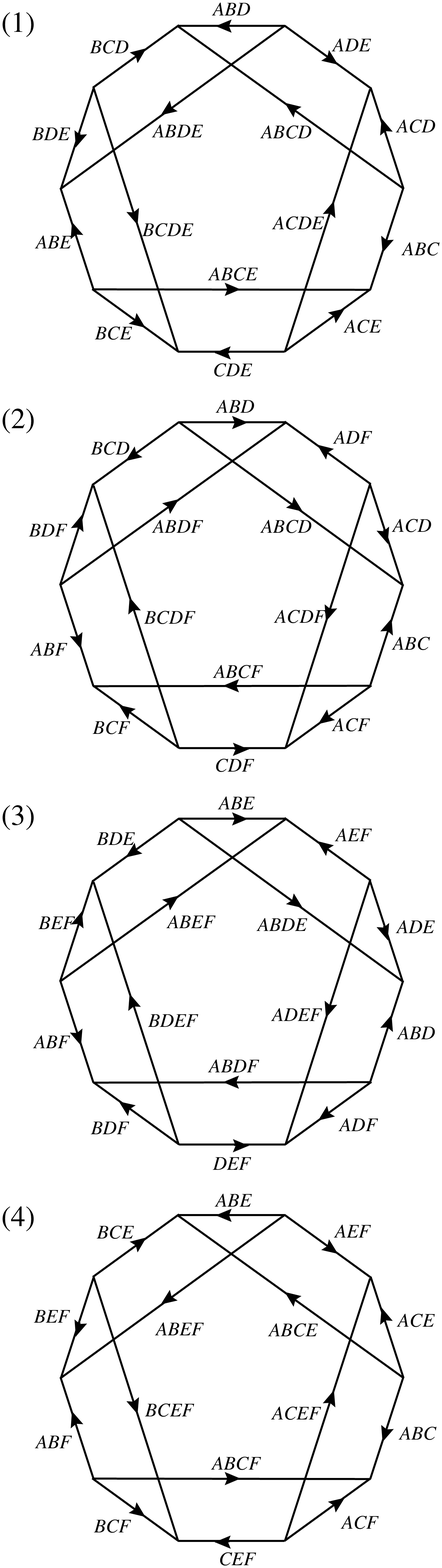}
  \end{minipage}
  & ~ = ~ &
  \begin{minipage}[c]{.45\textwidth}
   \epsfxsize=\textwidth \epsfbox{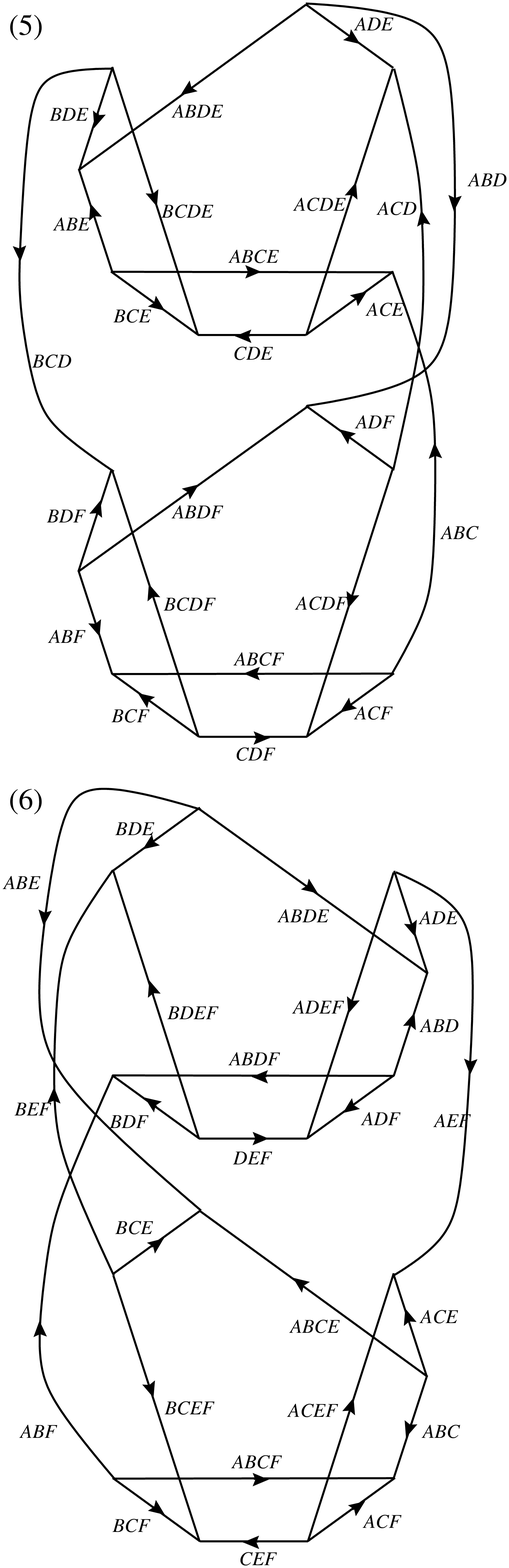}
  \end{minipage}
  \\
 \end{array}
\]
\[
 \begin{array}{cccc}
  = &
  \begin{minipage}[c]{.5\textwidth}
   \epsfxsize=\textwidth \epsfbox{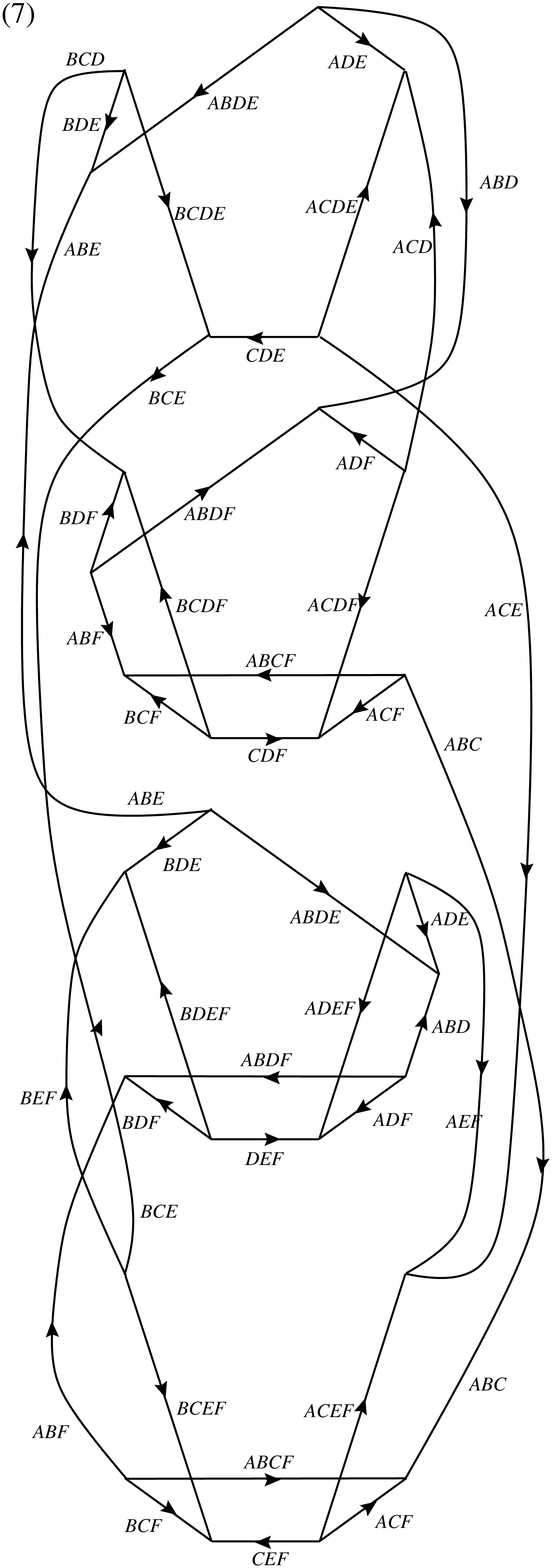}
  \end{minipage}
  & ~ = ~ &
  \begin{minipage}[c]{.5\textwidth}
   \epsfxsize=\textwidth \epsfbox{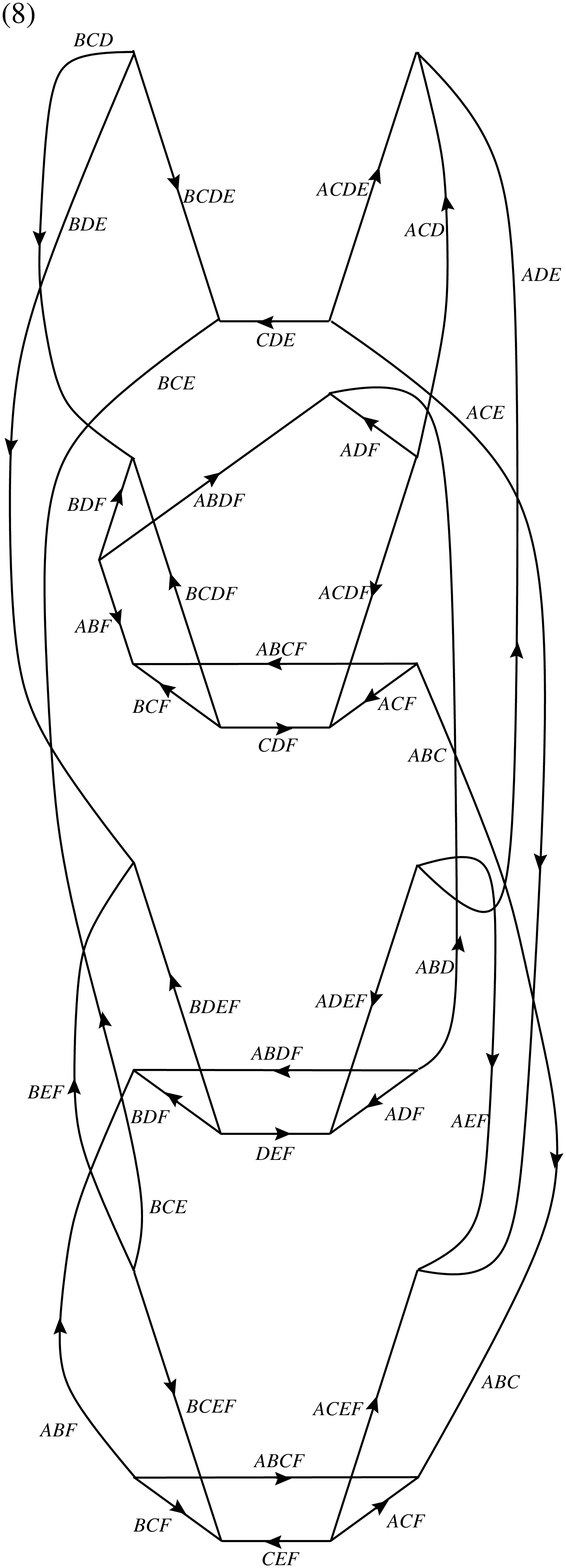}
  \end{minipage}
  \\
 \end{array}
\]
\[
 \begin{array}{cccc}
  = &
  \begin{minipage}[c]{.5\textwidth}
   \epsfxsize=\textwidth \epsfbox{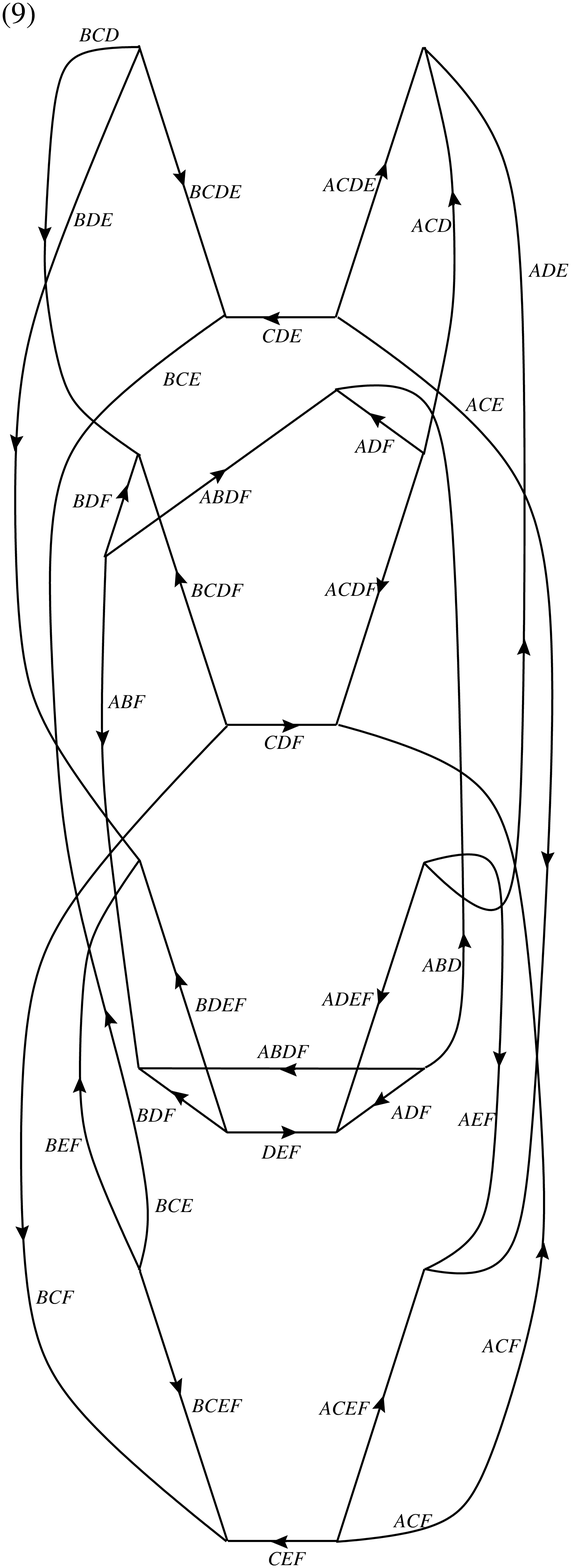}
  \end{minipage}
  & ~ = ~ & \Lambda \cdot
  \begin{minipage}[c]{.5\textwidth}
   \epsfxsize=\textwidth \epsfbox{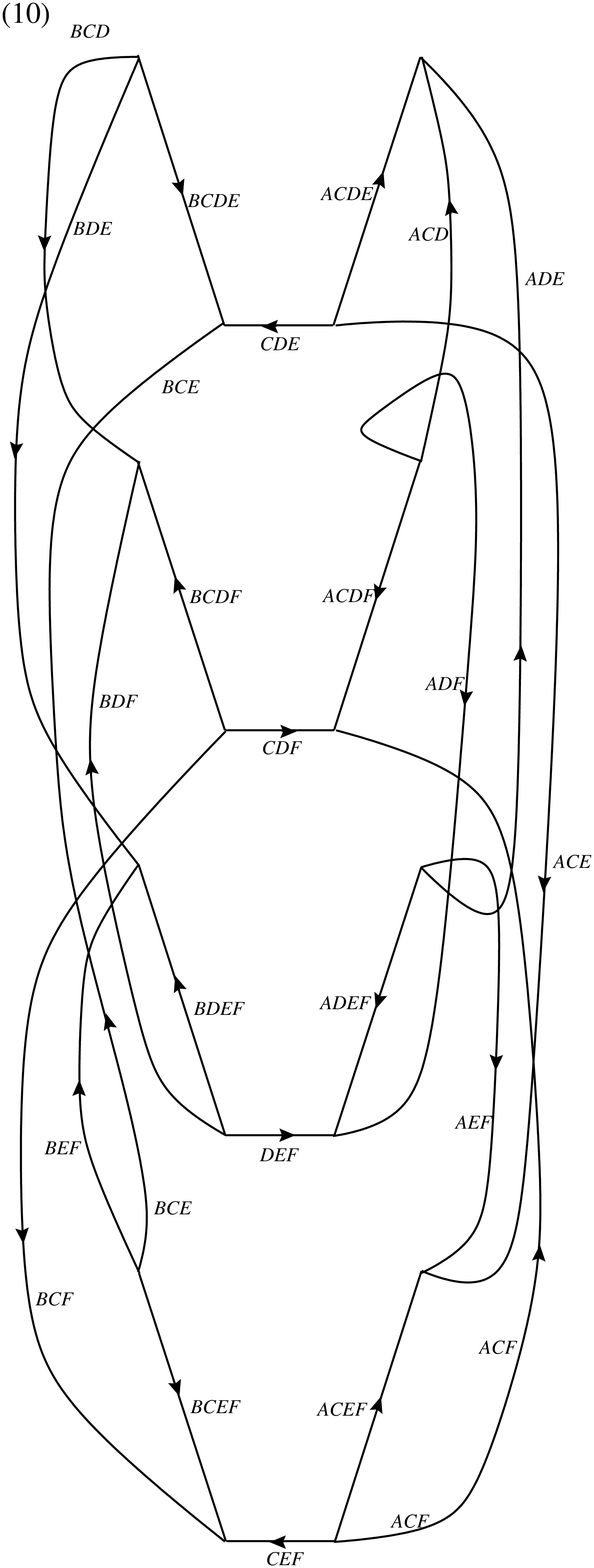}
  \end{minipage}
  \\
 \end{array}
\]
\[
 \begin{array}{cccc}
  = & \Lambda \cdot
  \begin{minipage}[c]{.42\textwidth}
   \epsfxsize=\textwidth \epsfbox{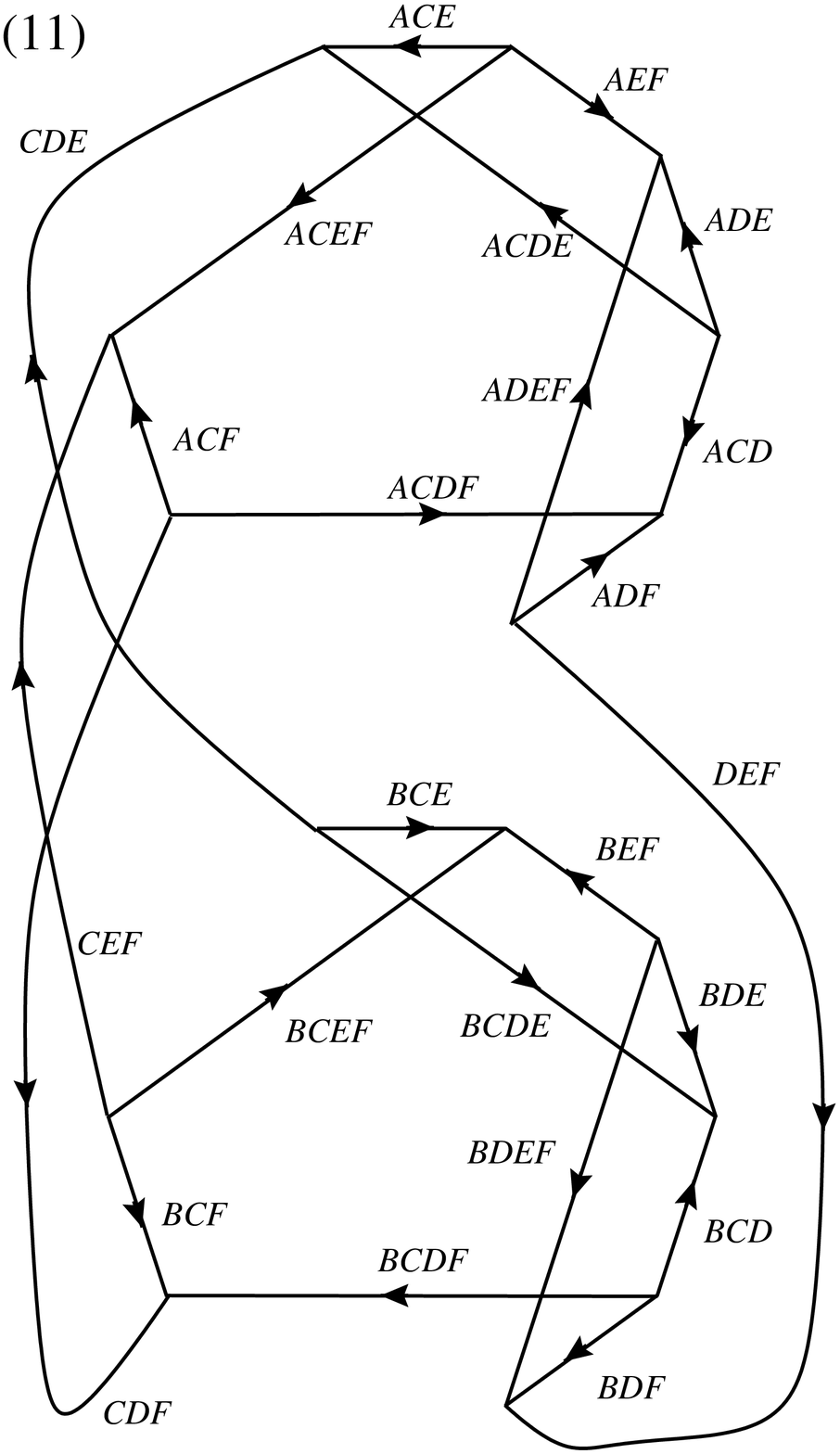}
  \end{minipage}
  & ~ = ~ & \Lambda \cdot
  \begin{minipage}[c]{.37\textwidth}
   \epsfxsize=\textwidth \epsfbox{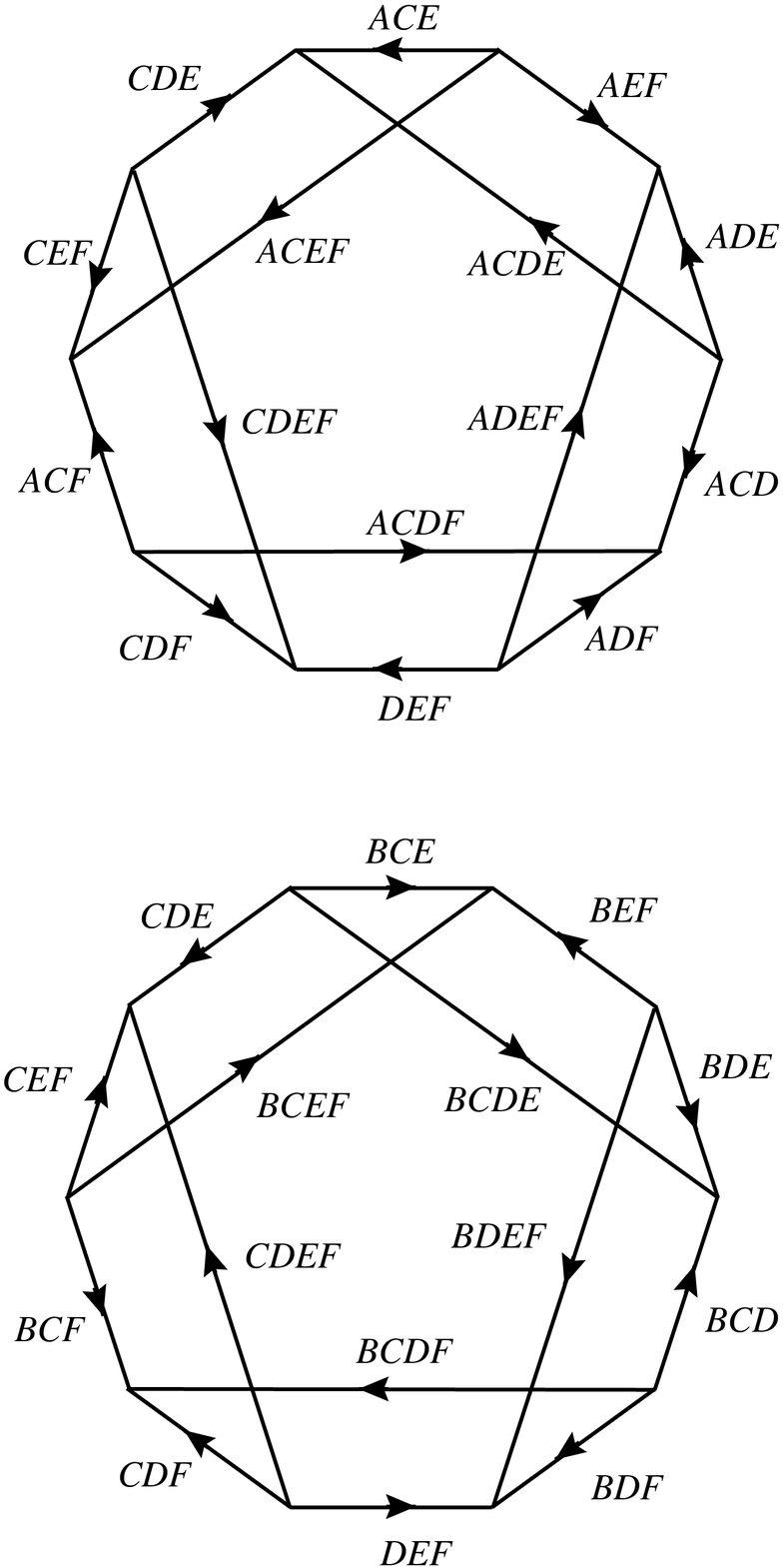}
  \end{minipage}
  \\
 \end{array}
\]

In the first equality we have used the formula (\ref{move_formula1}) 
for the internal lines of tetrahedra $ABCD$ and $ABEF$ to combine 
two diagrams (1) and (2) into (5) and (3) and (4) into (6), 
respectively. 
In the second equality we have again used the formula 
(\ref{move_formula1}) for the common internal lines $ABCE$ 
to combine the diagrams (5) and (6) into (7). 
In the third equality, we have used the formula (\ref{move_formula2})
to get rid of $ABE$ and then $ABDE$ successively in (7)
and reached the diagram (8).
Similarly in the forth equality
$ABC$ and $ABCF$ are removed from (8) into (9).
In the fifth equality we have used the relations 
(\ref{box1}) and (\ref{box2}) since there is a closed loop with two 
internal tetrahedra; $ABDF\rightarrow ABD\rightarrow 
ABDF\rightarrow ABF\rightarrow ABDF$ in the diagram (9).
In the sixth equality we rearranged the diagram (10)
in the form of diagram (11). 
In the last equality we have used again the formula (\ref{move_formula1}) 
in the inverse order to reproduce the internal line of tetrahedron $CDEF$.

We have thus shown that the product of initial four decuplet diagrams
is equivalent to the product of the final two decuplet diagrams
multiplied by the regularized constant $\Lambda$. 
Thus the product of the 15-$j$ symbols of the partition function 
(\ref{z_15j}) can be decomposed into another product by the 
4-2 move. 
The proof of 5-1 and 3-3 move invariance of the partition function 
proceeds quite parallel to the proof of 4-2 move. 
In the treatment of 5-1 move, we actually obtain
the regularization factor $\Lambda^4$,
while 3-3 move does not reproduce any regularization factor. 

We first note that in 4-2 move
there is a common link for the initial four 4-simplexes 
while there is a common site for the initial five 4-simplexes
in 5-1 move. 
We then ask how many powers of $\Lambda$ we need to be compatible 
with the Pachner moves. 
We first count the change of the number of $i$-simplexes
$\Delta N_i$ before and after the Pachner moves. 
We list the result in Table \ref{table:delta N}. 
\begin{table}[htbp]
\begin{center}
 \begin{tabular}{|c|c|c|c|c|c|} \hline
  & $\Delta N_0$ & $\Delta N_1$ & $\Delta N_2$ & $\Delta N_3$ & $\Delta N_4$\\
  \hline
  5-1 move & 1 & 5 & 10 & 10 & 4 \\  \hline
  4-2 move & 0 & 1 & 4 & 5 & 2\\  \hline
  3-3 move & 0 & 0 & 0 & 0 & 0 \\  \hline
 \end{tabular}
\end{center}
 \caption{The difference of the number of simplexes
 before and after the Pachner moves.}
 \label{table:delta N}
\end{table}

Since we have obtained $\Lambda^4$ in 5-1 move and $\Lambda$ in 
4-2 move, we need to obtain the following $a_i$ satisfying: 
\begin{equation}
\begin{array}{rcl}
 a_0 + 5a_1 + 10 a_2 +10 a_3 + 4 a_4 &=& 4, \\
 a_1 + 4 a_2 +5 a_3 + 2 a_4 &=& 1.
\end{array}
\label{arelations}
\end{equation}
Then the total power of $\Lambda$ of the 4-dimensional
simplicial manifold in consideration will be given by 
\begin{equation} 
\Lambda^{\sum^4_{i=0} a_i N_i}.
\end{equation} 
There are the following well known relations among the total 
number of $i$-simplex $N_i$ in 4-dimensional simplicial 
manifold\cite{TDLee}: 
\begin{eqnarray}   
N_0-N_1+N_2-N_3+N_4 &=& \chi, \nonumber \\
2 N_1-3 N_2+4 N_3-5 N_4 &=& 0, \label{Nrelations} \\
2N_3 &=& 5N_4, \nonumber 
\end{eqnarray} 
where $\chi$ is Euler number. 
Solving $N_2, N_3$ and $N_4$ in terms of $N_0$ and $N_1$ in 
(\ref{Nrelations}), we obtain the power of $\Lambda$ as 
\begin{equation}
\sum^4_{i=0} a_iN_i = N_0(a_0-10a_2-15a_3-6a_4)
                     +N_1(a_1+4a_2+5a_3+2a_4)
                     +\chi(10a_2+15a_3+6a_4).
\end{equation}
As far as $a_i$ satisfies the relation (\ref{arelations}), 
we can arbitrarily take the integer $a_i$. 
Taking the choice $a_2 = a_3 = a_4 =0$, we obtain Euler number 
independent solution of the $\Lambda$ power dependence, 
\begin{equation} 
\Lambda^{-N_0 + N_1}.
\end{equation} 

To summarize we obtain the Pachner move invariant partition function,
\begin{equation}
 Z_{LBF} =\sum_{\{J_i\}}
  \prod_{\hbox{\scriptsize site}} \Lambda^{-1}
  \prod_{\hbox{\scriptsize link}} \Lambda
  \prod_{\hbox{\scriptsize triangle}} (2J+1)
  \prod_{\hbox{\scriptsize tetrahedron}} (2J+1)
  \prod_{\hbox{\scriptsize 4-simplex}} \{ \hbox{15-$j$} \}.
\end{equation}

Finally we give arguments to determine the sign factors of the 15-$j$ 
symbols.
We first note that 
the decuplet graph of our 15-$j$ symbol with
any given sign factor configuration
is equal to the one that has reversed all sign factors, 
\begin{equation}
 \begin{minipage}[c]{.3\textwidth}
  \epsfxsize=\textwidth \epsfbox{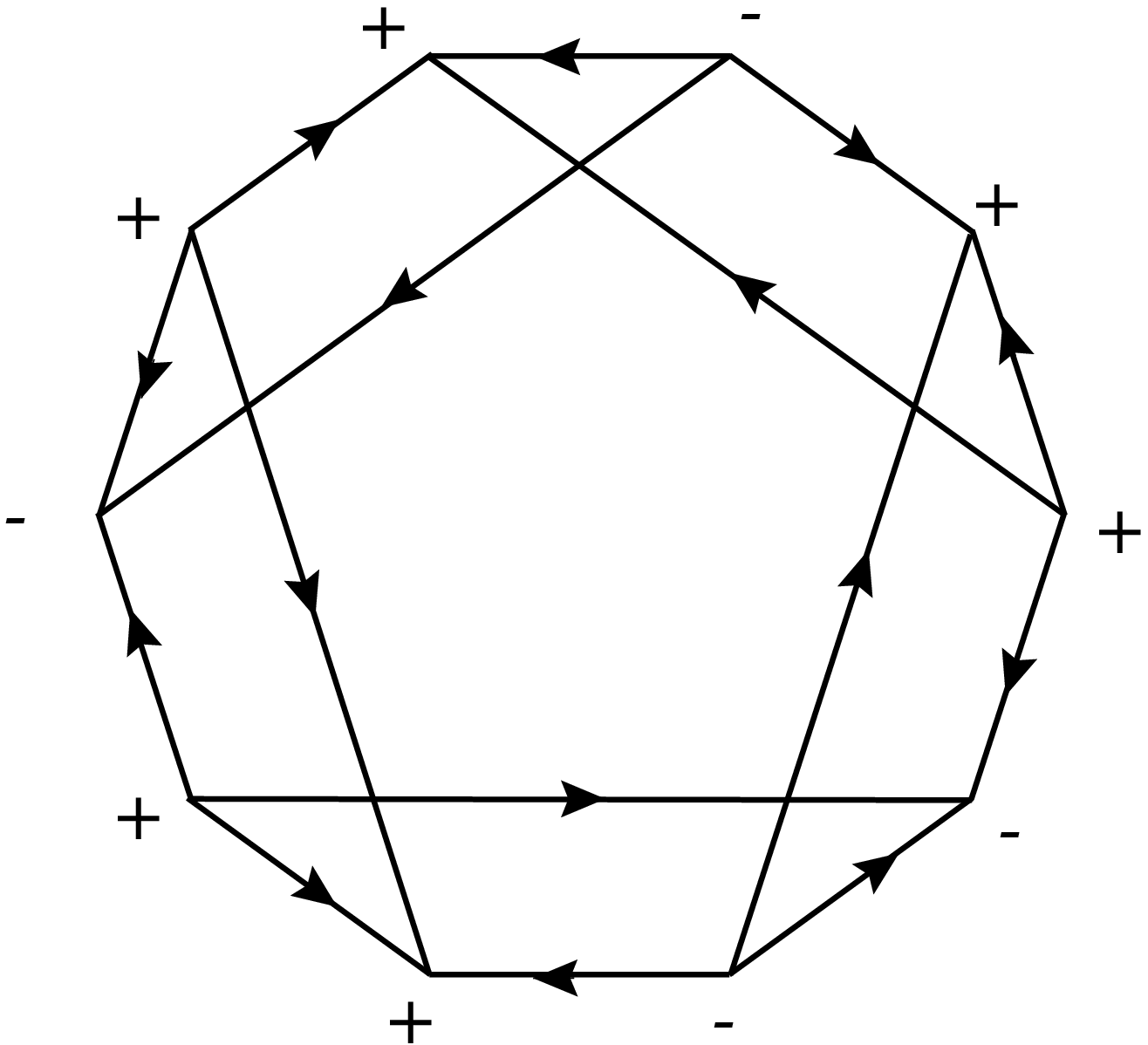}
 \end{minipage}
\quad = \quad
 \begin{minipage}[c]{.3\textwidth}
  \epsfxsize=\textwidth \epsfbox{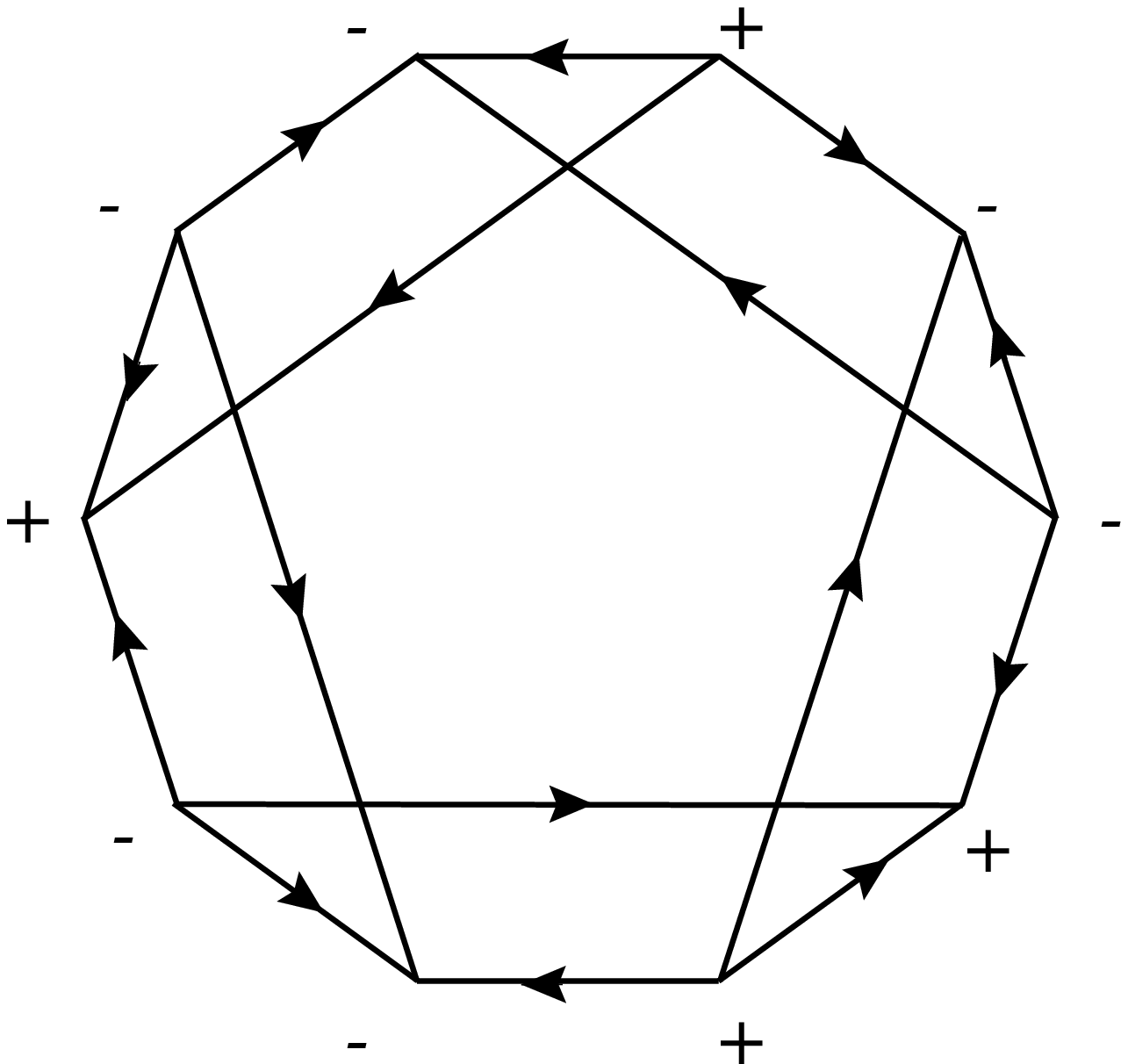}
 \end{minipage}.
\label{sign-reverse}
\end{equation}
We can thus take the sign factor of one particular vertex $+$
without loss of generality. 
We then take the following special choice of one sign factor for the 
initial four 4-simplexes of 4-2 move:
\begin{center}
\begin{minipage}[b]{.8\textwidth}
  \epsfxsize=\textwidth \epsfbox{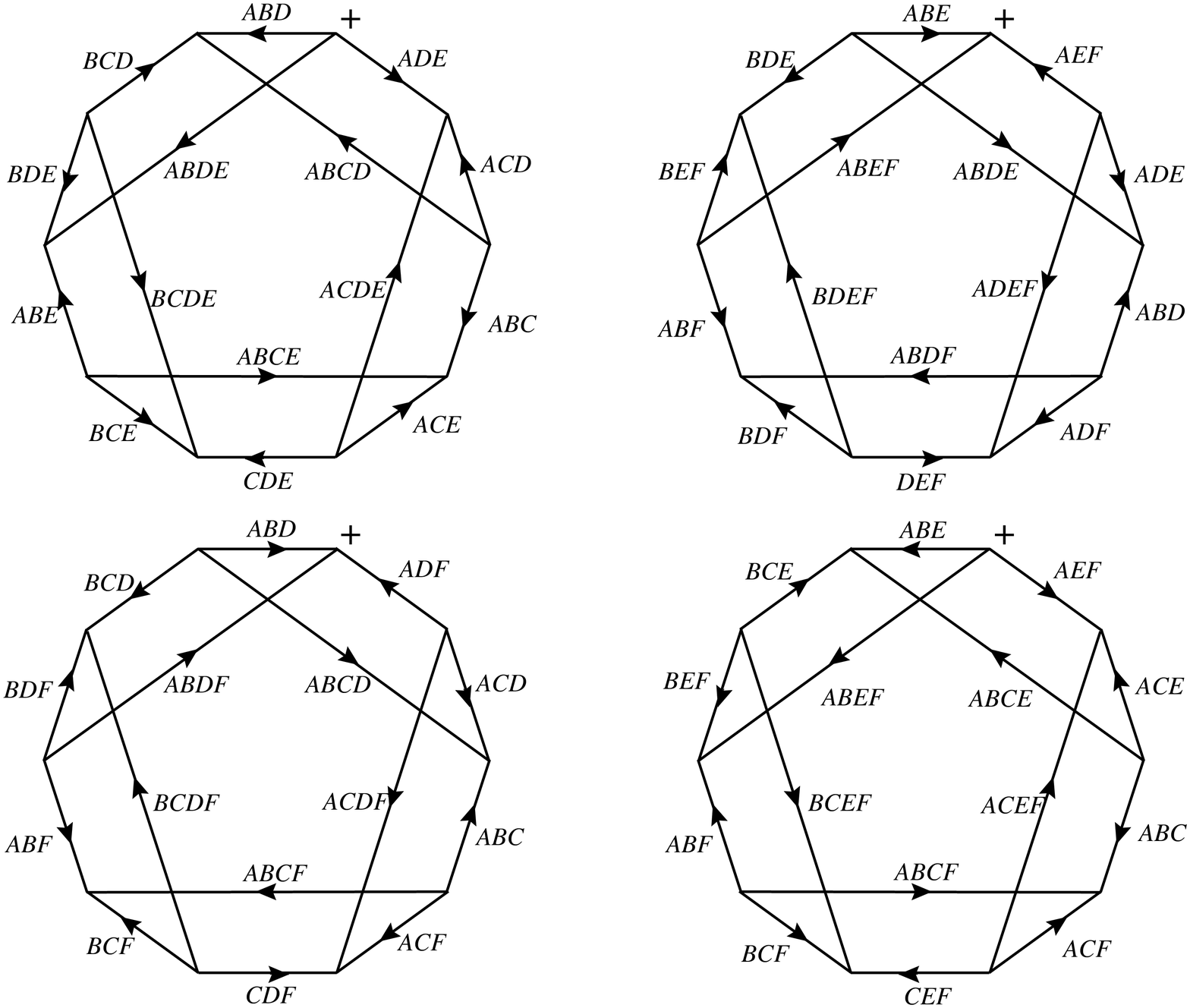}
\end{minipage}.
\end{center}

When we consider an arbitrary 4-dimensional simplicial manifold composed 
of 4-simplexes, any of two neighboring 4-simplexes have 
one boundary tetrahedron in common. 
In the corresponding expression of this simplicial manifold by the 
generalized 15-$j$ symbols, there are thus always the same pair of 
internal lines in the decuplet diagrams. 
As we have already mentioned in (\ref{I_BCDE1}) and (\ref{I_BCDE2}), 
the sign factors of 
two vertices associated to the same tetrahedron should be the same. 
Using this fact, we can determine the other sign factors of initial 
four 4-simplexes of 4-2 move up to the five unknown sign ambiguities 
denoted by $\alpha, \beta, \gamma, \delta$ and $\epsilon$   
\begin{center}
\begin{minipage}[b]{.8\textwidth}
  \epsfxsize=\textwidth \epsfbox{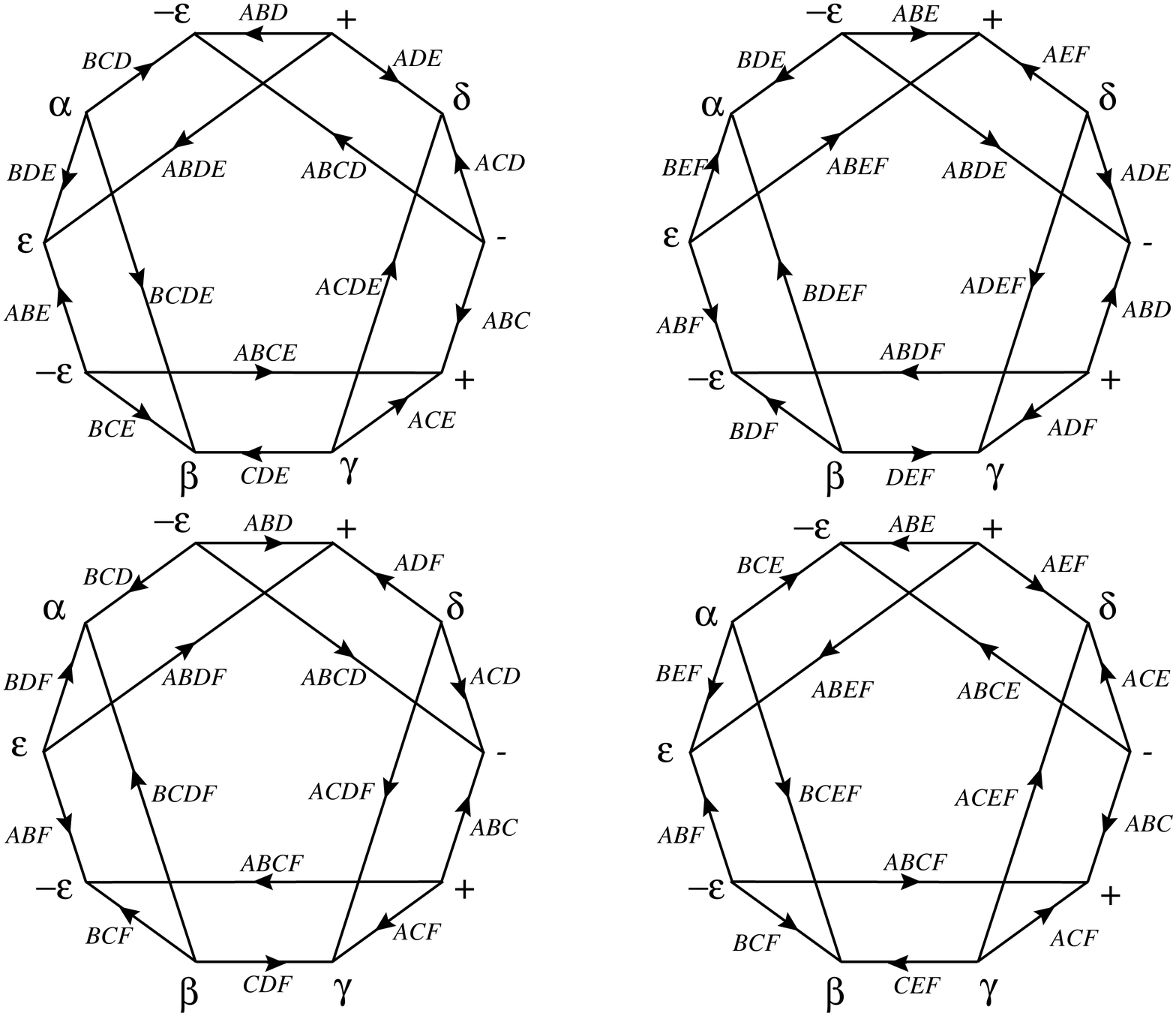}
\end{minipage}.
\end{center}

The final possible sign configuration of two 4-simplexes in 4-2 move can 
be found by tracing the graphical manipulations
and identifying the signs of common vertices, 
\begin{equation}
\begin{minipage}[c]{.8\textwidth}
 \epsfxsize=\textwidth \epsfbox{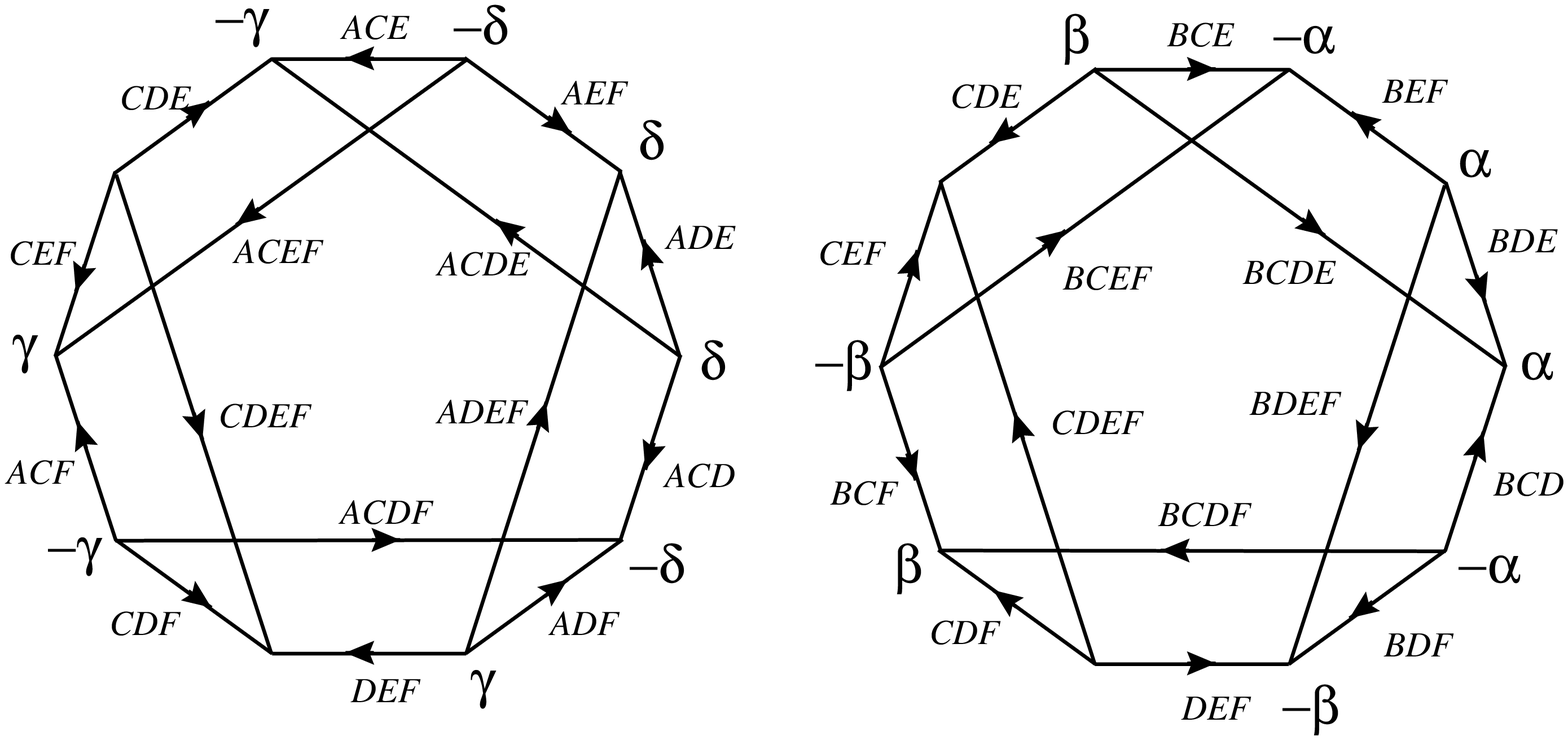}
\end{minipage}.
\end{equation}

The sign configuration of the decuplet graphs before and after the 
4-2 move should coincide since the overall boundary of the final 
two 4-simplexes have the same boundary tetrahedra as that of the 
initial four 4-simplexes as shown in (\ref{boundary_relation1}).
This requirement leads to the following relations: 
\begin{equation}
 \alpha = \delta = -, \quad \beta = -\gamma = -\epsilon. 
\end{equation}
We have thus obtained several constraints from the consistency of 
4-2 move. 

We can use the similar arguments as 4-2 move for 5-1 and 3-3 moves 
to get constraints on the sign ambiguity. 
It turns out that 4-2 move and 5-1 move eventually give the same 
constraint, 
\begin{equation}
 \begin{minipage}[c]{.3\textwidth}
  \epsfxsize=\textwidth \epsfbox{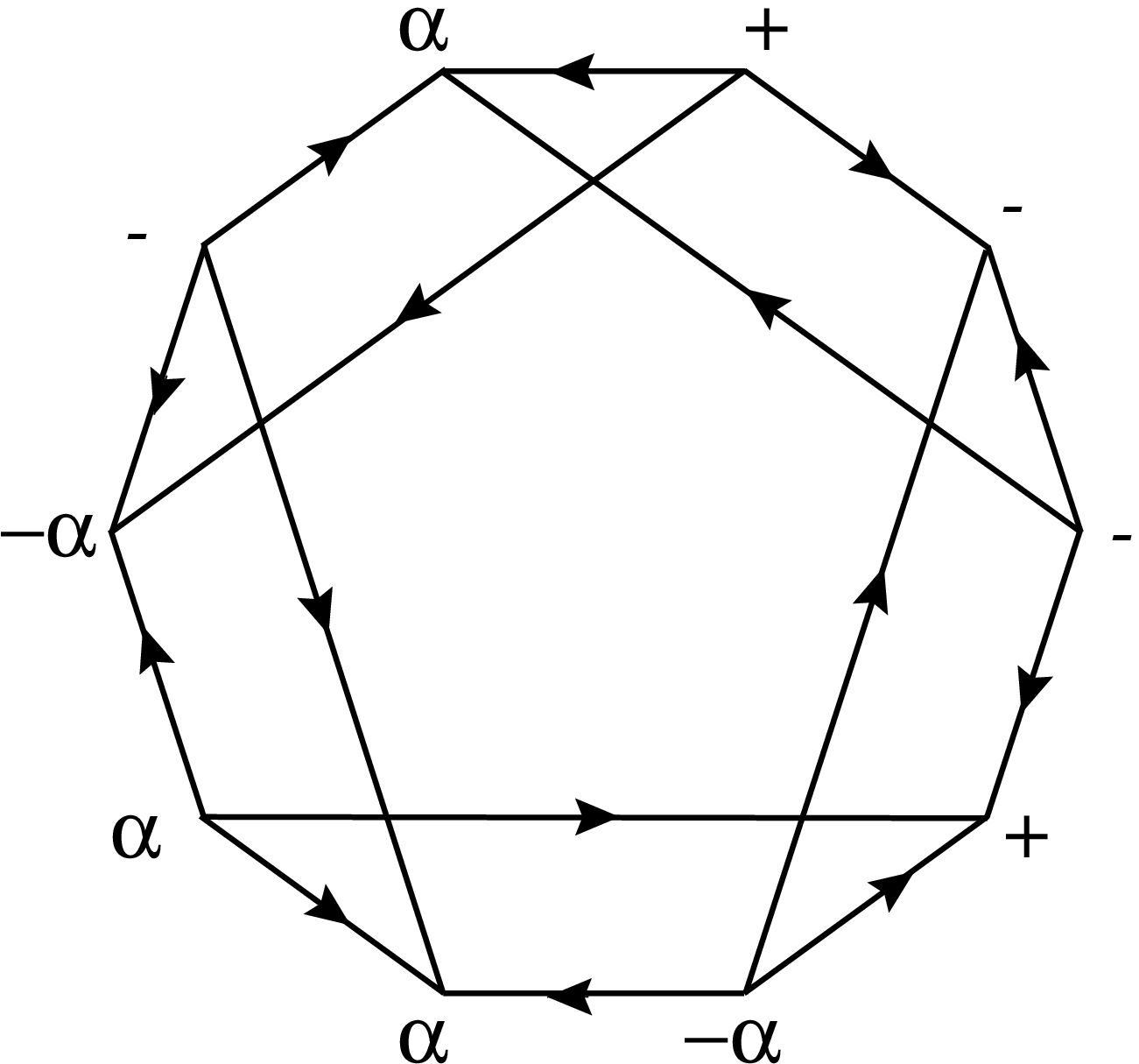}
 \end{minipage} ,
\label{signof42_51}
\end{equation}
while 3-3 move give the following constraint: 
\begin{equation}
 \begin{minipage}[c]{.3\textwidth}
  \epsfxsize=\textwidth \epsfbox{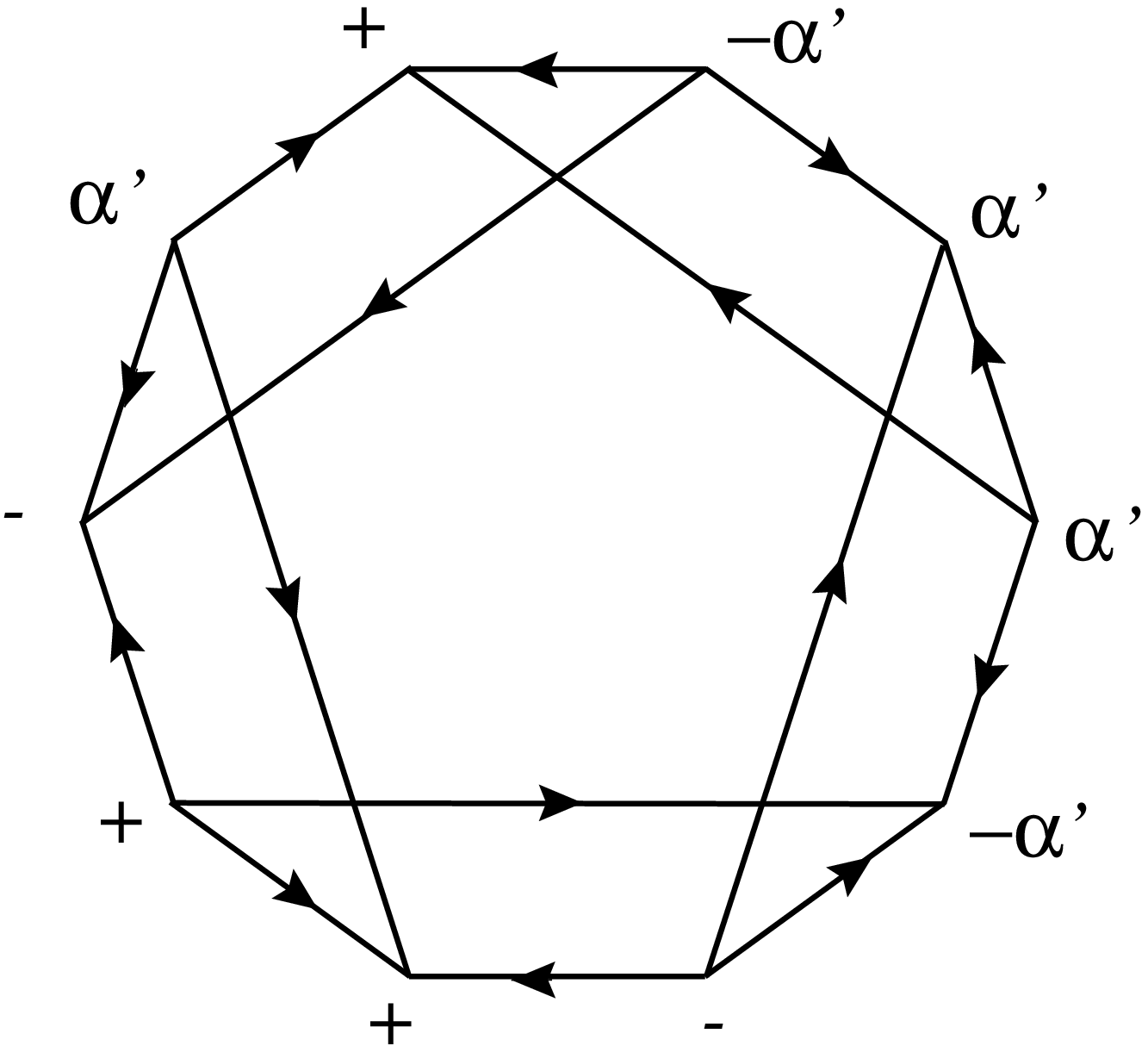}
 \end{minipage}.
\label{signof33}
\end{equation}
There are two consistent solutions satisfying (\ref{signof42_51}) 
and (\ref{signof33})
\begin{equation}
\begin{minipage}[c]{.3\textwidth}
  \epsfxsize=\textwidth \epsfbox{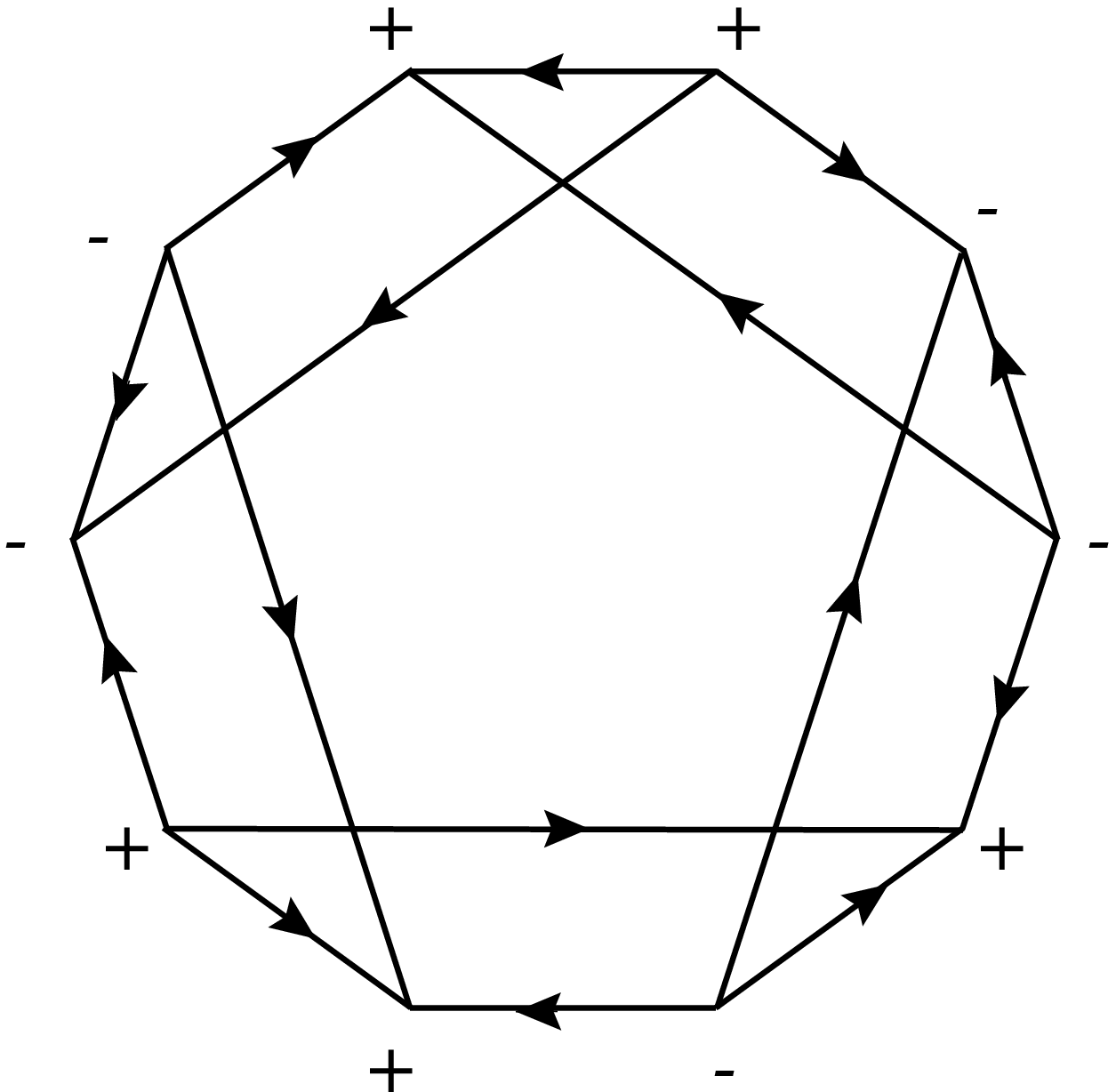}
\end{minipage}
\quad , \quad
\begin{minipage}[c]{.3\textwidth}
  \epsfxsize=\textwidth \epsfbox{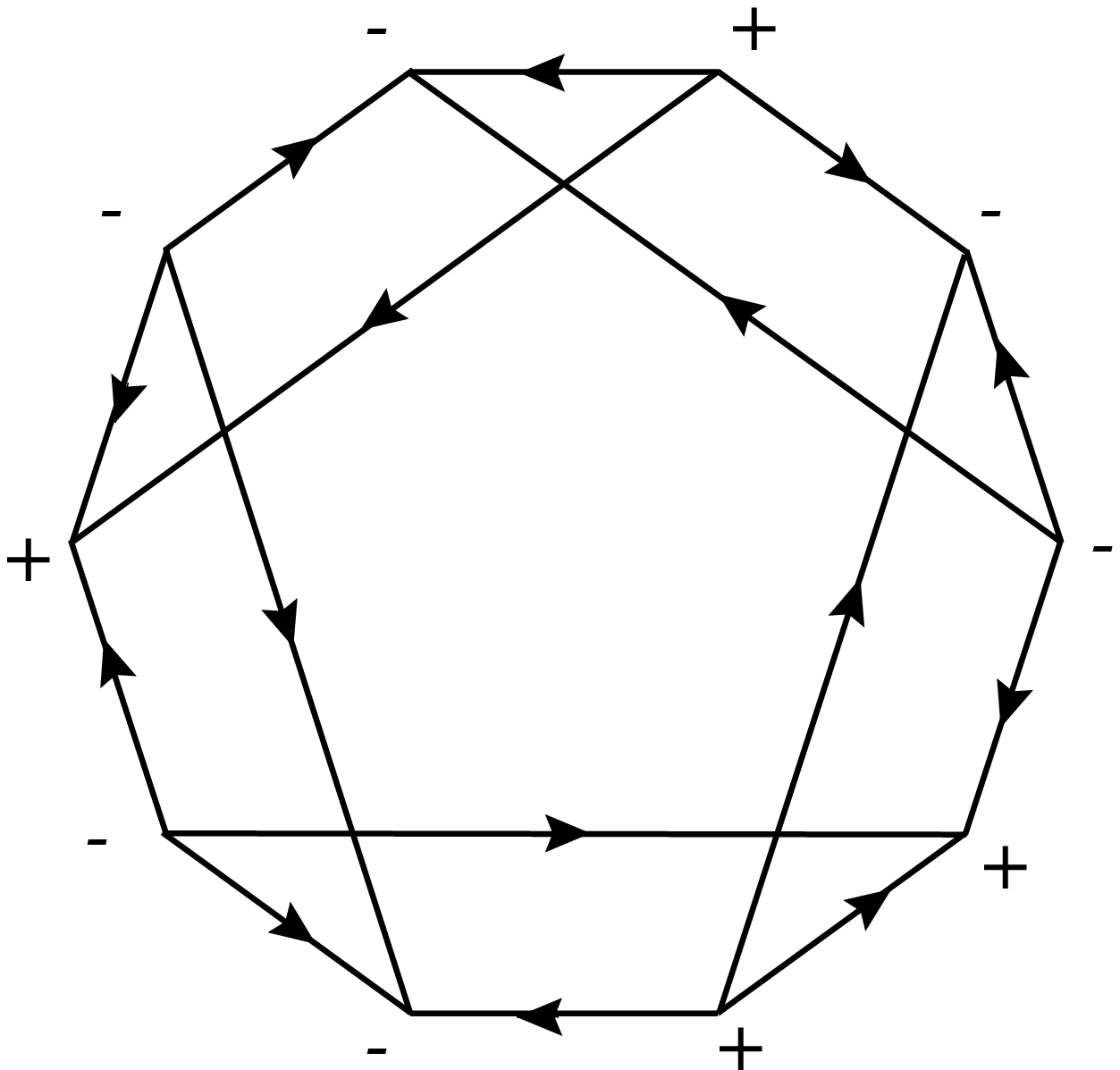}
\end{minipage},
\label{final-sign}
\end{equation}
where the first solution corresponds to $\alpha = +$ and $\alpha' = -$
while the second solution is obtained by taking 
$\alpha = -$ and $\alpha' = +$ with totally reversed sign factors.
At this stage we point out that our generalized 15-$j$ symbol
is different in sign factors from the one 
given by Crane and Yetter\cite{Crane-Yetter}.

We have thus obtained complete expression of the generalized 
15-$j$ symbol with the correct sign factors. 
Then the full analytic expression of the Pachner move invariant 
partition function is  
\begin{equation}
 Z_{LBF} = \sum_{J}
  \prod_{\hbox{\scriptsize site}} \Lambda^{-1}
  \prod_{\hbox{\scriptsize link}} \Lambda
  \prod_{\hbox{\scriptsize triangle}} (2J+1)
  \prod_{\hbox{\scriptsize tetrahedron}} (2J+1)
  \prod_{\hbox{\scriptsize 4-simplex}}
 \left\{
  \begin{array}{ccccc}
   J_1 & J_2 & J_3 & J_4 & J_5 \\
   J_6 & J_7 & J_8 & J_9 & J_{10} \\
   J_{11} & J_{12} & J_{13} & J_{14} & J_{15} 
  \end{array}
 \right\},
\end{equation}  
where we have used the first sign convention of (\ref{final-sign}).
Here the 15-$j$ symbol can be defined by
\begin{eqnarray}
 \left\{
  \begin{array}{ccccc}
   J_1 & J_2 & J_3 & J_4 & J_5 \\
   J_6 & J_7 & J_8 & J_9 & J_{10} \\
   J_{11} & J_{12} & J_{13} & J_{14} & J_{15} 
  \end{array}
 \right\}
  &=&
  \sum_{\hbox{\scriptsize all } m_i}
  (-)^{\sum_{i=1}^{15}(J_i-m_i)}
  \threej{J_1}{J_7}{J_6}{m_1}{m_7}{m_6}
  \threej{J_3}{J_8}{J_7}{-m_3}{-m_8}{-m_7} 
 \nonumber \\ &\times&
  \threej{J_4}{J_8}{J_9}{m_4}{m_8}{m_9}    
  \threej{J_1}{J_9}{J_{10}}{-m_1}{-m_9}{-m_{10}}
  \threej{J_2}{J_{11}}{J_{10}}{m_2}{m_{11}}{m_{10}} 
 \nonumber \\ &\times&
  \threej{J_4}{J_{11}}{J_{12}}{-m_4}{-m_{11}}{-m_{12}}
  \threej{J_5}{J_{13}}{J_{14}}{m_5}{m_{13}}{m_{14}}
  \threej{J_2}{J_{14}}{J_{13}}{-m_2}{-m_{14}}{-m_{13}}
 \nonumber \\ &\times&
  \threej{J_3}{J_{14}}{J_{15}}{m_3}{m_{14}}{m_{15}}
  \threej{J_5}{J_{15}}{J_6}{-m_5}{-m_{15}}{-m_6},  
\end{eqnarray}
which is graphically equivalent to 
$$
 \begin{minipage}[c]{.3\textwidth}
  \epsfxsize=\textwidth \epsfbox{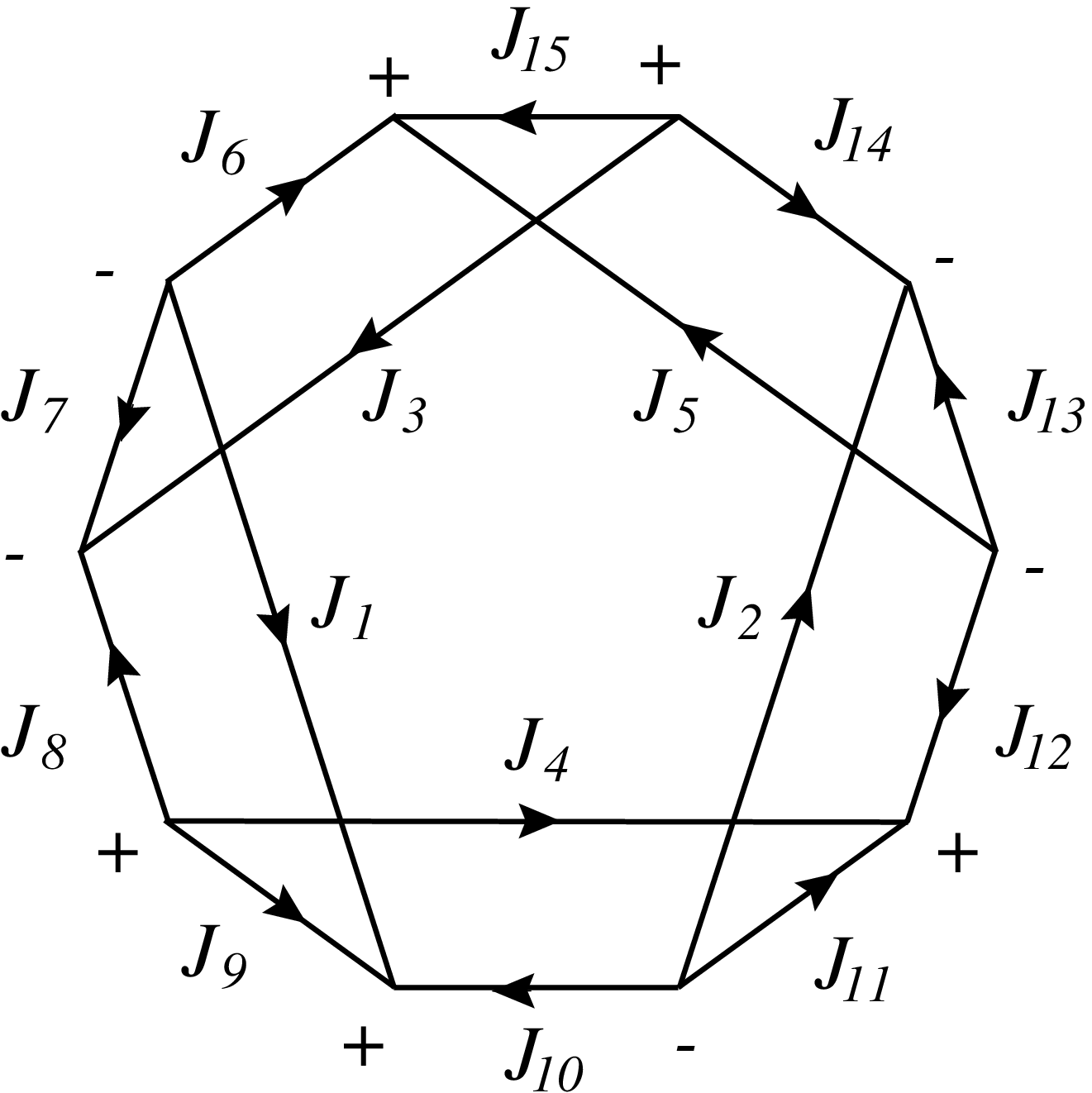}
 \end{minipage}.
$$
Another sign convention in (\ref{final-sign})
can be equally used to express the generalized 15-$j$ symbol.

\setcounter{equation}{0}
\renewcommand{\theequation}{\arabic {section}.\arabic{equation}}

\section{Conclusion and Discussions}

We have shown that the partition function of the lattice version 
of $BF$ action with vanishing holonomy constraint leads to the product of 
15-$j$ symbols with some particular 
combinations of angular momentum factors and regularized constants. 
We have then explicitly proved by graphical method that the partition 
function is Pachner move invariant if we take particular choice of 
sign assignments for the trivalent vertices of decuplet diagrams 
corresponding to the 4-simplexes. 
This means that the partition 
function is independent of how the simplicial manifold is divided by 
4-simplexes. 
This is to do with the fact that the partition function of the lattice 
version of $BF$ action is diffeomorphism invariant or even topological 
invariant in a stronger sense. 
Since the partition function is invariant how much finely we divide the 
4-dimensional simplicial manifold in consideration, the continuum limit 
of the lattice $BF$ action is expected to approach naively to the continuum 
$BF$ action. 
The vanishing holonomy constraint can be then interpreted as a gauge fixing 
condition to align the $B^a$ field and the curvature $F^a$. 

There are some desirable features in this formulation. 
Firstly we have found the lattice formulation of ``4-dimensional gravity'' 
which is first order formalism and is thus described by local 
variable $B$ and spin connection $U=e^{i\omega}$. 
This is the first 4-dimensional example of lattice gravity model which 
specifies the 
location of the local fields $B$ which could possibly be related to the metric 
and the lattice spin connection $U=e^{i\omega}$ on the 4-dimensional 
simplicial manifold. 
This lattice formulation is a beautiful unification of the idea of 
lattice QCD and the formulation of Regge calculus in the sense that 
the curvature generated by the product of link variables are concentrated 
on the simplexes where Regge calculus claims
as the location of gravitational curvature.
Secondly the magnitude of the $B$ field is naturally discretized due to 
the logarithmic form of our lattice $BF$ action. 
Thus the discreteness of the area of triangles on the original lattice is a 
natural consequence of the action.
It should be compared with the result 
of Rovelli and Smolin who claimed to get the discreteness of the area 
of triangles in the form $\sqrt{J(J+1)}$ from the 
analysis of an area operator\cite{Rovelli-Smolin}.

As we have already pointed out that the identification of 
$B=*(e\wedge e)$ renders the $BF$ action into the Palatini type of 
Einstein-Hilbert action. 
On the other hand in our $BF$ action we have taken the gauge algebra as 
one of the chiral counter part of $SO(4)$ algebra in the decomposition 
of the Euclidean local Lorentz group, $SO(4) \simeq SU(2)_L\times SU(2)_R$. 
There are several possibilities how the chiral decomposition of the 
local Lorentz group be related to the realistic Einstein-Hilbert 
action\cite{Plebanski}.
Here we simply comment an interesting possibility that the action 
$\int_M B^+_aF^+_a$ with the gauge group as 
one of the chiral partner $SU(2)_L$ will be formulated 
on the original lattice as in 
this paper while the action $\int_M B^-_aF^-_a$ with the other chiral 
partner $SU(2)_R$ gauge group will be formulated on the dual lattice. 
This situation can be symbolically be written 
\begin{equation}
Z_{LBF}[SO(4)] = Z_{LBF}[SU(2)_L] \cdot Z_{LBF}[SU(2)_R]. \nonumber
\end{equation}
In order that this theory be realistic lattice gravity model, we need 
some natural constraint to relate the two chiral $BF$ models. 
As we have mentioned in the end of section 3, the vanishing holonomy 
constraint (\ref{constraint1}) includes 9 relations for the chiral 
partner of the $SO(4)$ algebra, which is just the necessary irreducible 
degrees of freedom for the 1-form gauge parameter. 
In order to count the number of the constraint, we have assumed the chiral 
nature even for the spacetime suffix in addition to the local Lorentz 
gauge suffix. 
In fact there are 9 other constraints in the vanishing holonomy constraint, 
which may correspond to the number of constraints of the spacetime suffix of 
the other chiral partner. 
It is interesting to note that the vanishing holonomy constraint 
(\ref{constraint1}) may include both constraint in one relation, which 
might give some clue to find a realistic lattice gravity model. 

Our proposed lattice $BF$ action (\ref{LBF}) can be generalized into 
arbitrary dimensions. 
In fact $ISO(3)$ Chern-Simons gravity was formulated by the 3-dimensional 
version of lattice $BF$ action. 
We claim that $D$-dimensional lattice $BF$ action has the universal structure 
\begin{equation}
 S_{LBF}
  = \sum_{x} \tr \left( X(x) \Bigl[ \ln \plaq \Bigr] \right), \nonumber
\end{equation}
where $X(x)$ is ($D-2$)-form located on the center of 
($D-2$)-simplex, $x$, in the $D$-dimensional simplicial manifold. 
The lattice curvature term is defined in the similar way as the 
3- and 4-dimensional cases. 
It should be noted that the center of the dual plaquette $\tilde{P}$ 
coincides with the center of the simplex $x$. 
The vanishing holonomy constraint may have the form either 
the type of (\ref{econstraint}) or the type (\ref{constraint1}) 
depending on the suffix structure of gauge group for $X$. 
It is worth to recognize at this stage that this lattice $BF$ action 
is the lattice version of the leading term in the generalized Chern-Simons 
action in $D$ dimensions. 

In our lattice gravity formulation in 3 and 4 dimensions, 
1- and 2-form fields are, respectively, introduced on 
1- and 2-simplexes in the simplicial manifold: $e^a$ and $B^a$ are located 
on the link (1-simplex) and the triangle (2-simplex) of original lattice, 
respectively, while $U=e^{i\omega}$ is located on the dual link (1-simplex). 
There is thus natural correspondence between the form degrees and the simplex 
numbers. 
The generalized gauge theory which is formulated to include all form degrees 
may thus very naturally be fitting to the formulation of gauge theory on the 
simplicial manifold. 

We know that there is no dynamical graviton in the topological 
model while we need the graviton as a dynamical field in the realistic 
4-dimensional Einstein gravity. 
We believe that matter fields should be introduced to accommodate 
the dynamical 
degrees of freedom and thus would change the topological nature of the 
model. 
We know by now that the ghosts introduced by the 
quantization will be changed into fermionic matter fields via twisting 
mechanism\cite{Witten-TFT}.
In fact we found a mechanism that $R$ symmetry of $N=2$ supersymmetry in 
the twist is essentially related to the Dirac-K\"ahler fermion 
formulation\cite{K-T}. 
The Dirac-K\"ahler fermion needs all form degrees of freedom and 
can naturally be put on the simplicial manifold. 

It will be thus interesting to investigate lattice gravity formulation with 
matter fields on the simplicial manifold. 
We know that weak boson gauge field and Higgs field can be very 
naturally accommodated into the generalized gauge theory formulation 
and thus this kind of formulation possibly leads to the unified model 
on the simplicial manifold\cite{Nishi}.  

In our formulation we have introduced cut off dependent regularized constant. 
It is well known by the work of Turaev and Viro that the regularization of 
the Ponzano-Regge model can be naturally accommodated if we introduce the 
$q$-deformed formulation of 3-$j$ symbol\cite{Turaev-Viro}.
We didn't mention the $q$-deformed version of the present formulation 
in this paper although it is the main subject among mathematicians 
to obtain the 4-dimensional topological invariant
by the $q$-deformed formulation\cite{Crane-K-Yetter}. 
Since our 15-$j$ symbol is explicitly related to the 3-$j$ symbol which has 
the $q$-deformed counterpart, it will be straightforward to generalize our 
formulation into $q$-deformed version, which is expected to correspond with 
$BF$ gravity with cosmological term\cite{Ooguri-Sasakura}\cite{Baez}.  

\vskip 1cm

\noindent{\Large{\bf Acknowledgements}} \\
We thank L. Crane and H.B. Nielsen for useful discussions. 
This work is supported in part by Japanese Ministry of Education, Science, 
Sports and Culture under the grant number 09640330.

\vspace*{1cm}

%
%

\end{document}